\documentclass[a4paper,11pt]{article}
\usepackage{jheppub} 
\usepackage{lineno}
\usepackage{amsfonts}
\usepackage{amssymb}
\usepackage{upgreek}

\usepackage[colorinlistoftodos]{todonotes}
\usepackage{multicol}
\usepackage{graphicx}
\usepackage{subcaption}
\usepackage{wrapfig}
\usepackage[export]{adjustbox} 

\usepackage{natbib} 

\newcommand{\sign}{\text{sign}}

\def\beq{\begin{equation}}
\def\eeq{\end{equation}}
\def\bea{\begin{eqnarray}}
\def\eea{\end{eqnarray}}
\def\nn{\nonumber}

\title{\boldmath Enhancement of Harvesting Vacuum Entanglement in Cosmic String Spacetime}

\author[1]{Willy Izquierdo}
\affiliation[1]{Instituto de Física Teórica, Universidade Estadual Paulista\\ Rua Dr. Bento Teobaldo Ferraz, 271 - Bloco II, São Paulo - SP, CEP 01140-070, Brazil}

\author[3]{J. Beltran}
\affiliation[2]{Facultad de Ciencias, Universidad Nacional de Ingenieria\\
Av. Túpac Amaru 210, Rímac - Lima 1301, Perú.}

\author[2]{Enrique Arias}
\affiliation[3]{Instituto Politécnico, Universidade do Estado do Rio de Janeiro\\Rua Bonfim 25, Vila Amélia, Nova Friburgo - RJ, CEP 28.625-570, Brazil.}

\emailAdd{earias@iprj.uerj.br}

\abstract{We analyze the entanglement generation in a pair of qubits that experience the vacuum fluctuations of a scalar field in the Cosmic String spacetime. 
The qubits are modeled as Unruh-DeWitt detectors coupled to a massless scalar field. 
We introduce a Heisenberg $XY$-interaction between the qubits that enhances the generation of quantum correlations. 
It is supposed that the qubits 
begin at a general mixed state described by a density operator with no entanglement while the field stays at its vacuum state.  
In this way, we find the general properties and conditions to create entanglement between the qubits by exploiting the field vacuum fluctuations. 
We quantify the qubits entanglement using the Negativity measure based on the Peres-Horodecki positive partial transpose criterion. We find that the Cosmic String would increase the entanglement harvesting when both qubits are near the Cosmic String. When the qubits locations are far from the Cosmic String we recover the usual results for Minkowski space. The Heisenberg $XY$-interaction enhance the entanglement harvesting irrespective of the coupling nature (ferromagnetic or anti-ferromagnetic). When the qubits are far apart from each other we find a maximum entanglement harvesting at the resonance points between the Heisenberg coupling constant and the qubits energy gap.

}

\begin{document}
\maketitle
\flushbottom

\section{Introduction}
Quantum entanglement is a fundamental phenomenon that lies at the heart of Quantum Mechanics. Since it was established as a real consequence of the superposition principle, by Schrodinger in 1935, this concept has defied the classical intuitions of locality and reality \cite{RevModPhys.81.865}. Nevertheless, after a first period of confusion, we can say that, in the modern days, entanglement is a well-accepted aspect of the physical world at the quantum level.  Its counterintuitive properties have contributed to the development of a new paradigm for technology development. In this way, quantum entanglement is at the basis of quantum information, quantum communication, and quantum metrology. Quantum computers can outperform classical computers by using qubits, instead of bits, to process vast amounts of information simultaneously \cite{Preskill2018}.  Entanglement can be seen as a resource for protocols of quantum teleportation and quantum key distribution that provide secure transmission of data \cite{BENNETT20147}. Also, entanglement can enhance sensor precision measurements and can be exploited to simulate complex systems efficiently \cite{quant_simu_RevModPhy2014}. Hence, the study of entanglement in different scenarios is of great importance in order to understand the fundamentals of quantum mechanics and also to exploit its practical applications.

%

The analysis of quantum entanglement becomes even more intriguing within the context of curved spacetime. In this case, the interplay between gravity and quantum mechanics can provide some new and significant results about the entanglement dynamics and also for the properties of spacetime itself. In this sense, gravitational decoherence would be a critical line of study in order to ensure the use of quantum technologies in space \cite{grav_decohe_BLHU2022}. On the other side, Hawking radiation is related to the creation of entangled particle-antiparticle pairs at the event horizon and the black hole Bekenstein entropy could be related to entanglement entropy \cite{BH_entanglement_entropy}.  Furthermore, it has been suggested in the literature that a connection between wormholes and entanglement could be at the core of the causal structure of spacetime \cite{maldacena2013}. It is also worth noting that the spatial boundary effect could alter the entanglement dynamics of a system \cite{arias2016}. From this perspective, the geometry and topology of spacetime can be analyzed by investigating the entanglement dynamic in this background.

Before discussing the case of curved spacetime, it is worth noting that by using quantum field theory in flat spacetime and the Unruh effect for Rindler observers, some works have pointed out that the quantum field vacuum is an entangled state for causally disconnected partners \cite{SUMMERS1985,reznik_2003}.
In other words, the vacuum state represents a potential source of entanglement for two causally disconnected probes. This led to the possibility of extracting entanglement from the quantum vacuum, which was called \emph{Entanglement Harvesting} \cite{MartinMartinez_2012}.
Entanglement harvesting mainly consists of setting up two causally disconnected probes (Unruh-De Witt detectors) in a nonentangled state, that interact with the vacuum of a free quantum field for a finite time. Then, it is found that the final state of the detectors becomes entangled due to vacuum fluctuations, and its degree of entanglement is measured \cite{PhysRevA.71.042104,PhysRevA.79.012304}.
There are different factors that affect entanglement harvesting such as the curvature of space-time \cite{PhysRevD.79.044027,MartinMartinez_2012,Mart_n_Mart_nez_2016,MartinMartinez_2014}, the temperature of bath reservoir \cite{PhysRevD.79.044027,PhysRevA.72.062324}, the state of motion of the detectors \cite{PhysRevD.74.085031}, as well as its coupling with the quantum field and also the type of field spin with which the detectors are coupled.

In this paper, we focus on the effects that spacetime curvature and non-trivial topology can produce on the entanglement harvesting process. We consider as a gravitational background the spacetime of a cosmic string. The cosmic string is a topological defect that could have formed during a symmetry-breaking phase transition in the early universe \cite{NIELSEN197345,Kibble1,smith1990classical,cosmic_string_book,Kibble2}. The spacetime generated by this cosmic string is a flat space with a non-trivial topology characterized by an angle deficit around an infinitely long, straight cosmic string, which when embedded in Cartesian space looks like a conical space \cite{PhysRevD.23.852,PhysRevD.31.3288,anderson2015mathematical}. We analyze the entanglement dynamics of two Unruh-De Witt detectors that interact with the quantum vacuum of a scalar field defined over this spacetime. We consider also that the qubits (detectors) interact directly between them by a Heisenberg $XY$-interaction term.
We explore the influence of the Heisenberg interaction to maximize the degree of entanglement extracted from the vacuum. It is noted that a phase transition occurs for some definite value of the interaction constant that resonates with the qubits energy gap. This possibility of enhanced entanglement harvesting has not been explored so far.

The plan of this paper is according to the following structure. In Section 2, we briefly review the quantum field theory in the cosmic string curve spacetime, mainly determining the field modes and the Wightman function. In Section 3, we introduce the model of Unruh-De Witt detectors, and the effective dynamics of the qubits in interaction with quantum vacuum is elaborate. In Section 4, we develop in detail the entanglement measurement through the use of Negativity \cite{PhysRevA.58.883}.
Finally, in Section 5, we analyze the negative results for the cases in which the qubits are stationary in defined locations. 
We study the symmetric cases where the qubits' relative distance is axial, angular, or radial with respect to the cosmic string. In all these scenarios we analyze the entanglement dependence on the parameters that characterized the dynamics as interaction time, coupling constant, and relative distance between the qubits or to the cosmic string, for example. We consider a system of units such that $\hbar=c=1$.

\section{Quantum Field Theory in Cosmic String Spacetime}
\label{sec:qft_in_cosmic_string}

Quantum field theory (QFT) in curved spacetime is a semiclassical approximation of what a complete theory of Quantum Gravity would be. In this approach, quantum fields propagate over some curve spacetime that is described by the classical gravitational field equations of general relativity \cite{HOLLANDS20151, Fulling}.
In the process of formulating this theory, for general curved backgrounds, some key problems arise.
For example, the physical interpretation of the notion of particles through a unique definition of the vacuum state becomes a problematic, ambiguous, and observer-dependent concept. This is because, in Minkowsky space the vacuum is defined as an invariant state under the Poincare transformations. However, due to the generality of curved spacetimes, general manifolds do not always have symmetries that enable us to define a vacuum state \cite{Parker,Birrell,Fulling,WaldQFTCST}.
Nonetheless, for some kinds of spacetimes, we can still define univocally a vacuum state and the notion of the particle is possible.  
This is the case of a time-independent static metric, over which
the general formalism of QFT can be elaborated.

Although the vacuum state is well-defined for a static space. This space could present an unusual topology, which can give rise to different types of phenomena such as vacuum polarization \cite{Birrell,PhysRevD.34.1918}, the Casimir-Polder effect \cite{saharian_kotanjyan_2011}, the alteration of the entangled state of two qubits \cite{he_yu_hu_2020,huang_he_2020, huang_2021}, among other interesting phenomena \cite{Svaiter_1994,bilge_hortacsu_ozdemir_1998,PhysRevD.51.2591,PhysRevD.92.084062,PhysRevD.93.084028,PhysRevD.97.045007,PhysRevD.94.024039}. 

In particular, the space produced by a cosmic string has a non-trivial topology, locally flat but not globally, since it has an angular deficit around the string that gives it a conical shape as seen from an embedded diagram. 
The line element that describes the space produced by a straight cosmic string is given in cylindrical coordinates by \begin{equation}\label{cosmic string line element}
    ds^2=dt^2-dr^2-dz^2-r^2d\varphi^2,
\end{equation} where $0<r<\infty$,  $0<\varphi <2\pi b$ and $-\infty<z<\infty$. The parameter $b$ is defined by 
\begin{equation*}
    b=1-4\mu G,
\end{equation*} 
where $\mu$ is the linear mass density of the cosmic string and $G$ is the Newton gravitational constant. We describe a massless quantum field over this spacetime through the non-minimal coupling that leads us to establish the Klein-Gordon equation as \begin{equation}
    \left(\Box+\xi R\right)\phi(x)=0.
\end{equation}
Since the space of the cosmic string is locally flat, the Ricci scalar curvature $R$ is identically null. Thus, using the metric given in eq. \eqref{cosmic string line element}, the Klein-Gordon equation can be written as 
\begin{equation}
\label{Klein-Gordon with founts equation for cosmic string}
    \left(\partial^2_t-K\right)\phi(t,\textbf{x})=0,
\end{equation} 
where we have defined the spatial operator
\begin{equation}\label{operator for spectral transformation}
    K=\partial^2_r+\frac{1}{r}\partial_r+\frac{1}{r^2}\partial^2_\varphi+\partial^2_z.
\end{equation}
In a similar way, the Green functions of the quantum field must satisfy
\begin{equation}\label{definition of Green function for cosmic string}
    \left(\partial^2_t-K\right)G(t,\textbf{x},t',\textbf{x}')=\frac{1}{\sqrt{-g}}\delta(t-t')\delta(\textbf{x}-\textbf{x}'),
\end{equation}
where $g$ is the determinant of the metric shown in eq. \eqref{cosmic string line element}. In order to find the Green function, we introduce the spectral transformation of operator $K$. Let it be the eigenvalue equation for the operator $K$ \begin{equation}\label{eigenfuncion equation}
    K\psi_{\textbf{k}}(\textbf{x})=-\omega^2_{\textbf{k}}\psi_{\textbf{k}}(\textbf{x}),
\end{equation} where $\omega^2_{\textbf{k}}$ are the eigenvalues and $\psi_{\textbf{k}}$ are orthonormal eigenfunctions of the operator $K$, with the inner product define as follows \begin{equation}\label{ortonormality definition for spectral of K}
    \int d\mu(\textbf{k})\,\,\psi_{\textbf{k}}(\textbf{x})\psi^*_{\textbf{k}}(\textbf{x}')=\frac{\delta(\textbf{x}-\textbf{x}')}{\sqrt{h(\textbf{x})}},
\end{equation} where $d\mu(\textbf{k})$ is some measure on the space of $\textbf{k}$ and $h$ is the determinant of the metric induced on the spatial part of the stationary spacetime metric, see eq. \eqref{cosmic string line element}.

In this way, the spectral transformation of a function $f(\textbf{x})$ is given as the integral transformation of $f$ with orthonormal eigenfunctions of $K$ as Kernel, that is \begin{equation}
    f(\textbf{x})=\int d\mu(\textbf{k})\,\,\psi_{\textbf{k}}(\textbf{x})\Tilde{f}(\textbf{k}),
\end{equation} and the inverse transformation given by \begin{equation}
    \Tilde{f}(\textbf{k})=\int d\textbf{x} \sqrt{h(\textbf{x})}\,\,\psi^*_{\textbf{k}}(\textbf{x})f(\textbf{x}).
\end{equation} 

Hence, by taking the Fourier transformation on $t$ and the spectral transformation on $\textbf{x}$, we can write the Green function as \begin{equation}\label{Green function for a cosmic string 1}
    G(t,\textbf{x},t'\textbf{x}')=\frac{1}{\sqrt{2\pi}}\int d\omega\, e^{-i\omega t}\int d\mu(\textbf{k})\,\psi_{\textbf{k}}(\textbf{x})\tilde{G}(\omega,\textbf{k},t',\textbf{x}'),
\end{equation} 
after replacing this into eq. \eqref{definition of Green function for cosmic string}, using the orthonormality relation eq. \eqref{ortonormality definition for spectral of K} and the Fourier transformation for the Dirac delta function, we find \begin{equation}\label{Green function for a cosmic string 2}
    G(t,\textbf{x},t'\textbf{x}')=-\frac{1}{2\pi}\int d\mu(\textbf{k})\int d\omega\,\,\frac{\psi_{\textbf{k}}(\textbf{x})\psi^*_{\textbf{k}}(\textbf{x}')}{\omega^2-\omega^2_{\textbf{k}}} e^{-i\omega (t-t')},
\end{equation}
eq. \eqref{Green function for a cosmic string 2}, give us all the Green function of the scalar field in the cosmic string spacetime, as long as the specific boundary conditions are fulfilled. 
In this work, we are interested in calculating the Wightman function of positive frequency. This Wightman function, $G_+$, is defined as $i$ times the integral, in the complex plane of $\omega$, of eq. \eqref{Green function for a cosmic string 2}, over the clockwise contour enclosing the positive pole $\omega_{\textbf{k}}$, therefore
\begin{equation}\label{Wightman function for a cosmic string with positive frecuency}
    G^+(t,\textbf{x},t',\textbf{x}')=\int d\mu(\textbf{k})\,\,\frac{e^{-i\omega_{\textbf{k}}(t-t')}}{2\omega_{\textbf{k}}}\psi_{\textbf{k}}(\textbf{x})\psi^*_{\textbf{k}}(\textbf{x}').
\end{equation}

On the other hand, going back to the solutions of the eigenvalue problem in the eq.
\eqref{eigenfuncion equation}, we use a Fourier series to expand $\psi_{\textbf{k}}(r,\varphi,z)$, since it is periodic in $\varphi$, with period $2\pi b$. Then, by using the separation of variables method, we can show that the normalized solutions are given by
\begin{equation}\label{autofuntion of K operator2}
    \psi_{\textbf{k}}(r,\varphi,z)=\sqrt{\frac{p}{2\pi b}}J_{|\lambda l|}(p\,r)e^{i  \lambda l\varphi}e^{i \kappa z},
\end{equation} 
which is index by $l=\{0,\pm1,\pm2,...\}$, $p\in(0,\infty)$ and $\kappa\in(-\infty,\infty)$. In eq. \eqref{autofuntion of K operator2}, we find that $J_l$ is the usual Bessel function of order $l$, and the fundamental frequency of the Fourier series is $\lambda=1/b$.
In the massless field case, the eigenvalues are given by
 $\omega^2_{\textbf{k}}=p^2+\kappa^2$. 
 Furthermore one can identify the appropriate measure of the spectral transformation, given by
 \begin{equation}
    \int d\mu(\textbf{k})=\sum^\infty_{l=-\infty}\int_{-\infty}^{\infty} \frac{d\kappa}{2\pi}\int^\infty_0 \!\!\!dp.
\end{equation}
Therefore, the Wightman function is explicitly given by \begin{equation}\label{Wightman function for cosmic string}
    G^+(t,\textbf{x},t',\textbf{x}')=\frac{1}{8\pi^2rr'\sinh u}\frac{\lambda\sinh(\lambda u)}{\cosh(\lambda u)-\cos\left(\lambda(\varphi-\varphi')\right)},
\end{equation} where \begin{equation}\label{coshu definition for cosmic string}
    \cosh u=\frac{r^2+r'^2+(z-z')^2-(t-t'-i\epsilon)^2}{2rr'}.
\end{equation} 
However, in \cite{ALIEV1989142} the authors have shown that a charged particle moving freely in the space of the cosmic string emits radiation due to the topological properties of spacetime. This problem is solved by postulating the quantization of the cosmic string fundamental frequency, i.e. $\lambda=n$, with $n\in \mathbb{N}$. This integer is identified as the topological charge of the cosmic string. Hence, by this assumption, the radiation becomes identically null, and therefore the Wightman function can be written as \begin{align}
    \nonumber G^+(t,\textbf{x},t',\textbf{x}')=&\frac{1}{8\pi^2rr'}\sum^{n-1}_{k=0}\frac{1}{\cosh u-\cos(\varphi-\varphi'+2\pi k/n)}\\
    =&-\frac{1}{4\pi^2}\sum^{n-1}_{k=0}\frac{1}{(\Delta t-i\epsilon)^2-\left(\Delta r^2+\Delta z^2+4rr'\sin^2(\pi k/n+\Delta\varphi/2)\right)},
    \label{whithmann in cosmic string}
\end{align} where the qubits' spatial position are given in cylindrical coordinates by $\textbf{x}=(r,\varphi,z)$ and $\textbf{x}'=(r',\varphi',z')$. We also denote $\Delta t=t-t'$, $\Delta r=r-r'$, $\Delta z=z-z'$ and $\Delta \varphi=\varphi-\varphi'$. 

\section{Effective Dynamics for Qubits in Quantum Vacuum}
\label{sec twoqubits}

Let us model a qubit as a two-level system and consider a pair of identical qubits. We consider that the Hamiltonian that describes the free qubits is given by
\beq
{\cal H}_0=\omega\left(\tilde\sigma\otimes\mathbb{I}+\mathbb{I}\otimes\tilde\sigma\right)+
j\left(\sigma^+\otimes\sigma^-+\sigma^-\otimes\sigma^+\right),
\label{hamiltoniano libre qubits}
\eeq
where $\omega$ is the energy gap of each qubit and $j$ is the coupling constant of the Heisenberg $XY$-interaction between the qubits \cite{PhysRevA.64.012313}. This direct interaction guides the qubits to ``align" themselves and it is the cause of ferromagnetic properties in two-dimensional spin models. In this paper, we investigate the role of this Heisenberg interaction in the entanglement harvesting process. We suppose that each qubit has a ground state, $|g\rangle$, with zero energy, and an excited state, $|e\rangle$, with energy $\omega$.
Therefore, in eq. \eqref{hamiltoniano libre qubits}, we have defined the projector operator $\tilde\sigma=|e\rangle\langle e|$ and the leading operators $\sigma^+=|e\rangle\langle g|$ and $\sigma^-=|g\rangle\langle e|$, and also the identity operator $\mathbb{I}$, for the two-dimensional Hilbert space of each qubit.

If we use the canonical basis of the
two-qubits states, i.e., $\{|ee\rangle, |eg\rangle, |ge\rangle, |gg\rangle\}$, for the matrix representation of the qubits free Hamiltonian, one can show that
the matrix representation of ${\cal H}_0$ is no diagonal. This is because the canonical basis 
does not coincide with the energy eigenstates. It is not difficult to show that the energy states are given by
\begin{align}
|E_1\rangle&=|ee\rangle,\nn\\
|E_2\rangle&=(|eg\rangle+|ge\rangle)/\sqrt{2},\nn\\
|E_3\rangle&=(|eg\rangle-|ge\rangle)/\sqrt{2},\nn\\
|E_4\rangle&=|gg\rangle,
\label{qubits energy states}
\end{align} 
consequently, the energy levels are  $E_1=2\omega$, $E_2=\omega+j$,
$E_3=\omega-j$ and $E_4=0$. 
By using the energy states, 
$\{|E_1\rangle,|E_2\rangle,|E_3\rangle,|E_4\rangle\}$, as basis for the 
Hilbert space of the qubits, one gets a diagonal matrix representation for the 
qubits free Hamiltonian
\beq
{\cal H}_0=
\left(
\begin{array}{cccc}
2\omega & 0 & 0 & 0\\
0 & \omega + j & 0 & 0\\
0 & 0 &\omega -j & 0\\
0 & 0 & 0 & 0
\end{array}
\right).
\label{Hdiag}
\eeq

Let us now describe the interaction between the qubits and the massless quantum scalar field. We consider that the qubits possess a quantum internal structure, just described above. Nonetheless, we also assume that the qubits have a well-defined classical trajectory in spacetime. The interaction between the qubits and the scalar field is local.
In this way, depending on the specific qubits' law of motion they would perceive the vacuum state of the quantum field either as a system completely devoid of particles (inertial motion) or as a thermal reservoir (non-inertial accelerated motion), for example.
This model of qubits as two-level systems coupled locally with the scalar field is known as the Unruh-De Witt detector and it has been shown that it captures the main features of the radiation-matter interaction. Hence, we consider that the  interaction between the qubits and the scalar field is given by the Hamiltonian interaction 
\beq
{\cal H}_{int}=g\left({\cal M}_1\,\phi(\chi_1(\tau))+\mathcal{M}_2\,\phi(\chi_2(\tau))\right),
\label{Hint}
\eeq 
where the field is evaluated at the qubits spacetime trajectories, $\chi_n(\tau)$, with $n=\{1,2\}$.
The parameter $g$ is a small coupling constant of the interaction 
and the ${\cal M}_n$ are the monopole operators of the two-qubits system given by
\begin{align}
    \mathcal{M}_1&=m\otimes\mathbb{I},\nonumber\\
    \mathcal{M}_2&=\mathbb{I}\otimes m, 
\end{align}
where the monopole operator of each individual qubit is
$m=|e\rangle\langle g|+|g\rangle\langle e|$. Now, considering the whole qubits-field system, the total Hamiltonian that leads the time evolution is
\begin{equation}
    \mathbb{H}=\mathbb{H}_0+\mathcal{H}_{int},
\end{equation} 
where the free part of the system is composed of the free qubits Hamiltonian eq. \eqref{hamiltoniano libre qubits} and the Klein-Gordon Hamiltonian of the massless scalar field $\mathbb{H}_0=\mathcal{H}_0+\mathcal{H}^{KG}_0$. The free field Hamiltonian, $\mathcal{H}^{KG}_0$, is responsible for establishing the eq. \eqref{Klein-Gordon with founts equation for cosmic string}.

In order to find the effect of the quantum field in its vacuum state over the qubit dynamics, we use perturbation theory. So, let us now work on the Dirac interaction picture instead of the Schrodinger picture. This would simplify the calculations without losing the generality of the results. Therefore, we need to express the interaction Hamiltonian of the eq. \eqref{Hint} in the interaction picture. In order to do this, we perform a pull-back over the interaction Hamiltonian using the operator $\mathbb{U}_0(\tau,\tau_0)$.
This operator would be the time evolution operator of the total system, from time $\tau_0$ until time $\tau$, if there were no interactions between the qubits and the scalar field. Therefore, we can write $\mathbb{U}_0(\tau,\tau_0)={\cal U}_0^{field}(\tau,\tau_0)\,{\cal U}_0(\tau,\tau_0)$, where ${\cal U}_0^{field}$ is the time evolution operator of the free scalar field and ${\cal U}_0(\tau,\tau_0)$ corresponds to the free time evolution operator of the qubits from the initial time $\tau_0$, when the interaction begins, until time $\tau$. Since the free Hamiltonian of the field and the qubits do commute $[{\cal H}_0^{KG},{\cal H}_0]=0$, then we have that in the interaction picture we can write the interaction Hamiltonian as
\begin{align}
   {\cal H}^I_{int}=g\left({\cal M}^I_1\phi_I(\chi_1(\tau))+\mathcal{M}^I_2\phi_I(\chi_2(\tau))\right),
\end{align}
where the monopole and field operators in the interaction picture are respectively given by
\begin{align}    \mathcal{M}^I_n(\tau)&=\mathcal{U}^{-1}_0(\tau,\tau_0)\,\mathcal{M}_n\,\mathcal{U}_0(\tau,\tau_0),\\
\phi_I(\chi_n(\tau))&=\left(\mathcal{U}^{field}_0(\tau,\tau_0)\right)^{-1}\phi(\chi_n(\tau))\,\mathcal{U}^{field}_0(\tau,\tau_0).
\end{align}
By taking into account that the qubits in the free Hamiltonian eq. \eqref{hamiltoniano libre qubits} is time-independent, we have for the time evolution operator of the free qubits
${\cal U}_0(\tau,\tau_0)=e^{-i{\cal H}_0(\tau-\tau_0)}$. Also, since the quantum scalar field has a free Hamiltonian given by the Klein-Gordon Hamiltonian, then $\mathcal{U}^{field}_0(\tau,\tau_0)=e^{-i\mathcal{H}^{KG}_0(\tau-\tau_0)}$, is the free field time evolution operator.\\

On the other hand, by considering the basis of energy states, $\{|E_1\rangle,|E_2\rangle,|E_3\rangle,|E_4\rangle\}$, the matrix representation of the time evolution operator is diagonal and given by
\beq
{\cal U}_0(\tau,\tau_0)=
\left(
\begin{array}{cccc}
e^{-i2\omega(\tau-\tau_0)} & 0 & 0 & 0\\
0 &  e^{-i(\omega+j)(\tau-\tau_0)} & 0 & 0\\
0 &  0 & e^{-i(\omega-j)(\tau-\tau_0)} & 0\\
0 &  0 & 0 & 1
\end{array}
\right),
\eeq
with this, one can show that in the interaction picture, the monopole operators are given by
\beq
{\cal M}^I_n(\tau)=\frac{1}{\sqrt{2}}\left(
\begin{array}{cccc}
0 & e^{i(\omega-j)\delta\tau} & (-1)^ne^{i(\omega+j)\delta\tau} & 0\\
e^{-i(\omega-j)\delta\tau}&0 &0 & e^{i(\omega+j)\delta\tau}\\
(-1)^n e^{-i(\omega+j)\delta\tau} & 0 & 0 & (-1)^{n+1}e^{i(\omega-j)\delta\tau}\\
0 & e^{-i(\omega+j)\delta\tau}& (-1)^{n+1}e^{-i(\omega-j)\delta\tau} & 0
\end{array}
\right),
\label{Monopolos interaccion}
\eeq 
where $n=\{1,2\}$ and $\delta\tau=\tau-\tau_0$. It is worth realizing that in the interaction picture, the operators evolve only taking into account the free Hamiltonian, whereas the states evolve only considering the interaction Hamiltonian. Therefore, in the interaction picture, the field operators $\phi_I(\chi)$ satisfy the free Klein-Gordon equation, and its vacuum expectation values would give us the usual free Green function of the scalar field. In this way, we have that
\beq
G^+(\chi,\chi')=\langle0|\phi_I(\chi)\phi_I(\chi')|0\rangle,
\eeq is the Wightman function of the free scalar field, where we have to denote $|0\rangle$ as the quantum vacuum state of the scalar field.

We are interested in the influence of the vacuum fluctuations over the two-qubits state. Therefore, we suppose the qubits begin the interaction at a very general mixed state described by the density operator 
\beq
\rho_{in}=
\left(
\begin{array}{cccc}
p_1 & 0 & 0 & \alpha\\
0 & p_2 & \beta & 0\\
0 & \beta^* & p_3 & 0 \\
\alpha^* & 0 & 0 & p_4 
\end{array}
\right),
\label{rhoin}
\eeq
this density operator is of the kind of the $X$-state. We focus on this kind of state since there is strong numerical evidence suggesting that all mixed states of two qubits are equivalent to the X states through a single unitary transformation that preserves entanglement, see \cite{2013arXiv1310.7038H_Xstate}. Here, it is worth emphasizing that the matrix representation of eq. \eqref{rhoin} is realized in the qubits energy basis. We also consider that 
the quantum field remains in its quantum vacuum state $|0\rangle$. 
Therefore, the initial state density operator of the total qubits-field system is given by
\beq
\varrho_{in}=\rho_{in}\otimes|0\rangle\langle 0|.
\eeq
By using this initial state and the interaction Hamiltonian one can obtain
the density operator of the qubits-field system  $\varrho_I(\tau)$, in the interaction picture at any time $\tau>\tau_0$.
In order to do that, we must solve the quantum Liouville equation for the total density operator
\beq
i\frac{\partial\varrho_I(\tau)}{\partial\tau}=[{\cal H}_{int}^I(\tau),\varrho_I(\tau)],
\eeq
that satisfy the initial condition $\varrho_I(\tau_0)=\varrho_{in}$.
Hence, we perform a perturbation Dyson expansion assuming a small interacting
coupling constant $g$, in such a way that
\begin{align}
    \varrho_I(\tau)=\varrho_I(\tau_0)-i\int^\tau_{\tau_0} d\uptau \left[\mathcal{H}^I_{int}(\uptau),\varrho_I(\tau_0)\right]-\int^\tau_{\tau_0}d\uptau\int^{\uptau}_{\tau_0}d\uptau'\left[\mathcal{H}^I_{int}(\uptau),\left[\mathcal{H}^I_{int}(\uptau'),\varrho_I(\tau_0)\right]\right]+\cdots
    \label{dyson expansion}
\end{align}
Finally, since we are concerned only with the final qubits state, over the total density operator $\varrho_I(\tau)$ we perform a partial trace operation over the field degrees of freedom to obtain 
the final density operator associated only with the qubits 
$\rho_I(\tau)=\mathrm{Tr}_{field}\,\,\varrho_I(\tau)$. This is possible due to the 
independence between the qubits and the field in the absence of the interaction.
In this way, at some time $\tau$ after the interaction has begun, the state of the qubits
would be modified by the vacuum fluctuations in such a way that its density operator
is corrected and given by
\beq 
\rho_I(\tau)=\rho_{in}+g^2\sum^2_{n,p=1}\int_{\tau_0}^\tau d\uptau\int_{\tau_0}^{\tau} d\uptau'\delta\rho_{np}(\uptau,\uptau')G^+_{n p}(\uptau,\uptau'),
\label{rhofinal}
\eeq
where the matrices inside the integral above are defined by
\begin{align}
\delta\rho_{np}(\uptau,\uptau')={\cal M}^I_p(\uptau')\rho_{in}{\cal M}^I_n(\uptau)
-\Theta(\Delta\uptau){\cal M}^I_n(\uptau){\cal M}^I_p(\uptau')\rho_{in}
-\Theta(-\Delta\uptau)\rho_{in}{\cal M}^I_n(\uptau){\cal M}^I_p(\uptau'),
\nonumber
\end{align} 
and $\Delta\uptau=\uptau-\uptau'$, the Heaviside function is denoted by $\Theta$ and 
the Wightman function of the free massless scalar field evaluated 
at the qubits trajectory points at two different proper times is denoted by
\beq
G^+_{np}(\uptau,\uptau')=G^+(\chi_n(\uptau),\chi_p(\uptau')),
\label{Gnp}
\eeq
where $n,p=\{1,2\}$. For simplicity, we  consider the following qubits initial density operator which is diagonal in energy basis, i.e.
\begin{equation}\label{initial condition for density matrix}
    \rho_{in}=\left(\begin{matrix}
    p_1 & 0 & 0 & 0 \\
    0 & p_2 & 0 & 0 \\
    0 & 0 & p_3 & 0\\
    0 & 0 & 0 & p_4
    \end{matrix}\right).
\end{equation}
Therefore, using the expressions of eq.(\ref{Monopolos interaccion}) and eq.(\ref{initial condition for density matrix}), the terms inside the integral of eq. (\ref{rhofinal}), give us the corrections caused by the vacuum fluctuation over the qubits density operator are \beq
\rho_I(\tau)=
\left(
\begin{array}{cccc}
p_1+\delta p_1(\tau)& 0 & 0 & \delta\alpha(\tau) \\
0 & p_2+\delta p_2(\tau) & \delta\beta(\tau) & 0\\
0 & \delta \beta^*(\tau) & p_3+\delta p_3(\tau) & 0\\
\delta \alpha^*(\tau) & 0 & 0 & p_4+\delta p_4(\tau)\\
\end{array}
\right),
\eeq 
where the corrections to the matrix elements are given in the form 
\begin{align}
    \delta \mathcal{A}=\frac{g^2}{2}\sum^2_{n,m=1}\int^\tau_{\tau_0}d\uptau\int^\tau_{\tau_0}d\uptau'\delta\rho^{\mathcal{A}}_{nm}(\uptau,\uptau')G^+_{nm}(\uptau,\uptau'),
\end{align} 
being $\mathcal{A}$ some member of the set $\{p_1,p_2,p_3,p_4,\alpha,\beta,\alpha^*,\beta^*\}$ and for each term we have the respective factor $\delta\rho^{\mathcal{A}}_{nm}$, that must be inside the integral together with the Wightman function.
In this way we have 
\begin{align}
   \nn \delta\rho^{p_1}_{nm}=&-p_1\left[e^{i(\omega-j)\Delta\uptau}+(-1)^{n+m}e^{i(\omega+j)\Delta\uptau}\right]+p_2e^{-i(\omega-j)\Delta\uptau}+(-1)^{n+m}p_3e^{-i(\omega+j)\Delta\uptau},\\
\nn \delta\rho^{p_2}_{nm}=& p_1e^{i(\omega-j)\Delta\uptau}-p_2\left[e^{-i(\omega-j)\Delta\uptau}+e^{i(\omega+j)\Delta\uptau}\right]+p_4e^{-i(\omega+j)\Delta \uptau},\\
\nn \delta\rho^{p_3}_{nm}=&(-1)^{n+m}\left[p_1e^{i(\omega+j)\Delta\uptau}-p_3\left(e^{-i(\omega+j)\Delta\uptau}+e^{i(\omega-j)\Delta\uptau}\right)+p_4e^{-i(\omega-j)\Delta\uptau}\right],\\
\nn \delta\rho^{p_4}_{nm}=&p_2e^{i(\omega+j)\Delta\uptau}+(-1)^{n+m}p_3e^{i(\omega-j)\Delta\uptau}-p_4\left[e^{-i(\omega+j)\Delta\uptau}+(-1)^{n+m}e^{-i(\omega-j)\Delta\uptau}\right],\\
\nn \delta\rho^{\alpha}_{nm}=&e^{i\omega(T-2\tau_0)}\Big[p_2e^{ij\Delta\uptau}-(-1)^{n+m}p_3e^{-ij\Delta\uptau}\\
\nn &\qquad \qquad \  +\left(p_1\Theta(-\Delta\uptau)+p_4\Theta(\Delta\uptau)\right)\left((-1)^{n+m}e^{ij\Delta\uptau}-e^{-ij\Delta\uptau}\right)\Big],\\
\nn \delta\rho^{\beta}_{nm}=& e^{ij(T-2\tau_0)}\Big[(-1)^n\left(p_1e^{i\omega\Delta\uptau}-p_4e^{-i\omega\Delta\uptau}\right)\\
\nn & \qquad \qquad \ +(-1)^m\left(p_2\Theta(-\Delta\uptau)+p_3\Theta(\Delta\uptau)\right)\left(e^{i\omega\Delta\uptau}-e^{-i\omega\Delta\uptau}\right)\Big],\\
\label{correction density matrix}
\end{align} 
where we have define the variables $T=\uptau+\uptau'$ and $\Delta\uptau=\uptau-\uptau'$. It is important to note here that these corrections maintain the unitarity and normalization of the density operator.

\section{Qubits Entanglement Measurement}
\label{sec:entanglement_generation}

The entanglement measurement of a quantum state is a fundamental tool that characterizes the collective non-local quantum behavior of multi-particle systems. In general, for the case of mixed states of systems with several subsystems, one has that the measurement of entanglement is a rather difficult task due to certain minimization processes. However, in some cases, such procedures can be avoided through the use of simpler measurements that are monotonically related to the entanglement formation. For a pair of two-level systems, we have that quantum entanglement can be completely determined by the Positive Partial Transpose (PPT) criterion developed by Peres \cite{PhysRevLett.77.1413_Peres_criterio}.
Under this criterion, the density matrix of the system will be separable if the resulting matrix, after applying partial transpose on the components of one of its subsystems, has a non-negative spectrum.

In order to measure the entanglement of the two-qubit system after the interaction with the quantum vacuum in the cosmic string spacetime, we use the \textit{Quantum Negativity} to quantify the degree of entanglement.
This entanglement measurement was introduced by Zyczkowski \textit{et al.} \cite{PhysRevA.58.883} and derived from the PPT separability criterion. It is worth noting that this entanglement measurement is of the kind monotonically under LOCC (Local Operation and Classical Communication), satisfying the principal requirement of any entanglement measurement, see \cite{PhysRevA.65.032314}.


Let us define the Negativity of a quantum state described by the density operator $\rho(\tau)$, at time $\tau$, as follows
\begin{equation}\label{negativity definition}
    \mathcal{N}(\tau)=2\left|\sum_{\lambda_i<0}\lambda_i\right|=\left(\sum_i|\lambda_i|-\lambda_i\right),
\end{equation} 
where $\lambda_i$ are the eigenvalues of the partial transpose $\rho^{PT}(\tau)$ of the density operator. In the expression above the first sum is restricted to negative eigenvalues only, while the second sum considers all the eigenvalues.

In our case, in order to know the initial entanglement between the qubits before the interaction with the quantum vacuum,
we need to calculate the Negativity associated with the initial state $\rho_{in}$, given in eq. \eqref{initial condition for density matrix}. For this initial state, the only eigenvalue of the partial transpose that could be negative is $\lambda_{in}=(p_1+p_4-\sqrt{(p_2-p_3)^2+(p_1-p_4)^2})/2$. Hence, we have the initial negativity given by
\begin{equation}
    \mathcal{N}_{in}=-\Theta(-\lambda_{in})\left(p_1+p_4-\sqrt{(p_2-p_3)^2+(p_1-p_4)^2}\right).
\end{equation}
In order to observe the generation of entanglement produced by the interaction between the qubits with the quantum vacuum, we must consider that at the beginning of the interaction, there is no qubits entanglement. For this, it is necessary to establish that $\lambda_{in}\geq 0$, which is guaranteed as long as the following condition is met
\begin{equation}
    |p_2-p_3|\leq2\sqrt{p_1p_4}.
    \label{condition null initial negativity}
\end{equation} 

From this expression, we note that we can tune up the initial values of the populations in such a way as to ensure
a null degree of entanglement in the initial state eq. \eqref{initial condition for density matrix}.
One can realize that if we choose the initial population to be such that $p_1=p_4$ and $p_2=p_3$ then the condition eq. \eqref{condition null initial negativity} is satisfied and there would be zero entanglement between the qubits at the initial state.

Then, using this particular choice for the initial populations, one can analyze the entanglement generation between the qubits due to the Unruh effect. Therefore, here we study the case where the initial state, at time $\tau=\tau_0$, has zero entanglements and is of the form
\beq\label{initial density matrix}
\rho_{in}=
\left(
\begin{array}{cccc}
{ p} & 0 & 0 & 0\\
0 & { 1/2-p} & 0 & 0\\
0 & 0 & { 1/2-p} & 0 \\
0 & 0 & 0 & { p} 
\end{array}
\right),
\eeq
where the single probability that defines the initial state has to be $0\leq p\leq1/2$. After the interaction with the vacuum of the cosmic string, the final qubit density state is of the form
\beq
\rho(\tau)=
\left(
\begin{array}{cccc}
{ p}+\delta p_1 & 0 & 0 & \delta\alpha \\
0 & { 1/2-p} +\delta p_2& \delta\beta & 0\\
0 & \delta\beta^* & { 1/2-p}+\delta p_3 & 0 \\
\delta\alpha^* & 0 & 0 & { p} +\delta p_4
\end{array}
\right),
\label{rhofinA}
\eeq
here it is worth noting, that even if the initial state is diagonal, the final state is not, and there exist induced coherent terms $\delta\alpha$, $\delta\alpha^*$, $\delta\beta$ and $\delta\beta^*$
generated by the interaction with the field at its vacuum state. Due to the symmetry of the Wightman function evaluated on the space-time location of static qubits, we can show that the corrections to the diagonal term of the density operator that give us the probabilities can be written as
\begin{equation}
    \delta p_k=g^2\int^{\mathcal{T}/2}_{-\mathcal{T}/2}d\uptau \int^{\mathcal{T}/2}_{-\mathcal{T}/2}d\uptau' \left(\delta p^D_k( G^+_{11}(\uptau,\uptau')+G^+_{22}(\uptau,\uptau'))+\delta p^A_k G^+_A(\uptau,\uptau')\right),
    \label{probability corrections}
\end{equation} 
where $k=\{1,2,3,4\}$ and we have denoted the Wightman function evaluated at different qubits trajectories as $G^+_A(\uptau,\uptau')=G^+_{12}(\uptau,\uptau')$. From eq. \eqref{correction density matrix} we
can regroup some terms and define the factor that appears inside the integrals in eq. \eqref{probability corrections}, as follows
\begin{subequations}\label{diagonal correction terms for density matrix}\begin{align}
    \delta p^D_1=&-\left(e^{ij\Delta\uptau}+e^{-ij\Delta\uptau}\right)\left(p\left(e^{i\omega\Delta\uptau}+e^{-i\omega\Delta\uptau}\right)-\frac{1}{2}e^{-i\omega\Delta\uptau}\right),\\
    \delta p^A_1=&2\left(e^{ij\Delta\uptau}-e^{-ij\Delta\uptau}\right)\left(p\left(e^{i\omega\Delta\uptau}-e^{-i\omega\Delta\uptau}\right)+\frac{1}{2}e^{-i\omega\Delta\uptau}\right),\\
    \delta p^D_2=&\delta p^A_2/2=\left(e^{i\omega\Delta\uptau}+e^{-i\omega\Delta\uptau}\right)\left(p\left(e^{ij\Delta\uptau}+e^{-ij\Delta\uptau}\right)-\frac{1}{2}e^{ij\Delta\uptau}\right),\\
    \delta p^D_3=&-\delta p^A_3/2=\left(e^{i\omega\Delta\uptau}+e^{-i\omega\Delta\uptau}\right)\left(p\left(e^{ij\Delta\uptau}+e^{-ij\Delta\uptau}\right)-\frac{1}{2}e^{-ij\Delta\uptau}\right),\\
    \delta p^D_4=&-\left(e^{ij\Delta\uptau}+e^{-ij\Delta\uptau}\right)\left(p\left(e^{i\omega\Delta\uptau}+e^{-i\omega\Delta\uptau}\right)-\frac{1}{2}e^{i\omega\Delta\uptau}\right),\\
    \delta p^A_4=&-2\left(e^{ij\Delta\uptau}-e^{-ij\Delta\uptau}\right)\left(p\left(e^{i\omega\Delta\uptau}+e^{-i\omega\Delta\uptau}\right)-\frac{1}{2}e^{i\omega\Delta\uptau}\right).
\end{align}\end{subequations} 
In a similar way we can write the coherence terms induced by interaction with the quantum vacuum as follows
\begin{align}
    \delta\alpha=g^2\int^{\mathcal{T}/2}_{-\mathcal{T}/2}d\uptau \int^{\mathcal{T}/2}_{-\mathcal{T}/2}d\uptau' \left(\delta\alpha^D\left(G^+_{11}(\uptau,\uptau')+G^+_{22}(\uptau,\uptau')\right)+\delta\alpha^AG^+_A(\uptau,\uptau')\right), 
\end{align} 
where the factor inside the integral are give by
\begin{subequations}\label{alpha correction terms for density matrix}\begin{align}
    \delta\alpha^D&=\frac{1}{2}e^{i\omega(T+\mathcal{T})}\left(e^{ij\Delta\uptau}-e^{-ij\Delta\uptau}\right),\\
    \delta\alpha^A&=\left(1-4p\right)e^{i\omega(T+\mathcal{T})}\left(e^{ij\Delta\uptau}+e^{-ij\Delta\uptau}\right).
\end{align}\end{subequations}
Additionally, the other coherent term induced by vacuum fluctuations can be expressed as \begin{align}
    \delta\beta=g^2\int^{\mathcal{T}/2}_{-\mathcal{T}/2}d\uptau \int^{\mathcal{T}/2}_{-\mathcal{T}/2}d\uptau' \delta{\beta}^D\left(G^+_{22}(\uptau,\uptau')-G^+_{11}(\uptau,\uptau')\right),
\end{align} where the factors inside the integral above is given by
\begin{align}\label{beta correction terms for density matrix}
    \delta\beta^D=\frac{1}{2}e^{ij(T+\mathcal{T})}\left(e^{i\omega\Delta\uptau}-e^{-i\omega\Delta\uptau}\right).
\end{align}
All the above integral corrections give us the final state of the qubits after the interaction with the vacuum fluctuations has initialized. In these integrals, we now suppose that the initial time where the interaction begins is $\tau_0=-\mathcal{T}/2$ and that we are observing the qubits final state at time $\tau=\mathcal{T}/2$, in such a way that $\mathcal{T}$ has the meaning of total interaction time between the qubits and the field vacuum state.

In order to obtain explicitly the effect over the qubits due to the quantum field we need to work out all the corrections shown previously. We can show that the probability corrections, $\delta p_k$, depend upon the pair of integrals given by
\begin{align}
    \begin{split}\label{F function}
    F(\Omega,\mathcal{T},r_n)=&\int^{\mathcal{T}/2}_{-\mathcal{T}/2}d\uptau \int^{\mathcal{T}/2}_{-\mathcal{T}/2}d\uptau' e^{i\Omega\Delta\uptau}G^+_{nn}(\uptau,\uptau'),
    \end{split}\\\label{H function}
    H(\Omega,\mathcal{T},r_1,r_2,\Delta z, \Delta \varphi)=&\int^{\mathcal{T}/2}_{-\mathcal{T}/2}d\uptau \int^{\mathcal{T}/2}_{-\mathcal{T}/2}d\uptau' e^{i\Omega\Delta\uptau}G^+_A(\uptau,\uptau'),
\end{align}
where $n=\{1,2\}$. With these equations, we find it useful to define the functions 
\begin{align}
    \mathcal{I}(\Omega)=&\frac{F(\Omega,\mathcal{T},r_1)+F(\Omega,\mathcal{T},r_2)}{2}+H(\Omega,\mathcal{T},r_1,r_2, \Delta z,\Delta \varphi),\label{I function}\\
    \mathcal{J}(\Omega)=&\frac{F(\Omega,\mathcal{T},r_1)+F(\Omega,\mathcal{T},r_2)}{2}-H(\Omega,\mathcal{T}, r_1, r_2, \Delta z,\Delta \varphi).
    \label{J function}
\end{align}
it is clear that these functions depend on a series of variables i.e. $\mathcal{I}=\mathcal{I}(\Omega,\mathcal{T},r_1,r_2, \Delta z,\Delta \varphi)$ and $\mathcal{J}=\mathcal{J}(\Omega,\mathcal{T},r_1,r_2, \Delta z,\Delta \varphi)$. However, in the equations above we make explicit only the dependence on the gap energy $\Omega$,
this is because by using different energy gaps on these functions we can compute the corrections of the probabilities terms in the following manner
\begin{subequations}\label{diagonal correction terms for density matrix2}\begin{align}
    \delta p_1&=g^2\left[-p\left(\mathcal{I}(\omega-j)+\mathcal{J}(\omega+j)\right)+(1/2-p)\left(\mathcal{I}(-\omega+j)+\mathcal{J}(-\omega-j)\right)\right],\\
    \delta p_2&=g^2\left[p\left(\mathcal{I}(\omega-j)+\mathcal{I}(-\omega-j)\right)-(1/2-p)\left(\mathcal{I}(-\omega+j)+\mathcal{I}(\omega+j)\right)\right],\\
    \delta p_3&=g^2\left[p\left(\mathcal{J}(-\omega+j)+\mathcal{J}(\omega+j)\right)-(1/2-p)\left(\mathcal{J}(\omega-j)+\mathcal{J}(-\omega-j)\right)\right],\\
    \delta p_4&=g^2\left[-p\left(\mathcal{I}(-\omega-j)+\mathcal{J}(-\omega+j)\right)+(1/2-p)\left(\mathcal{I}(\omega+j)+\mathcal{J}(\omega-j)\right)\right].
\end{align}\end{subequations}

On the other side, to obtain the coherence terms induced by the quantum vacuum on the density operator, we need to compute the following integrals 
\begin{align}
  \begin{split}\label{Y0 function}      \mathcal{Y}_0(\omega,j,\mathcal{T},r_n)&=\int^{\mathcal{T}/2}_{-\mathcal{T}/2}d\uptau\int^{\mathcal{T}/2}_{-\mathcal{T}/2}d\uptau' e^{i\omega T}\left(e^{ij\Delta\uptau}-e^{-ij\Delta\uptau}\right)G^+_{nn}(\uptau,\uptau'),  
  \end{split}
    \\\label{Yi integral definition}
    \mathcal{Y}_I(\omega,j,\mathcal{T},r_1,r_2,\Delta z,\Delta\varphi)&=\int^{\mathcal{T}/2}_{-\mathcal{T}/2}d\uptau\int^{\mathcal{T}/2}_{-\mathcal{T}/2}d\uptau' e^{i\omega T}\left(e^{ij\Delta\uptau}+e^{-ij\Delta\uptau}\right)G^+_{A}(\uptau,\uptau'),
\end{align}
where $n=\{1,2\}$. Therefore, one can cast the induced coherence terms in the following manner \begin{align}\label{alpha correction terms for density matrix2}
    \delta\alpha&=\frac{g^2}{2}\,e^{i\omega\mathcal{T}}\left[\mathcal{Y}_0(\omega,j,\mathcal{T},r_1)+\mathcal{Y}_0(\omega,j,\mathcal{T},r_2)+2(1-4p)\mathcal{Y}_I(\omega)\right],\\\label{beta correction terms for density matrix2}
    \delta\beta&=\frac{g^2}{4}\,e^{ij\mathcal{T}}\left[\mathcal{Y}_0(j, \omega,\mathcal{T},r_2)-\mathcal{Y}_0(j, \omega,\mathcal{T},r_1)\right].
\end{align}
In the Appendix, we have worked out explicitly the expression of the finite time integrals, by using
the Wightman function of the massless scalar field in the spacetime generated by a cosmic string. 

With all these results, we can analyze the final entanglement of the state of the two qubits \eqref{rhofinA}. Thus, considering the definition of negativity \eqref{negativity definition} given by, after a time $\cal{T}$ in which both qubits interact with the vacuum state in the cosmic string spacetime, the eigenvalues $\lambda$ of the partial transpose of the final density matrix that can be negative are given by 
\begin{align}
\lambda_1&=\frac{1}{2}\left(2p+\delta p_1+\delta p_4-\sqrt{(\delta p_1-\delta p_4)^2+\left(\delta p_2-\delta p_3\right)^2-\left(\delta\beta-\delta\beta^*\right)^2}\right),\nonumber\\\label{eigenvalues of the partial traspose of the density matrix dependent of time}
\lambda_2&=\frac{1}{2}\Big(1-2p+\delta p_2+\delta p_3-\sqrt{\left(\delta\beta+\delta\beta^*\right)^2+4\delta\alpha\delta\alpha^*}\Big),
\end{align} 
therefore, we have that the negativity of the qubits, after some interaction time ${\cal T}$ with the vacuum, is given by 
\begin{equation}
\mathcal{N}(\tau)=|\lambda_1|-\lambda_1+|\lambda_2|-\lambda_2,
\label{negativity}
\end{equation}
by considering that the eigenvalues depend on the corrections to the density operator and that we have found all these corrections in terms of the integrals shown before, we are at position of understanding the dynamics of entanglement of the qubits that have interacted with the quantum vacuum fluctuations.
\section{Properties of Harvesting Entanglement in Cosmic String Spacetime}

Let us consider statics qubits located at general spacetime positions $\chi_1(\tau)=(\tau,r_1,\varphi_1,z_1)$ and $\chi_2(\tau)=(\tau,r_2,\varphi_2,z_2)$.
Therefore, the function $F(\Omega,\mathcal{T},r)$, that contributes to the corrections of the population terms of the density matrix and was
defined in eq. \eqref{F function}, was obtain in the appendix 
\eqref{function F} and given by
\begin{align}
\nonumber F(\mathcal{T},\Omega,r)=&\,\,\frac{1}{16}\left( 2\Omega\mathcal{T} \Theta (\Omega)+e^{-|\Omega |\mathcal{T}}\right)+\frac{\mathcal{T}^2}{16}e^{-|\Omega|\mathcal{T}}\sum^{n-1}_{k=1}\frac{1}{4r^2\sin^2(\pi k/n)+\mathcal{T}^2}\\
\label{function F 2}
    &+\Theta(\Omega)\frac{\mathcal{T}^3}{8}\sum^{n-1}_{k=1}\frac{\sin\left(2\Omega r \sin(\pi k/n)\right)}{2r\sin(\pi k/n)\left[\mathcal{T}^2+4r^2\sin^2(\pi k/n)\right]},
\end{align}
similarly the function $H(\Omega, \mathcal{T}, r_1,r_2,\Delta z,\Delta \varphi)$, define in eq. \eqref{H function}, was obtain worked out explicitly in the appendix, \eqref{function H}, and is given by
\begin{align}
    \label{function H 2}
     H(\mathcal{T},\Omega,d_{kn})&=\frac{\mathcal{T}^2}{16}e^{-|\Omega|\mathcal{T}}\sum^{n-1}_{k=0}\frac{1}{d^2_{kn}+\mathcal{T}^2}+\Theta(\Omega)\frac{\mathcal{T}^3}{8}\sum^{n-1}_{k=0}\frac{\sin(\Omega d_{kn})}{d_{kn}\left(d^2_{kn}+\mathcal{T}^2\right)},
\end{align}
where the distance between qubits is denoted as
\begin{equation}
    \label{gkn definition 2}
    d_{kn}(r_1,r_2,\Delta z,\Delta\varphi)=\sqrt{\Delta r^2+\Delta z^2+4r_1r_2\sin^2(\pi k/n+\Delta\varphi/2)},
    \end{equation}
which depends on $k=\{0,1,...,n-1\}$ and $n$ defined by the cosmic string.
On the other hand, the functions that define the coherence term corrections are given in terms of the functions $\mathcal{Y}_0(\omega,j,\mathcal{T},r)$  and $\mathcal{Y}_I(\omega,j,\mathcal{T},r_1,r_2,\Delta z, \Delta \varphi)$, define in eq. \eqref{Y0 function} and eq. \eqref{Yi integral definition} respectively, and which explicit form is obtained in the appendix  \eqref{Y0 function result2}  and \eqref{YI function result1} leading to the following results
\begin{align}
   \nonumber \mathcal{Y}_0(\omega, j,\mathcal{T}, r)=&\frac{1}{8}\mathcal{T}j(|\omega|\mathcal{T}+1)e^{-|\omega|\mathcal{T}}
   +\frac{\mathcal{T}^3e^{-|\omega|\mathcal{T}}}{8}
   \sum^{n-1}_{k=1}\frac{\sin\left(2jr\sin(\frac{\pi k}{n})\right)}{4r^2\sin^2\left(\frac{\pi k}{n}\right)(\mathcal{T}^2+ 4r^2\sin^2(\frac{\pi k}{n}))}\\
    &\times\left[2r\sin\left(\frac{\pi k}{n}\right)\cos\left(2|\omega|r\sin\left(\frac{\pi k}{n}\right)\right)+\mathcal{T}\sin\left(2|\omega|r\sin\left(\frac{\pi k}{n}\right)\right)\right],
    \label{Y0 function result3}
\end{align}
\begin{align}    
    &\mathcal{Y}_I(\omega,j, \mathcal{T},d_{kn})=
    \,\,\frac{\mathcal{T}^2}{8}\left(e^{j\mathcal{T}}\Theta(-j-|\omega|)+e^{-j\mathcal{T}}\Theta(j-|\omega|)\right)\sum^{n-1}_{k=0}\left(\frac{1}{d^2_{kn}+\mathcal{T}^2}\right)
    \nonumber\\
    &+\frac{\mathcal{T}^2e^{-|\omega|\mathcal{T}}}{16}\sum^{n-1}_{k=0}\frac{1}{d^2_{kn}(d^2_{kn}+\mathcal{T}^2)}\Bigg\{\gamma_+\bigg[\mathcal{T}\cos(j d_{kn})\Big(d_{kn}\sin(|\omega|d_{kn})-\mathcal{T}\cos(|\omega|d_{kn})\Big)\nonumber\\   &\,\,\,\,\,\,\,\,\,\,\,\,\,\,\,\,\,\,\,\,\,\,\,\,\,\,\,\,\,\,\,\,\,\,\,\,\,\,\,\,\,\,\,\,\,\,\,+d^2_{kn}+\mathcal{T}^2\bigg]+\mathcal{T}\sin(j d_{kn})\gamma_-\Big[d_{kn}\cos(|\omega|d_{kn})+\mathcal{T}\sin(|\omega|d_{kn})\Big]\Bigg\},
    \label{YI function result3}
\end{align}
where we have defined
\begin{align*}
    \gamma_\pm=\text{sign}(j+|\omega|)\pm\text{sign}(-j+|\omega|).
\end{align*}

With all these explicit results, we can determine some general properties of the entanglement between the two qubits after interacting for a time $\mathcal{T}$ with the quantum vacuum. For this purpose, it will be sufficient to analyze the behavior of the eigenvalues of the partial transpose of the density matrix given in eq. \eqref{eigenvalues of the partial traspose of the density matrix dependent of time}. We can see that these eigenvalues depend on the functions $F(\Omega,\mathcal{T},r)$, $H(\Omega, \mathcal{T}, r_1,r_2,\Delta z,\Delta \varphi)$, $\mathcal{Y}_0(\omega,j,\mathcal{T},r)$, and $\mathcal{Y}_I(\omega,j,\mathcal{T},r_1,r_2,\Delta z, \Delta \varphi)$. Therefore, many properties of the entanglement between the two qubits can be determined by analyzing these four functions only. Thus, we find the following general properties of harvesting vacuum entanglement in cosmic string spacetime.

\setcounter{subsubsection}{0} 

\subsubsection{Parity in qubits energy gap $\omega$}\label{prop 1}
By this property, we find that Negativity only depends on the absolute value of the qubits energy gap, i.e. $\mathcal{N}=\mathcal{N}(|\omega|)$. 
This can be seen from the parity under $\omega\rightarrow -\omega$ of the corrections \eqref{diagonal correction terms for density matrix2}. From there, it is clear that 
under the reflection $\omega\rightarrow -\omega$, the terms
$\delta p_2$, $\delta p_3$ and $\delta p_1+\delta p_4$ are even, while $\delta p_1-\delta p_4$ is odd, respectively . In the same manner the coherence terms corrections \eqref{alpha correction terms for density matrix2} and \eqref{beta correction terms for density matrix2} depend on the functions given at eq. \eqref{Y0 function result3} and eq. \eqref{YI function result3}, which are clearly even under $\omega\rightarrow -\omega$. Hence the eigenvalues \eqref{eigenvalues of the partial traspose of the density matrix dependent of time} and the Negativity \eqref{negativity} are even with respect to the qubits energy gap $\omega$. 


\subsubsection{Null  initial probability condition ($p=0$)}\label{prop 2}
In general, we find that the Negativity will be nonzero only for a very small initial probability $p\approx 0$. 
Indeed, by inspecting the eigenvalues $\lambda_1$ or $\lambda_2$ given in eq. \eqref{eigenvalues of the partial traspose of the density matrix dependent of time}, one notes that the eigenvalues would be negative only if the square roots overpass the leading terms $2p$ and $1-2p$, for $\lambda_1$ and $\lambda_2$, respectively. However, since the corrections to the populations terms $\delta p$ and to the coherence terms $\delta\alpha$ and 
$\delta\beta$ are proportional to $g^2$, with a very small coupling constant $g$, then the only possibility to have some negative eigenvalue in eq. \eqref{eigenvalues of the partial traspose of the density matrix dependent of time} is when the leading terms are very close to zero, $p\approx 0$ for $\lambda_1$ or very close to $p\approx 1/2$ for $\lambda_2$.
If we consider $p=1/2$, in order
to have the possibility to obtain a negative $\lambda_2$, then we need to analyze the coherence terms corrections.
However, since the corrections $\delta\alpha$ and $\delta\beta$ depend on the functions $\mathcal{Y}_0$ and $\mathcal{Y}_I$, which contain exponential decaying terms $e^{-|\omega|\mathcal{T}}$, at some instant of time, the term $\sqrt{(\delta\beta+\delta\beta^*)^2+4\delta\alpha\delta\alpha^*}$ will be always much smaller than the positive term $p_2(t)+p_3(t)=1-2p+\delta p_2+\delta p_3$, causing the eigenvalue $\lambda_2$ to be necessary positive. 
Thus, the only possibility for negativity to be nonzero, in general, is when the initial probability is very small $p\approx0$ or when it is very close to $p\approx1/2$, but this last case is valid only for very small interaction times $\mathcal{T}$. It is worth noting that considering the condition $p \approx 0$ in the expression for $\lambda_1$ results in the negativity function that depends only on $p$ having a linear form like $\mathcal{N}(p)\approx-c_1p + c_2$, with \( c_1,c_2 > 0 \). Thus, we observe that the maximum value of the negativity is given precisely at $p = 0$. This considerations are based on the validity of the perturbation approximation we are using. In the cases where the corrections are not small, we can easily obtain negative results in eq. \eqref{eigenvalues of the partial traspose of the density matrix dependent of time}, however, this would break the small perturbation hypotheses and more terms are needed in eq. \eqref{dyson expansion} to describe such cases.

From a physical point of view, the fact that negativity is nonzero for finite times only for $p\approx 0$ means that entanglement between the two qubits generally occurs when 
the initial density operator is a mixed state of being with equal probabilities in 
the states $|E_2\rangle$ and $|E_3\rangle$, defined in eq. \eqref{qubits energy states}. It should be noted that this result is obtained considering the initial density matrix shown at eq. \eqref{initial density matrix}.

\subsubsection{Finite Distance Minkowski Limit}\label{prop 3}
Let us consider the case where the qubits are far away from the cosmic string but at a finite relative distance between them.
Without loss of generality, let us consider the case where $\Delta\varphi=0$ and the distances $\Delta r$ and $\Delta z$  are finite, whereas the radial qubits positions $r_1$ and $r_2=r_1+\Delta r$ tend to infinity. 
By analyzing the Wightman function in the cosmic string spacetime \eqref{whithmann in cosmic string}, we see that when one radial distance goes to infinity the only term in the series that remains is $k=0$, while the other terms $k=\{1,2,...,n-1\}$ vanish. This is equivalent to setting $n=1$ in the series \eqref{whithmann in cosmic string}  and therefore to consider a spacetime without the angular deficit and no cosmic string at all. 
Then, one recognizes that by considering the qubits being far away from the cosmic string in eq. \eqref{whithmann in cosmic string} one recovers the Wightman function in Minkowski spacetime.
Hence, by considering the qubits being far away from the cosmic string, the previous results would reduce to the same correction we expect for the Minkowski spacetime and 
the functions $F^M, H^M,\mathcal{Y}^M_0$, and $\mathcal{Y}^M_I$ that give us the corrections to the matrix elements of the density operator in this case are given by\footnote{These functions are denoted with an index $M$ to indicate that they are functions for calculating vacuum corrections in Minkowski space.}   \begin{align*}
F^M(\mathcal{T} ,\Omega)&=  \frac{e^{-|\Omega|\mathcal{T}}}{16} +\frac{1}{8} \Omega\mathcal{T} \Theta (\Omega),\\
\mathcal{Y}^M_{0} (\omega ,j,\mathcal{T} )&=  \frac{1}{8}\mathcal{T} j(|\omega |\mathcal{T} +1)e^{-|\omega |\mathcal{T}},\\
H^M(\mathcal{T} ,\Omega,d)&= \frac{1}{16}\frac{\mathcal{T}^{2} e^{-|\Omega|\mathcal{T}}}{d^{2} +\mathcal{T}^{2}} +\frac{\mathcal{T}^{3}}{8}\frac{\sin (\Omega d)}{d(d^{2} +\mathcal{T}^{2} )} \Theta (\Omega),
\end{align*}
\begin{align*}
    &\mathcal{Y}^M_{I} (\omega ,j,\mathcal{T} ,d)= 
    \frac{\mathcal{T}^{2}}{8(d^{2} +\mathcal{T}^{2} )}\left( e^{j\mathcal{T}} \Theta (-j-|\omega |)+e^{-j\mathcal{T}} \Theta (j-|\omega |)\right)\\
    &+\frac{1}{16}\frac{\mathcal{T}^{2} e^{-|\omega |\mathcal{T}}}{d^{2} (d^{2} +\mathcal{T}^{2} )}\Biggl\{\biggl[\mathcal{T}\cos (jd)( d\sin (|\omega |d)-\mathcal{T}\cos (|\omega |d)) +d^{2} +\mathcal{T}^{2}\biggr] \gamma _{+}\\ &     \,\,\,\,\,\,\,\,\,\,\,\,\,\,\,\,\,\,\,\,\,\,\,\,\,\,\,\,\,\,\,\,\,\,\,\,\,\,\,\,\,\,\,\,\,\,\,\,\,\,\,\,\,\,\,\,\,\,\,\,+\mathcal{T}\sin (jd)\Big[ d\cos (|\omega |d)+\mathcal{T}\sin (|\omega |d)\Big] \gamma _{-}\Biggr\},
\end{align*} where the qubits relative distance is $d=\sqrt{\Delta z^2+\Delta r^2}$.

\subsubsection{Qubits upon the Cosmic String}\label{prop 4}
Let us now consider that both qubits are located upon the cosmic string. Hence, we have that in this situation $r_1=r_2=0$, and the distance \eqref{gkn definition 2} is given by the axial distance between the qubits, i.e., $d_{kn}=\Delta z$, for all $k=\{0,1,2,...,n-1\}$. Considering this in eqs. \eqref{function F 2}, \eqref{function H 2}, \eqref{Y0 function result3} and \eqref{YI function result3}, one can prove that in the case (where the qubits are upon the cosmic string) all the functions are proportional to the Minkowski functions $H=nH^M$, $F=nF^M$, $\mathcal{Y}_0=n\mathcal{Y}^M_0$, and $\mathcal{Y}_I=n\mathcal{Y}^M_I$, where the proportionality constant is the cosmic string topological charge, $n$. Therefore, we see that in the limit of zero distance between the two qubits and the cosmic string, the negativity will behave exactly the same as in Minkowski space but with a higher intensity depending on $n$, which corresponds to the cosmic string linear mass density.

\subsubsection{Linear Dependence with the Time}\label{prop 5}
In general, it is observed that the negativity increases linearly with $\mathcal{T}$ when the interaction time between the qubits and the quantum vacuum is long enough. This happens because 
the functions 
\eqref{function F 2}, \eqref{function H 2}, \eqref{Y0 function result3} and \eqref{YI function result3} increase linearly with time for very large values of ${\cal T}$. Thus, at very large interaction times, entanglement increases linearly with ${\cal T}$. Nonetheless, this growth can not be unlimited. We must take into account that for large values of time, the corrections $\delta p$, $\delta\alpha$, and $\delta\beta$ also grow with time and could be not considered small corrections anymore.  At this point, for large values of time, the breakdown of the perturbation approximations we are using happens.
We can see this upper limit for interaction time by analyzing the final population terms $p_i+\delta p_i$, with $i=\{1,..,4\}$, 
which must always be positive and less than one. This condition sets a limit in our theory for the amount of time in which we can interact with the system of two qubits with the vacuum of the cosmic string. If this limit is exceeded, it will be necessary to consider more perturbative terms in eq. \eqref{dyson expansion} to correctly describe such a situation.

\subsubsection{Equivalence of Ferromagnetic and Antiferromagnetic $XY$-Interaction}\label{prop 6}
If we consider that one qubit is located at any point in the cosmic string spacetime whereas the other qubit is very far away, we can prove that the nature of the $XY$-interaction (ferromagnetic or anti-ferromagnetic) between the qubits is irrelevant. This could be seen explicitly by analyzing the symmetry with respect to the inversion $j\rightarrow -j$, in this situation.
More precisely, if we consider that the first qubit is located at $r_1=r$ and $\varphi_1=z_1=0$, while the second qubit is at $r_2=r+\Delta r$, $z_2=\Delta z$ and $\varphi_2=\Delta\varphi$, and we also consider that $\Delta r$ and $\Delta z$ tend to infinity (in such a way that the second qubit is very far away from the first qubit), then the distance $d_{kn}\rightarrow\infty$ for all $k=\{0,..,n-1\}$, see eq. \eqref{gkn definition 2}. Therefore, we obtain that $H$ is identically null, see \eqref{function H 2}, and $\mathcal{I}=\mathcal{J}$, see eq. \eqref{I function} and eq. \eqref{J function}. This ensures the following parity properties
of the population corrections in eq. \eqref{diagonal correction terms for density matrix2}
 \begin{align*}
\delta p_{1}(-j) & =\delta p_{1}(j),\\
\delta p_{4}(-j) & =\delta p_{4}(j),\\
\delta p_{2}(-j) & =\delta p_{3}(j).
\end{align*} 
On the other side, since $d_{kn}\rightarrow\infty$, we have that 
the function defined in eq. \eqref{YI function result3} is null, $\mathcal{Y}_I=0$. Furthermore, it can be observed that the function $\mathcal{Y}_0(\omega, j,\mathcal{T}, r)$, given in eq. \eqref{Y0 function result3}, is even with $\omega$ and odd in $j$.
Then the coherence terms correction 
$\delta\alpha$ and $\delta\beta$, defined in eq. \eqref{alpha correction terms for density matrix2} and eq. \eqref{beta correction terms for density matrix2}, behave as follows under the inversion of the sign of the coupling constant $j$ \begin{align*}
    \delta\alpha(-j)&=-\delta\alpha(j),\\
    \delta\beta(-j)&=\delta\beta^*(j).
\end{align*} 
By introducing all these results into \eqref{eigenvalues of the partial traspose of the density matrix dependent of time}, it will be observed that the negativity is even in $j$. This means that when the distance between the qubits is sufficiently large, the entanglement harvesting will not distinguish the nature of the ferromagnetic or antiferromagnetic of the qubits $XY$-interaction. Furthermore, since this does not depend on the specific value of $r$, this fact will occur also regardless of the presence of the cosmic string.

\subsubsection{Coincidence of Qubits Locations}\label{prop 7}
Let us consider the case where both qubits are located at the same point in spacetime $\chi_1(\tau)=\chi_2(\tau)=(\tau,r,\varphi,z)$.
Considering the notation for the Wightman functions given at eq. \eqref{Gnp}, we see that in the case where the qubits coincidence in space, we have $G_A(\tau,\tau')=G_{12}(\tau,\tau')=G_{11}(\tau,\tau')=G_{22}(\tau,\tau')$. Therefore, from the definitions \eqref{F function} and \eqref{H function} we obtain that the functions that give us the corrections to the population terms are exactly the same, i.e. $F=H$. This leads us to $\mathcal{I}=2F$ and $\mathcal{J}=0$, see eq. \eqref{I function} and eq. \eqref{J function} and to the following expression for the corrections  of the diagonal terms
 \begin{align*}
\delta p_{1} & =2g^{2}[ -pF (\omega -j)+(1/2-p)F (-\omega +j)],\\
\delta p_{2} & =2g^{2}[ p(F (\omega -j)+F (-\omega -j)) -(1/2-p)(F (-\omega +j)+F (\omega +j))],\\
\delta p_{3} & =0,\\
\delta p_{4} & =2g^{2}[ -pF (-\omega -j)+(1/2-p)F (\omega +j)]. 
\end{align*}
Also, since the radial positions of the qubits are the same $r_1=r_2=r$, then the corrections to the second coherence term are identically null, $\delta\beta=0$, see eq. \eqref{beta correction terms for density matrix2}. 
If we want to obtain entanglement generation for a finite time, we need to consider that the initial probability $p$ is very small and specifically $p=0$ to obtain a maximum value of Negativity when the other variables are fixed, as explained in Property \ref{prop 2}. Therefore, in this case, the entanglement generation would depend on the eigenvalue $\lambda_1$, given at eq. \eqref{eigenvalues of the partial traspose of the density matrix dependent of time}, which can be written in the case where the qubits positions coincidence as \begin{equation*}
    \lambda_1^{0}=\frac{g^2}{2}\left[F(-\omega+j)+F(\omega+j)-\sqrt{\left[F(-\omega+j)-F(\omega+j)\right]^2+\left[F(-\omega+j)+F(\omega+j)\right]^2}\right].
\end{equation*}
From this general result, it can be observed that $\lambda_1^0$ is always negative, (given a finite amount of entanglement) except for the case \begin{equation}
    F(-\omega+j)=F(\omega+j),
\end{equation} where the eigenvalue and the negativity become zero. It is worth noting that all these previous results do not depend on the specific position where the qubits coincide.
\begin{figure}[h!]
\begin{subfigure}{0.5\textwidth}
\centering
     \includegraphics[scale=0.535,frame]{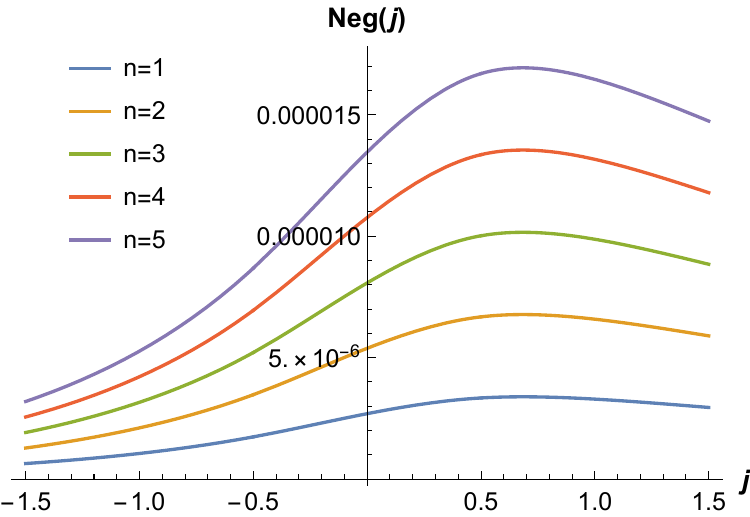}
          \caption{$\mathcal{T}$=1}
          \label{T1_omega05_j_p_}
\end{subfigure}
\begin{subfigure}{0.5\textwidth}
\centering
     \includegraphics[scale=0.55,frame]{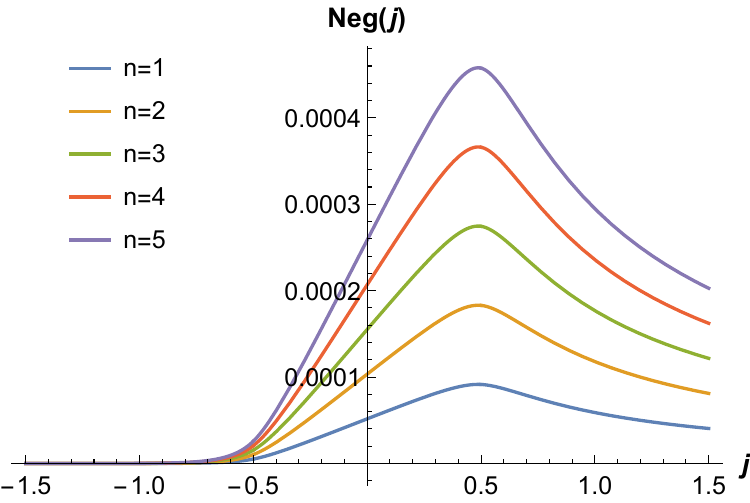}
          \caption{$\mathcal{T}$=10}
          \label{T10_omega05_j_p_}
\end{subfigure}

\centering
\begin{subfigure}{0.5\textwidth}
\centering
     \includegraphics[scale=0.55,frame]{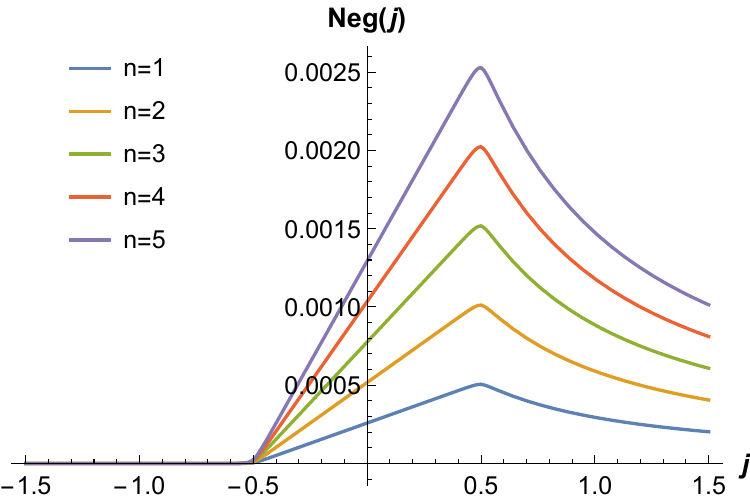}
          \caption{$\mathcal{T}$=50}
          \label{T50_omega05_j_p_}
\end{subfigure}

\caption{Figures (a), (b), and (c) show the graphs of negativity as a function of the interaction constant between the qubits $j$ in the interval [-1.5,1.5] at times $\mathcal{T}=1,10$, and $50$ respectively, evaluated with $\Delta z=\Delta r=\Delta \varphi=r=p=0$ and $\omega=0.5$. }
\label{T_omega05_j_p0_}
\end{figure}

In order to analyze more easily the different situations in which this condition is satisfied, we consider now the case $r=0$. 
In this specific situation the condition of null negativity $F(-\omega+j)=F(\omega+j)$ reduce simply to
\begin{align*}
     2(\omega +j)\mathcal{T} \Theta (\omega +j)+e^{-|\omega +j|\mathcal{T}}=2(-\omega +j)\mathcal{T} \Theta (-\omega +j)+e^{-|-\omega +j|\mathcal{T}},
\end{align*} 
this condition can be analyzed by the monotonous behavior of the function $2x\Theta(x)+e^{-|x|}$. Then, we can conclude that in the case where both arguments in the Heaviside functions are positive $j>\omega$, the only possibility to have null entanglement and negativity ($F(-\omega+j)=F(\omega+j)$) is just satisfied when $\omega=0$ or $\mathcal{T}=0$, which does not correspond to physically interesting cases to consider.

However, when both arguments in Heaviside functions are negative, $j<-\omega$, the conditions of null negativity can be approximately satisfied for very large interaction times since in this situation the exponential approach to zero. It should be noted that the same result will be obtained even when $r \neq 0$.
Thus, according to this result, when the distance between the qubits is very small and $j<-\omega$, and also the interaction time is sufficiently large, then the entanglement between the qubits will be almost zero, regardless of the position of the qubits with respect to the cosmic string.
This can be verified graphically, see figures (\ref{T1_omega05_j_p_}, \ref{T10_omega05_j_p_}, \ref{T50_omega05_j_p_}). 

\subsubsection{Infinite Distance Minkowski Limit}\label{prop 8}
It is important to analyze the behavior of the negativity when $r\rightarrow\infty$ with $\Delta\varphi\neq 0$. Under this limiting condition, we have $H=\mathcal{Y}_I=0$ and furthermore $F=F^M$ and $\mathcal{Y}_0=\mathcal{Y}^M_0$. Thus, for the eigenvalues of the eq. \eqref{eigenvalues of the partial traspose of the density matrix dependent of time}, taking into account these conditions yields
\begin{align}\label{eigenvalue lambda 1 with r infty}
\lambda _{1}^{M_{\infty}} = &  p+\frac{g^{2}}{16} \left( (1/2-2p)\left[ |\omega -j|\mathcal{T} +e^{-|\omega -j|\mathcal{T}} +|\omega +j|\mathcal{T} +e^{-|\omega +j|\mathcal{T}}\right] -\mathcal{T}\sqrt{\omega ^{2} +j^{2}}\right) ,\\
\nonumber\lambda _{2}^{M_{\infty}} = & \frac{1}{2}-p-\frac{g^{2}}{16}\Big( (1/2-2p)\left[ |\omega -j|\mathcal{T} +e^{-|\omega -j|\mathcal{T}} +|\omega +j|\mathcal{T}+e^{-|\omega +j|\mathcal{T}}\right]\\\label{eigenvalue lambda 2 with r infty}
&\qquad\qquad+\mathcal{T} |j|(|\omega |\mathcal{T} +1)e^{-|\omega |\mathcal{T}}\Big).
\end{align}

From this, it can be observed that due to the smallness of the coupling constant $g$, the negativity will be zero when the initial probabilities are different from 0 and $1/2$. Therefore, we are only interested in analyzing the cases when $p=0,1/2$.

For the case of $p=1/2$, the only eigenvalue that can be negative is $\lambda_2$. However, if we take into account that $|\omega |\mathcal{T} +1\leq e^{|\omega |\mathcal{T}}$, then by the triangular inequality \begin{equation*}
    \mathcal{T} |j|(|\omega |\mathcal{T} +1)e^{-|\omega |\mathcal{T}} \leq \mathcal{T} |j|\leq \frac{1}{2}[ |\omega +j|\mathcal{T} +|-\omega +j|\mathcal{T}],
\end{equation*} the eigenvalue $\lambda_2(p=1/2)$ will always be positive, resulting in the negativity also being zero for this case.

For the case of $p=0$, the eigenvalue that can be negative is $\lambda_1$. Now, if we take into account the inequality $|\omega ^{2} -j^{2} |\leq \omega ^{2} +j^{2}$, which is always satisfied, we can quickly see that $|\omega -j|+|\omega +j| \leq 2\sqrt{\omega^{2} +j^{2}}$ and therefore, the eigenvalue $\lambda_1(p=0)$ will not be positive as long as the time $\mathcal{T}$ is sufficiently large for the sum of the exponentials $e^{-|\omega +j|\mathcal{T}}$ and $e^{-|\omega -j|\mathcal{T}}$ to be negligible. Furthermore, $\lambda_1$ will always be positive when $j=0$\footnote{The eigenvalue $\lambda_1$ will also always be positive when $\omega=0$, but this case is not of interest as the two qubits become simply two identical systems occupying the same state.}. Thus, the Negativity will be nonzero when the interaction time is sufficiently large and $j\neq 0$.

This limiting condition $r\rightarrow\infty$ with $\Delta\varphi\neq 0$ is very important as it allows us to observe the behavior of the negativity in terms of $r$. In fact, since the Negativity is generally nonzero only for $p=0$, it will be sufficient to analyze the functions $F, H$, and $\mathcal{Y}_0$ to determine the behavior of the Negativity. Thus, from expressions \eqref{function F 2}, \eqref{Y0 function result3}, and \eqref{function H 2}, it is necessary to observe that these functions contain terms of the form $\sin(x)/x$ and $1/(x^2+\mathcal{T}^2)$, which decay to zero as $r$ increases. Additionally, since the function $\sin(x)$ is involved, this decay occurs in an oscillatory manner, with the number of oscillation peaks depending on $\omega$, $j$, and $\sin(\pi k/n)$. Therefore, at a fixed interaction time with $\Delta\varphi\neq 0$, the Negativity will reach a maximum value and then decrease, oscillating towards the Negativity value given by the eigenvalue $\lambda_1$ in eq. \eqref{eigenvalue lambda 1 with r infty}. Now, if $\Delta\varphi=0$, then the Negativity will decay to the Negativity value in Minkowski space. It is worth noting that the Negativity value given by the eigenvalue $\lambda_1$ in eq. \eqref{eigenvalue lambda 1 with r infty} corresponds to the Negativity value in Minkowski space with $d\rightarrow\infty$.

\subsubsection{Entanglement behavior in Energy Phase Space $(\omega, j)$}\label{prop 9}

Regarding the behavior of the Negativity with $\omega$, we can use the limiting condition from Property \ref{prop 8} and analyze the eigenvalue $\lambda_1$ with $p=0$. Thus, when $\omega\leq|j|$, the negativity increases with $\omega$, and when $\omega>|j|$, the negativity decreases with $\omega$. Analyzing the energy levels of the two-qubit system, we see that when $\omega\leq|j|$, the state $|E_3\rangle$ becomes the ground state if $\omega\leq j$, and if $\omega>j$, the state $|E_4\rangle$ becomes the ground state. 

\subsubsection{Entanglement as a function of the qubits distance $d_{kn}$.}\label{prop 10}

Finally, following the same reasoning as the Property \ref{prop 8}, we can also get an idea of the behavior of the negativity in terms of the distance between qubits $d_{kn}$. It will reach a maximum at a certain point with sufficiently small $d_{kn}$ and then decay, oscillating towards a negativity value that depends on a fixed $r$ and $n$. In particular, for $j=0$ with the limiting conditions $d_{kn}\rightarrow \infty$ and $p=0$, the eigenvalue $\lambda_1$ of \eqref{eigenvalues of the partial traspose of the density matrix dependent of time} is given by \begin{align*}
\lambda_{1} = \frac{g^2}{16} e^{-\omega \mathcal{T}}\left[ 1+\frac{\mathcal{T}^{2}}{2}\sum _{k=1}^{n-1}\left(\frac{1}{4r_{1}^{2}\sin^{2} (\pi k/n)+\mathcal{T}^{2}} +\frac{1}{4r_{2}^{2}\sin^{2} (\pi k/n)+\mathcal{T}^{2}}\right)\right],
\end{align*} from which it can be observed that it is always positive and therefore the negativity is zero. Thus, when $j=0$ at a fixed interaction time, the negativity as a function of $d_{kn}$ reaches a maximum when it is sufficiently small and then decreases to zero.

Before performing the numerical study, it is necessary to specify a range of values for the different variables in which the negativity will be evaluated. Recalling what was mentioned in the Property \ref{prop 2}, this range of values for $\mathcal{T}$, $\omega$, $j$, and $g$ must ensure that the corrections are very small. Thus, we will consider $g=0.01$, $0 \leq \mathcal{T} \leq 1000$, $0 < \omega \leq 1.5$, and $-1.5 \leq j \leq 1.5$. Additionally, although $r$ and $d_{kn}$ do not need to guarantee small corrections, we will consider $0 \leq r \leq 500$ and $0 \leq d_{kn} \leq 500$. Lastly, we could consider the entire valid range for initial probabilities, $0 \leq p \leq 0.5$. However, due to what was stated in Property \ref{prop 2} and as we will show in some graphs, the negativity takes the following form $\mathcal{N}(p)\approx -c_1p+c_2$ with $c_1,c_2>0$, for very small values of $p$, observing that its maximum value occurs when $p=0$. Therefore, we will consider only $p=0$.

\subsection{THE ANGULAR CASE}
\begin{figure}[h!]
\centering
     \begin{tikzpicture}[x=0.75pt,y=0.75pt,yscale=-1,xscale=1]

\draw    (331.55,13.99) -- (332,153) ;
\draw  [draw opacity=0][fill={rgb, 255:red, 74; green, 144; blue, 226 }  ,fill opacity=0.2 ] (252.3,36.15) -- (530.44,36.15) -- (411.24,130.84) -- (133.1,130.84) -- cycle ;
\draw    (331.77,83.49) -- (405.96,51.42) ;
\draw [shift={(407.8,50.63)}, rotate = 156.62] [fill={rgb, 255:red, 0; green, 0; blue, 0 }  ][line width=0.08]  [draw opacity=0] (12,-3) -- (0,0) -- (12,3) -- cycle    ;
\draw    (331.77,83.49) -- (399.95,111.47) ;
\draw [shift={(401.8,112.23)}, rotate = 202.31] [fill={rgb, 255:red, 0; green, 0; blue, 0 }  ][line width=0.08]  [draw opacity=0] (12,-3) -- (0,0) -- (12,3) -- cycle    ;
\draw  [draw opacity=0][fill={rgb, 255:red, 0; green, 0; blue, 0 }  ,fill opacity=0.5 ] (348.21,77.12) .. controls (350.17,79) and (351.3,81.23) .. (351.29,83.62) .. controls (351.27,85.78) and (350.32,87.8) .. (348.67,89.54) -- (331.59,83.49) -- cycle ; \draw  [draw opacity=0] (348.21,77.12) .. controls (350.17,79) and (351.3,81.23) .. (351.29,83.62) .. controls (351.27,85.78) and (350.32,87.8) .. (348.67,89.54) ;
\draw [color={rgb, 255:red, 208; green, 2; blue, 27 }  ,draw opacity=1 ] [dash pattern={on 4.5pt off 4.5pt}]  (407.8,50.63) .. controls (436.2,81.43) and (419.4,101.02) .. (401.8,112.23) ;

\draw (373.13,48.67) node [anchor=north west][inner sep=0.75pt]  [font=\Large]  {$r$};
\draw (355.78,100.29) node [anchor=north west][inner sep=0.75pt]  [font=\Large]  {$r$};
\draw (352.73,75.6) node [anchor=north west][inner sep=0.75pt]  [font=\normalsize]  {$\Delta \varphi $};

\end{tikzpicture}
      \caption{Qubits position in the angular case.}
      \label{angular case disposition}
\end{figure}
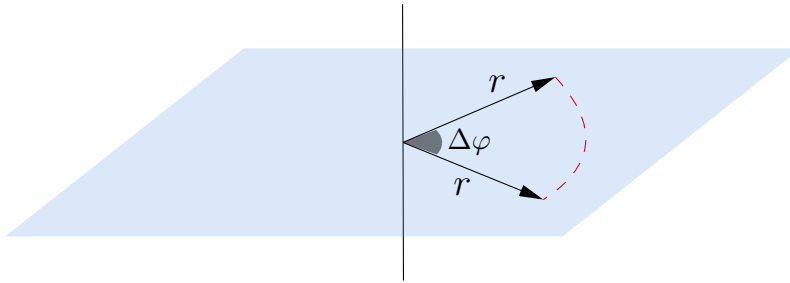

In this section, we analyze the entanglement dynamics between the qubits when they are localized on a plane (orthogonal to the cosmic string) and at the same radial distance to the string. In this case, the distance between the qubits is characterized only by the angular separation $\Delta \varphi$, while the distance to the cosmic string is given by  $r$, see figure \ref{angular case disposition}. Hence the qubits position in this case are given by

\begin{align}
 \chi_1(\tau)&=(\tau,r,\varphi_1,z),\nonumber\\
\chi_2(\tau)&=(\tau,r,\varphi_2,z).  
\end{align}
%

The Wightman function between the different qubits, denoted as $G^+_{12}(\tau,\tau')=G^+_{21}(\tau,\tau')=G^+(\chi_1(\tau),\chi_2(\tau'))$, is given in this case by \begin{equation}
    G^+_{12}(\tau,\tau')=-\frac{1}{4\pi^2}\sum^{n-1}_{k=0}\frac{1}{(\Delta\tau-i\varepsilon)^2-d^2_{kn}},
\end{equation} where, the distance $d_{kn}$ is defined here as 
\begin{equation}
    d^2_{kn}=4r^2\sin^2(\pi k/n+\Delta \varphi/2),
\end{equation}
and the angular distance is $\Delta\varphi=\varphi_1-\varphi_2$. Now, we proceed to analyze the entanglement generation on the pair of qubits as a function of the different parameters in this special angular configuration of qubits locations. We realize this by setting some values of the parameters fixed while allowing us to change others and plotting the negativity as a corresponding function.


    \begin{figure}[t!]
    \centering
     \includegraphics[scale=0.58,frame]{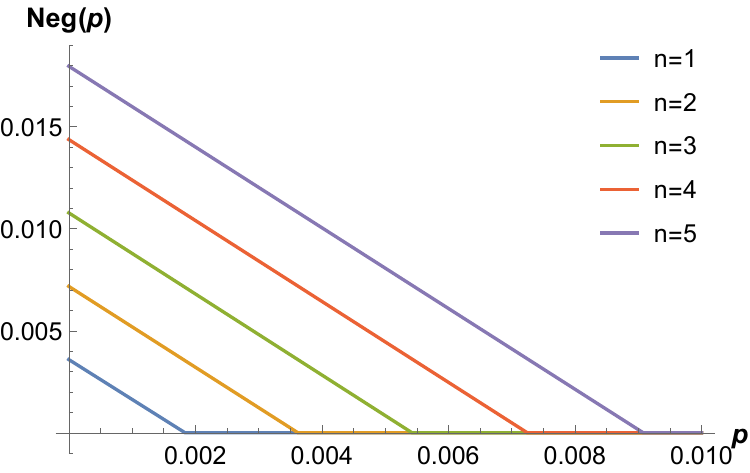}
\caption{Negativity as a function of the initial probability $p$ for different cosmic string quantization values $n$.  We consider $\mathcal{T}=1000$, $\omega=1$, $j=0.5$, $\Delta \varphi=\pi/n$, and $r=1$.}
\label{T100_omega1_j05_varphi314on_r10_p}
\end{figure}

\begin{itemize}
    \item 
        \textit{\underline{Negativity dependence on the initial probability} }\\
    As established before, the interaction of the qubit with the vacuum fluctuations begins at a zero entanglement density operator parametrized by the probability $0<p<1/2$, see eq. (\ref{initial density matrix}).
    The figure \ref{T100_omega1_j05_varphi314on_r10_p}, shows the negativity as a function of $p$. We see that a linear dependence on $p$ is manifest. This is expected since in the perturbation regime all the corrections to the density operator are linear in $p$, see eq. (\ref{diagonal correction terms for density matrix2}). From figure \ref{T100_omega1_j05_varphi314on_r10_p},    
    it can be noted that a non-null entanglement generation only occurs for the values of $p$ close to zero. Also, one can see that after some maximum critical value of initial probability, there occurs a \textit{sudden death} of entanglement. This is in accordance with what was stated previously in Property \ref{prop 2}. Therefore, for the following analysis, we only need to consider a null initial probability, $p=0$, in order to inspect the amount of entanglement harvesting.

\begin{figure}[th!]
\begin{subfigure}{0.5\textwidth}
\centering
     \includegraphics[scale=0.55,frame]{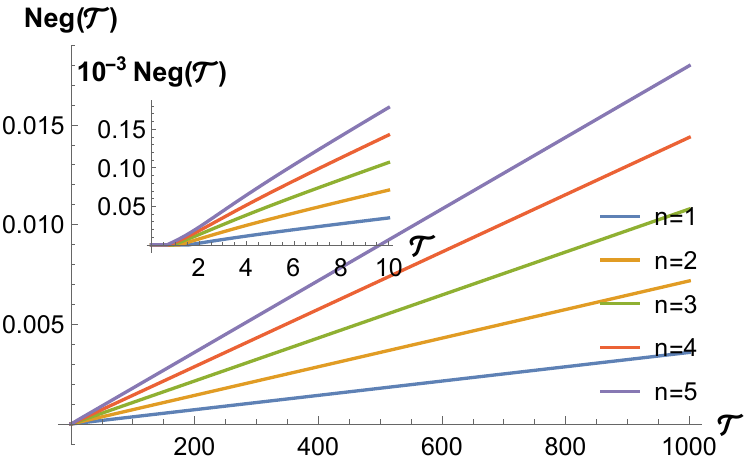}
          \caption{$r=1, \omega=1, j=0.5$}
\end{subfigure}
\begin{subfigure}{0.5\textwidth}
\centering
     \includegraphics[scale=0.55,frame]{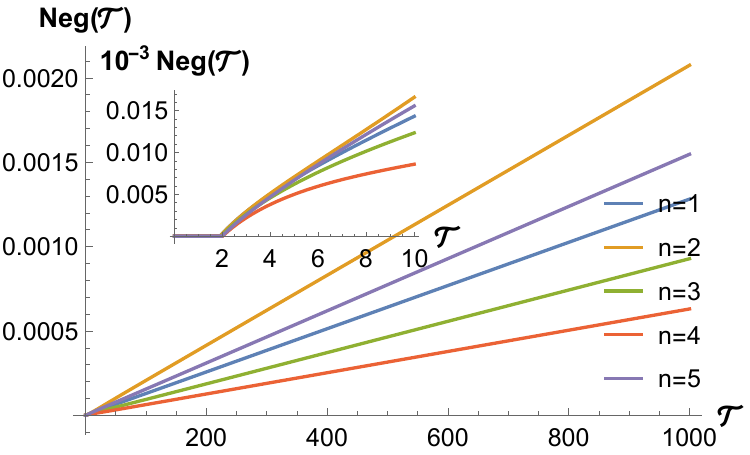}
        \caption{$r=10, \omega=1, j=0.5$}
\end{subfigure}

\begin{subfigure}{0.5\textwidth}
\centering
     \includegraphics[scale=0.55,frame]{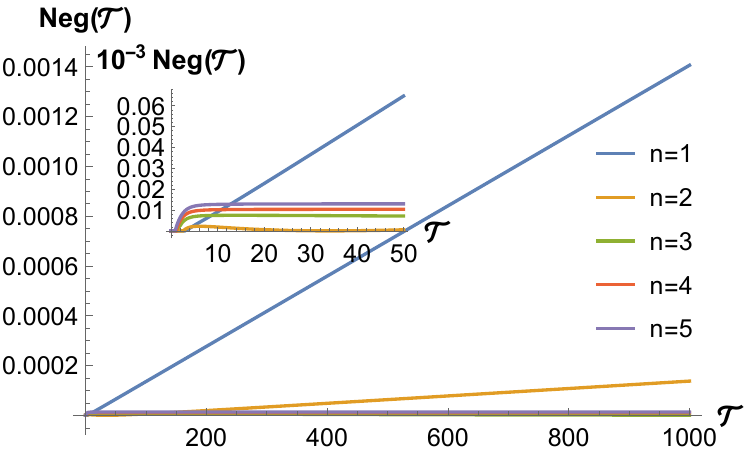}
          \caption{$r=1, \omega=0.5, j=-0.5$}
\end{subfigure}
\begin{subfigure}{0.5\textwidth}
\centering
     \includegraphics[scale=0.55,frame]{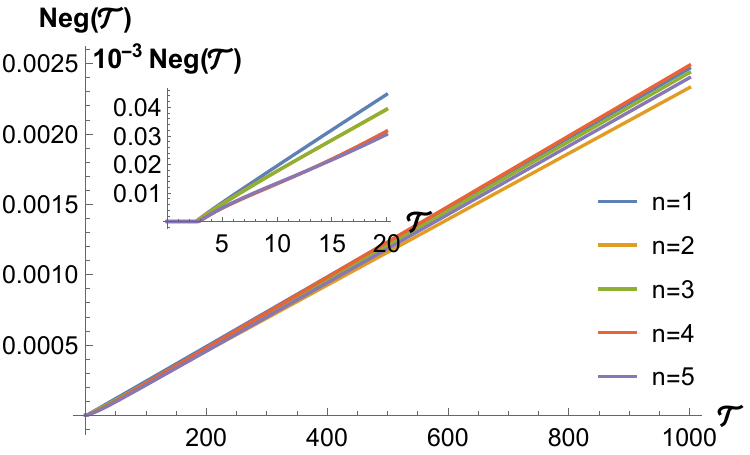}
        \caption{$r=10, \omega=0.5, j=-0.5$}
\end{subfigure}

\begin{subfigure}{0.5\textwidth}
\centering
     \includegraphics[scale=0.55,frame]{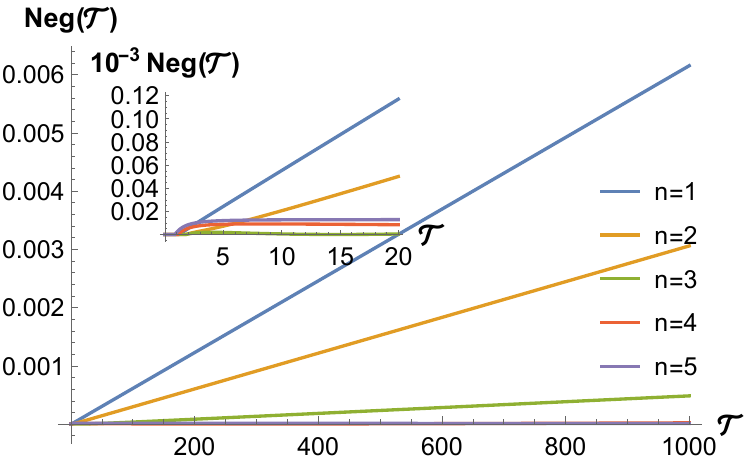}
          \caption{$r=1, \omega=1, j=-1$}
\end{subfigure}
\begin{subfigure}{0.5\textwidth}
\centering
     \includegraphics[scale=0.55,frame]{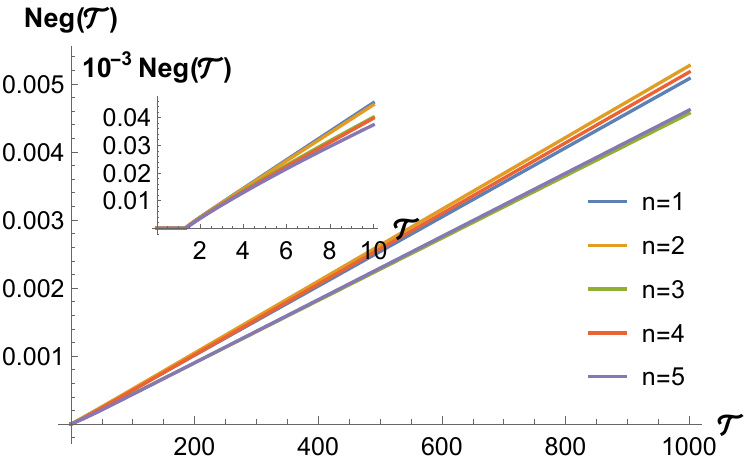}
        \caption{$r=10,  \omega=1, j=-1$}
\end{subfigure}

\caption{Negativity as a function of the interaction time $\mathcal{T}$. We consider a maximum angular distance of $\pi/n$ between the qubits for all the values of string energy density.}
\label{T_omega_j_r_phipiovern_n}
\end{figure}

    \item 
    
    \textit{\underline{Time evolution of entanglement harvesting}}\\
    We now look for the evolution in time of the entanglement harvesting process for the angular configuration of the qubits.
    In this way, figure \ref{T_omega_j_r_phipiovern_n} shows the behavior of the Negativity with respect to interaction time $\mathcal{T}$. From this, one sees that entanglement negativity is null at the initial time and it remains zero for some finite small amount of interaction time. After some minimum critical time, we see a \textit{sudden birth of entanglement}. After this, one notes that for large values of interaction time the dependence of negativity is linear on time. 
    This behavior agrees with the general property of Negativity stated previously at Property \ref{prop 5}. Then we see that negativity diverges as time goes to infinity. This 
    indicates that the perturbation theory we use here is valid only for small values of time and for longer values of ${\cal T}$ more terms in the Dyson series must be considered.
    Nonetheless this general behavior, we want to call attention to the specific cases shown in figures (\ref{T_omega_j_r_phipiovern_n}c) and (\ref{T_omega_j_r_phipiovern_n}d) when the interaction between the qubits is $j=-\omega$, and the distance to the cosmic string $r$ is small. When these conditions are fulfilled, one sees that the negativity does not diverge but has a constant asymptotic behavior for a sufficiently long interaction time $\mathcal{T}$ and high enough values of the string energy density $n$.

\begin{figure}[b!]
\begin{subfigure}{0.325\textwidth}
\centering
     \includegraphics[scale=0.38,frame]{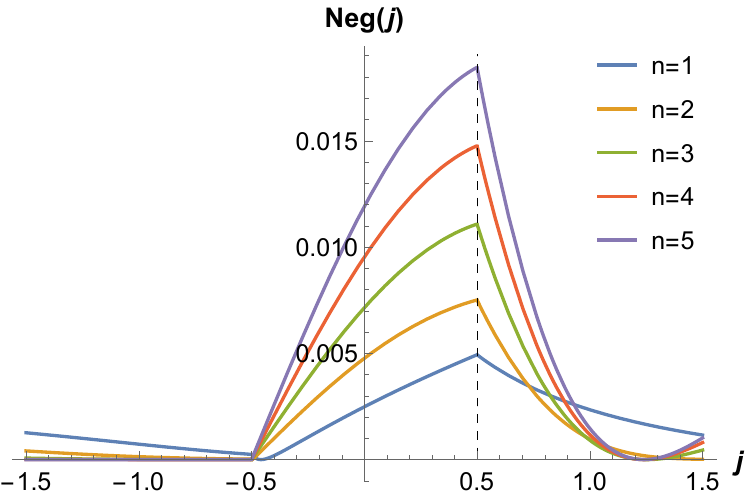}
          \caption{$r=1,\Delta \varphi=\pi/4n$}
\end{subfigure}
\begin{subfigure}{0.325\textwidth}
\centering
     \includegraphics[scale=0.38,frame]{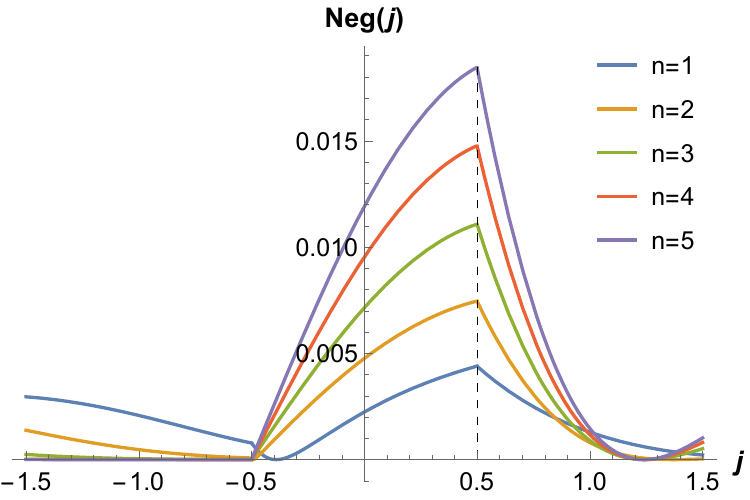}
        \caption{$r=1,\Delta \varphi=\pi/2n$}
\end{subfigure}
\begin{subfigure}{0.325\textwidth}
\centering
     \includegraphics[scale=0.38,frame]{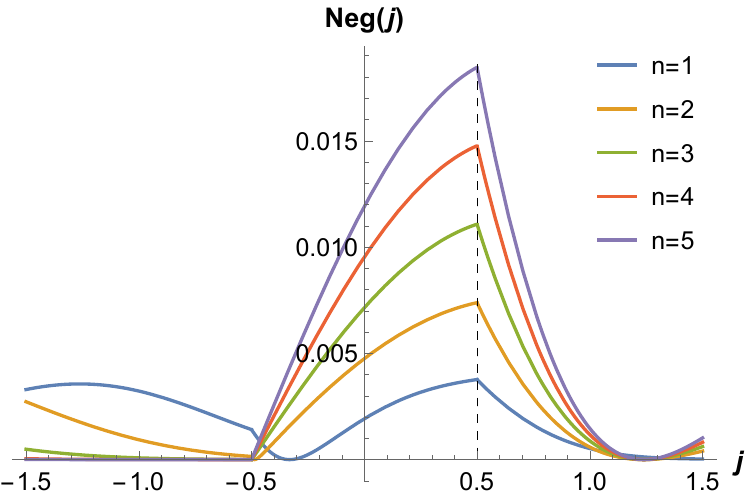}
        \caption{$r=1,\Delta \varphi=\pi/n$}
\end{subfigure}

\begin{subfigure}{0.325\textwidth}
\centering
     \includegraphics[scale=0.38,frame]{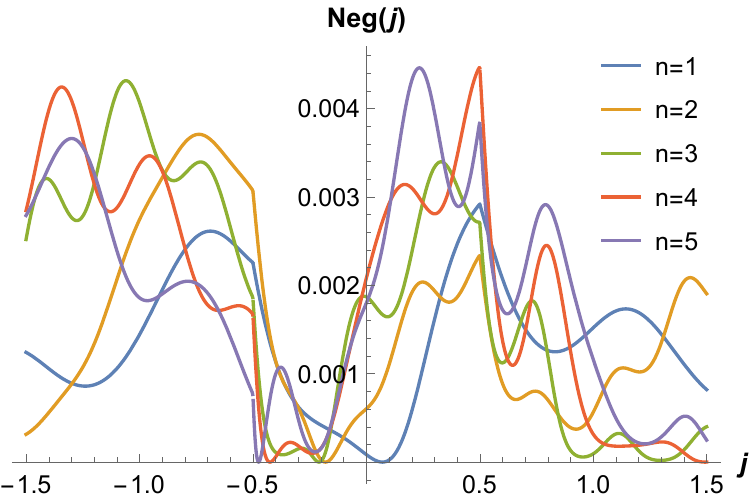}
          \caption{$r=10,\Delta \varphi=\pi/4n$}
\end{subfigure}
\begin{subfigure}{0.325\textwidth}
\centering
     \includegraphics[scale=0.38,frame]{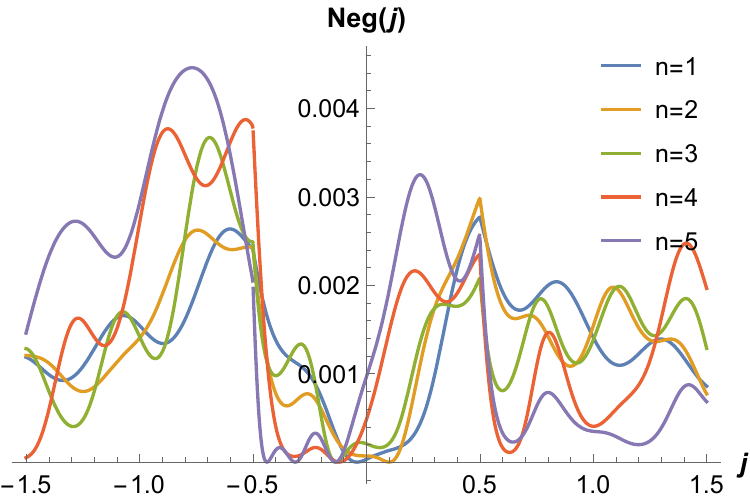}
        \caption{$r=10,\Delta \varphi=\pi/2n$}
\end{subfigure}
\begin{subfigure}{0.325\textwidth}
\centering
     \includegraphics[scale=0.38,frame]{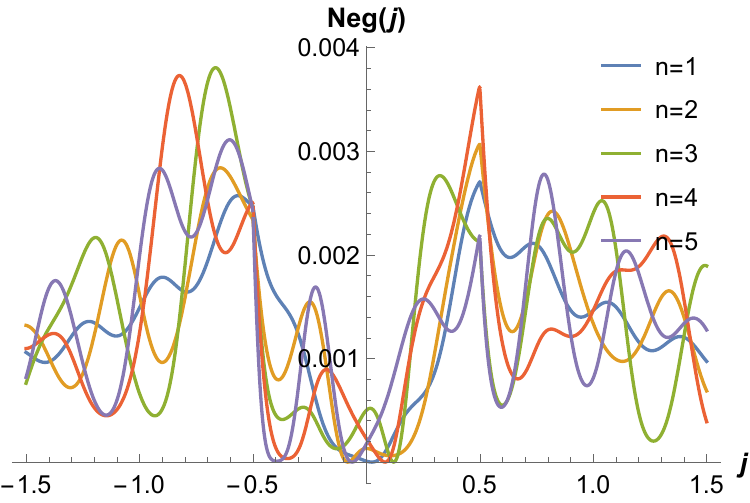}
        \caption{$r=10,\Delta \varphi=\pi/n$}
\end{subfigure}

\begin{subfigure}{0.325\textwidth}
\centering
     \includegraphics[scale=0.38,frame]{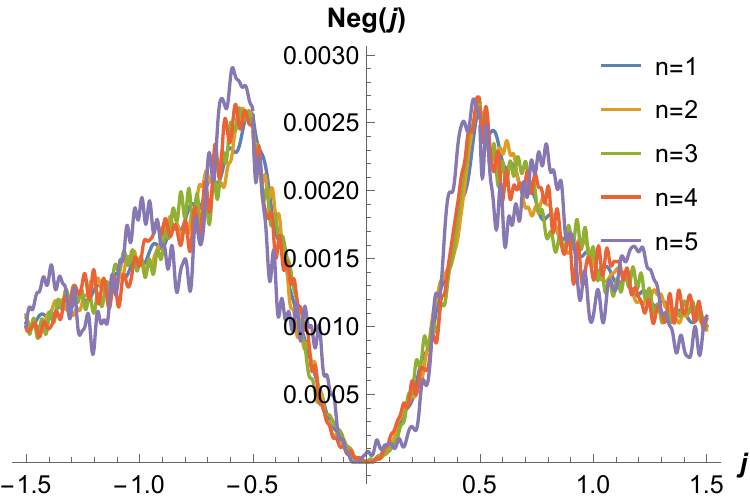}
          \caption{$r=100,\Delta \varphi=\pi/4n$}
\end{subfigure}
\begin{subfigure}{0.325\textwidth}
\centering
     \includegraphics[scale=0.38,frame]{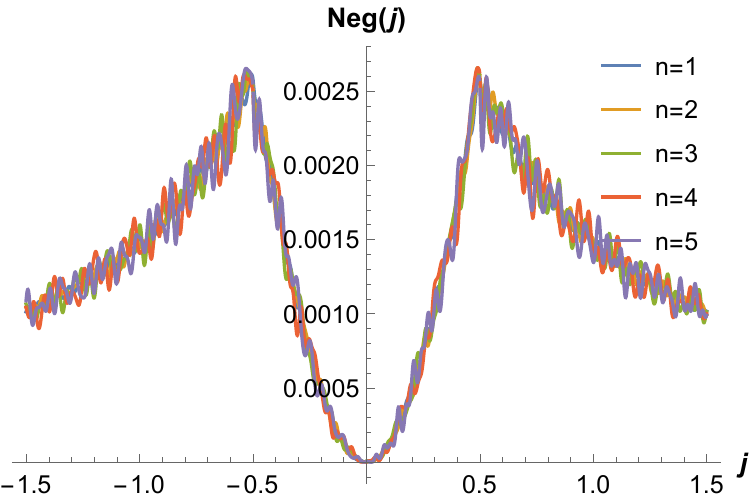}
        \caption{$r=100,\Delta \varphi=\pi/2n$}
\end{subfigure}
\begin{subfigure}{0.325\textwidth}
\centering
     \includegraphics[scale=0.38,frame]{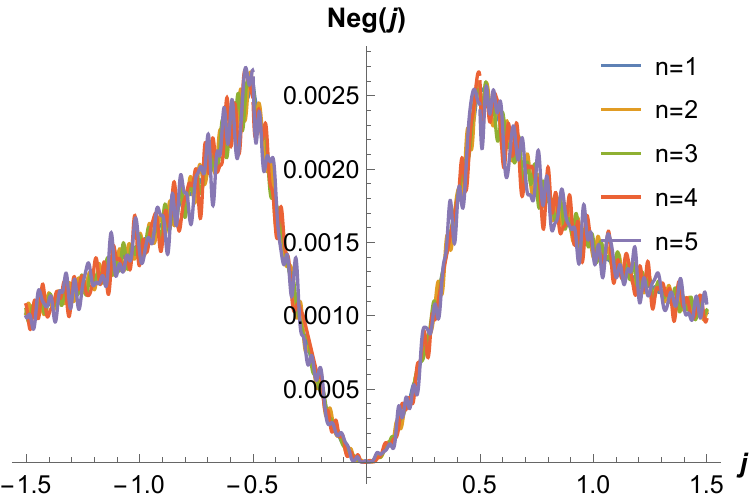}
        \caption{$r=100,\Delta \varphi=\pi/n$}
\end{subfigure}
\caption{Negativity as a function of the Heisenberg interaction coupling constant $j$ for different values of $r$ and $\Delta\varphi$. In all the graphs we set $\mathcal{T}=1000$ and $\omega=0.5$.}
\label{T100_omega05_j_r_phi}
\end{figure}
\item 
\underline{\textit{Negativity as function of $XY$-Heinsenberg interaction coupling}}\\
In this work, we consider a direct interaction between the qubits that brings the possibility of enhancement of entanglement harvesting. We consider this as a general $XY$-Heisenberg interaction term between the qubits that resembles the coupling between magnetic moments in planar magnetism. By inspection of the qubits Hamiltonian, eq. (\ref{hamiltoniano libre qubits}), we see that the interaction coupling $j<0$ refers to a ferromagnetic interaction while $j>0$ gives us an anti-ferromagnetic term. We want to explore the dependence of Negativity on this parameter and obtain how the presence and nature of this direct interaction affect entanglement generation between the qubits.
Hence, in figure \ref{T100_omega05_j_r_phi} it is shown the dependence of Negativity on $j$ for different values of $r$ and $\Delta\phi$. We first note that for small values of $r$, there appears a maximum value of entanglement at $j=\omega$. Then at these small distances, the presence of an anti-ferromagnetic interaction that resonates with the qubits energy gap is beneficial for the entanglement harvesting process. Still, for small distances, we note that the Negativity
behaves in the same manner for all values of string energy density $n$
and is proportional to $n$.
This is a verification of Property \ref{prop 4}, since for small $r$ both qubits are located on the cosmic string and the results are identical to the Minkowski limit (without cosmic string) but with a proportionality constant $n$.
Now, on the other hand, for intermediate distances an oscillatory dependence on Negativity on $j$ appears, nonetheless it is always possible to find some value $j$ that overcomes the entanglement harvesting at $j=0$, then showing an enhancement of harvesting. A symmetry between the ferromagnetic and anti-ferromagnetic interactions appears for large distances $r$. There we note that the Negativity is even with respect to $j$ and a maximum value of entanglement is obtained at the resonance values $j=\pm\omega$.
For these values of $j$, we notice that the qubits energy structure is degenerate and an abrupt change of the ground state occurs. This is an indication of a quantum phase transition of the qubits system at $j=\pm\omega$ leading to a maximum entanglement between the qubits at these points. It is worth noting that in the angular case, the distance between the qubits is proportional to the distance of them to the string, $r$, then for large $r$ we have a large separation between the qubits and we expect an equivalence of the ferromagnetic and antiferromagnetic interaction as established at Property \ref{prop 6}.

\begin{figure}[h!]
\begin{subfigure}{0.325\textwidth}
\centering
  \includegraphics[scale=0.38,frame]{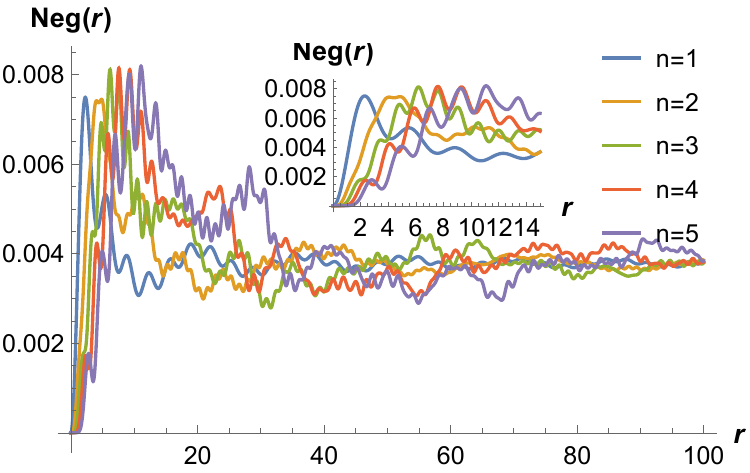}
          \caption{$j=-1.5; \Delta\varphi=\pi/4n$}
\end{subfigure}
\begin{subfigure}{0.325\textwidth}
\centering
\includegraphics[scale=0.38,frame]{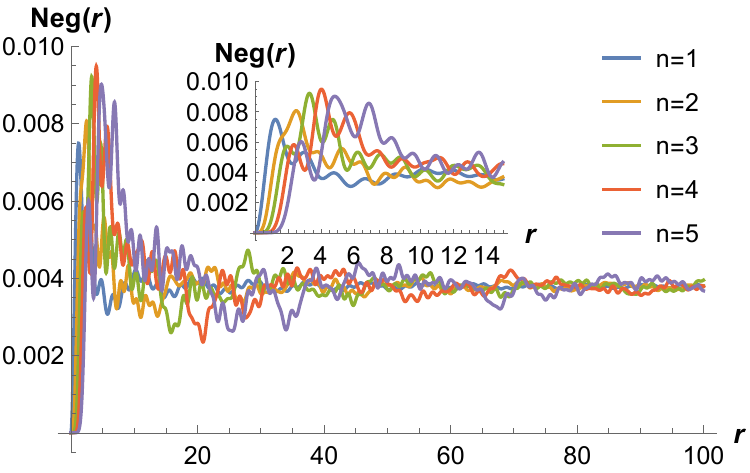}
        \caption{$j=-1.5; \Delta\varphi=\pi/2n$}
\end{subfigure}
\begin{subfigure}{0.325\textwidth}
\centering
     \includegraphics[scale=0.38,frame]{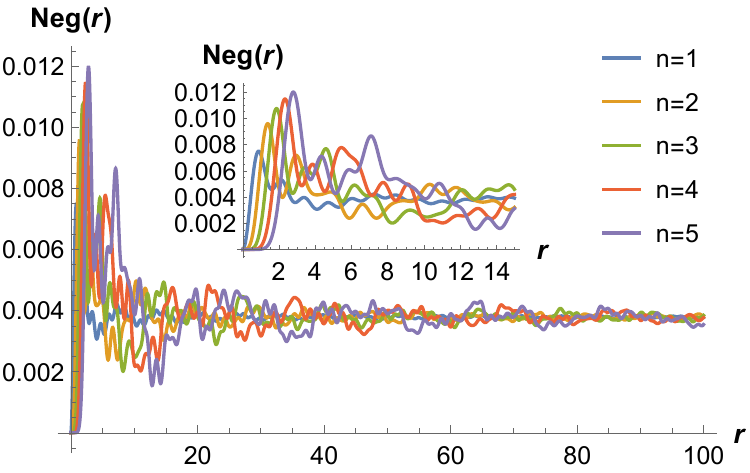}
        \caption{$j=-1.5; \Delta\varphi=\pi/n$}
\end{subfigure}

\begin{subfigure}{0.325\textwidth}
\centering
     \includegraphics[scale=0.38,frame]{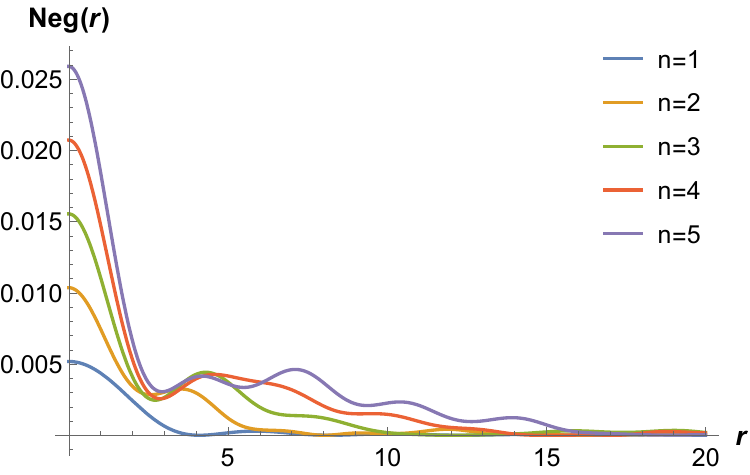}
          \caption{$j=0; \Delta\varphi=\pi/4n$}
\end{subfigure}
\begin{subfigure}{0.325\textwidth}
\centering
     \includegraphics[scale=0.38,frame]{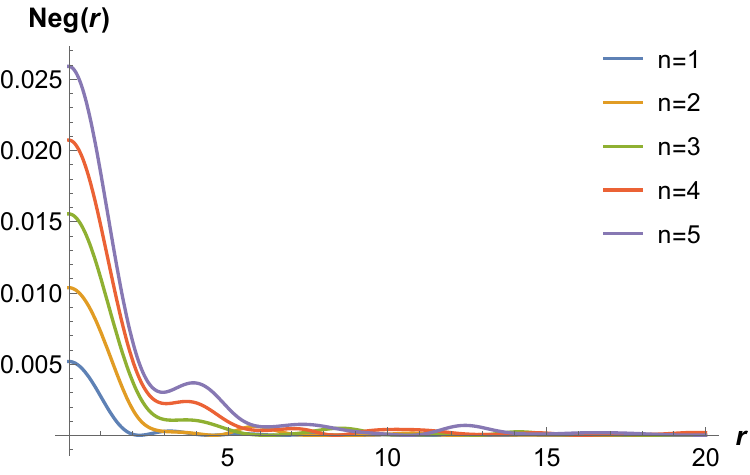}
        \caption{$j=0; \Delta\varphi=\pi/2n$}
\end{subfigure}
\begin{subfigure}{0.325\textwidth}
\centering
     \includegraphics[scale=0.38,frame]{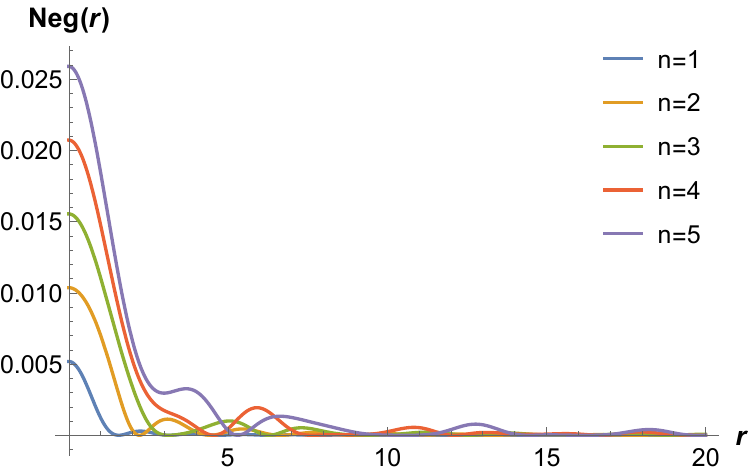}
        \caption{$j=0; \Delta\varphi=\pi/4n$}
\end{subfigure}

\begin{subfigure}{0.325\textwidth}
\centering
     \includegraphics[scale=0.38,frame]{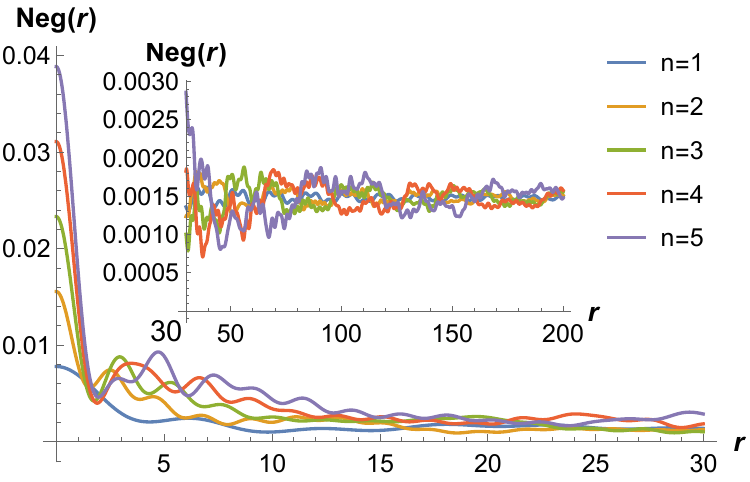}
          \caption{$j=0.5; \Delta\varphi=\pi/4n$}
\end{subfigure}
\begin{subfigure}{0.325\textwidth}
\centering
     \includegraphics[scale=0.38,frame]{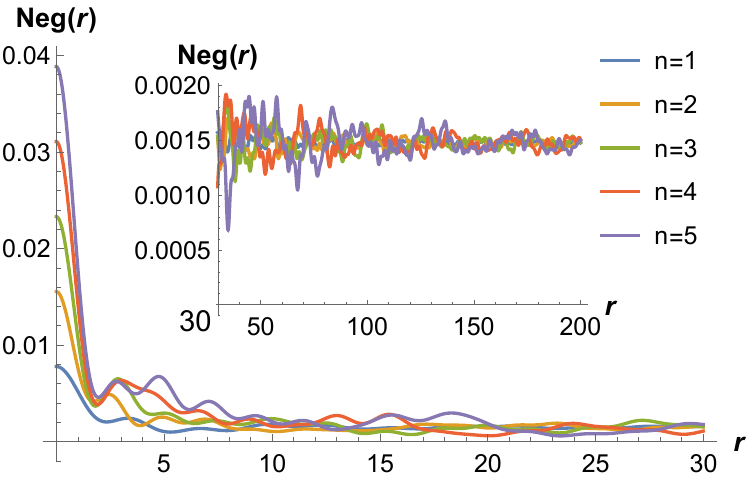}
        \caption{$j=0.5; \Delta\varphi=\pi/2n$}
\end{subfigure}
\begin{subfigure}{0.325\textwidth}
\centering
     \includegraphics[scale=0.38,frame]{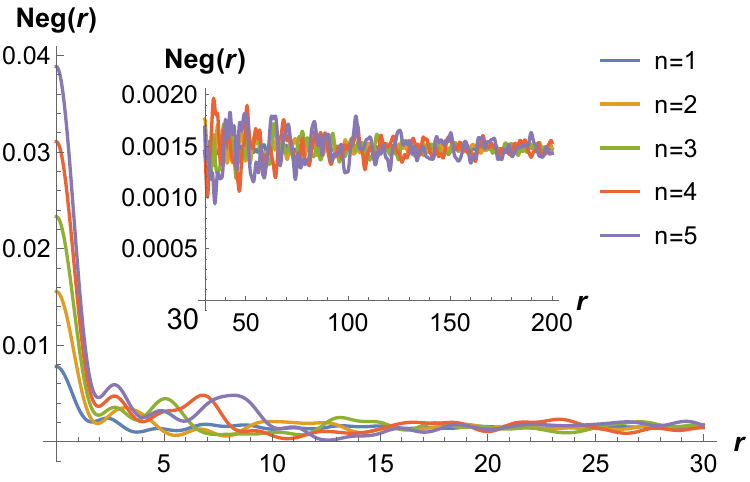}
        \caption{$j=0.5; \Delta\varphi=\pi/n$}
\end{subfigure}

\caption{Negativity as function of $r$ for different values of coupling constant $j$ and angular distance $\Delta\varphi$. We use $\mathcal{T}=1000$ and $\omega=1$.}
\label{T100_omega1_j_r_phi_n}
\end{figure}

\item  
\textit{\underline{Entanglement harvesting as a function of the distance from cosmic string}}\\
It is interesting to analyze the qubits dependence on the distance to the cosmic string. In order to do this, figure  \ref{T100_omega1_j_r_phi_n},
shows the Negativity as a function of $r$ for different values of coupling constant $j$ and angular distance $\Delta\varphi$.
There one notice that for ferromagnetic interaction, $j<0$, figures (\ref{T100_omega1_j_r_phi_n}a), (\ref{T100_omega1_j_r_phi_n}b) and (\ref{T100_omega1_j_r_phi_n}c),
the Negativity begins at zero for small distance, oscillates as $r$ increases and tends asymptotically to a finite value at large values of $r$. This is in contrast with the behavior in the absence of the $XY$-Heisenberg interaction, $j=0$. In this latter case, figures (\ref{T100_omega1_j_r_phi_n}d), (\ref{T100_omega1_j_r_phi_n}e) and (\ref{T100_omega1_j_r_phi_n}f), 
we note that the Negativity starts a finite maximum value, oscillates and decays asymptotically to zero.\\
For other side, for the antiferromagnetic interaction, $j>0$, we obtain, at figures
(\ref{T100_omega1_j_r_phi_n}g), (\ref{T100_omega1_j_r_phi_n}h) and (\ref{T100_omega1_j_r_phi_n}j),
that entanglement is maximum for small values of $r$, oscillates and tends to a finite value for $r$ very large. 
In general, as $r$ increases, the distance between the qubits also increases and the Negativity tends towards the value given by $\lambda^{M_\infty}_1$, see eq. (\ref{eigenvalue lambda 1 with r infty}), in accordance with Property \ref{prop 8}.\\
We conclude then, that a direct interaction (ferromagnetic or antiferromagnetic) increases the entanglement harvesting for large distances, leading to a finite amount of Negativity as $r$ tends to infinity. For the case without interaction, $j=0$, Negativity tends to zero for large distances from the cosmic string. This is in accordance with the general result indicated as Property \ref{prop 8}.\\
Now, by inspection of figure  \ref{T100_omega1_j_r_phi_n}, one notice that at very small distance $r\approx0$, there is a maximum value for the antiferromagnetic ($j>0$) and non-interacting ($j=0$) case, while Negativity is zero for small distances in the ferromagnetic case ($j<0$).
This is in accordance with what was shown in figures (\ref{T100_omega05_j_r_phi}a),
(\ref{T100_omega05_j_r_phi}b), and (\ref{T100_omega05_j_r_phi}c),
where we see that when both qubits are located very close to each other there is a maximum value of entanglement for the antiferromagnetic resonance case of $j=\omega$ while the ferromagnetic resonance value $j=-\omega$ leads us to zero Negativity. This is what was established in general terms at Property \ref{prop 7}.\\
Finally, the Negativity dependence on $d_{kn}$ 
is made explicit
in figures (\ref{T100_omega1_j_r_phi_n}d), 
(\ref{T100_omega1_j_r_phi_n}e), and 
(\ref{T100_omega1_j_r_phi_n}f), for $j=0$. There one can see that the negativity follows the behavior described in Property \ref{prop 10}.
 \begin{figure}[h!]
\begin{subfigure}{0.5\textwidth}
\centering
     \includegraphics[scale=0.58,frame]{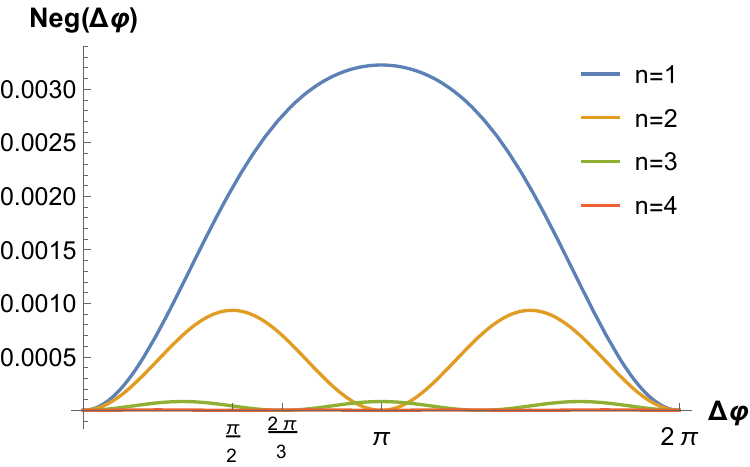}
          \caption{$r=1; j=-1$}
\end{subfigure}
\begin{subfigure}{0.5\textwidth}
\centering
     \includegraphics[scale=0.58,frame]{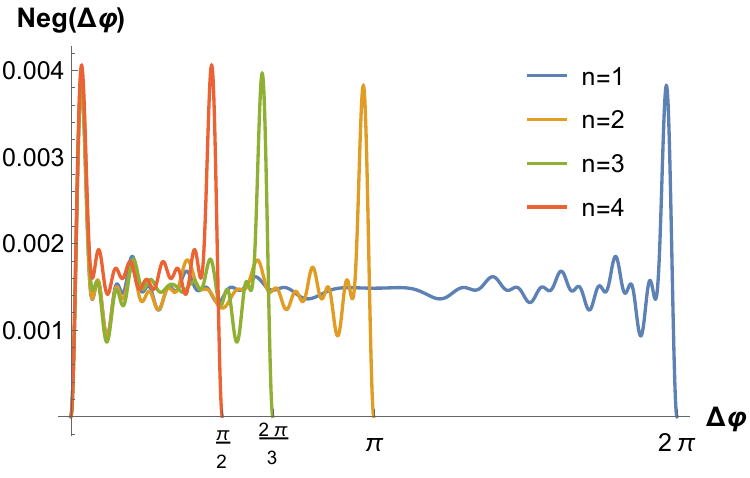}
          \caption{$r=25; j=-1$}
\end{subfigure}
\begin{subfigure}{0.5\textwidth}
\centering
     \includegraphics[scale=0.58,frame]{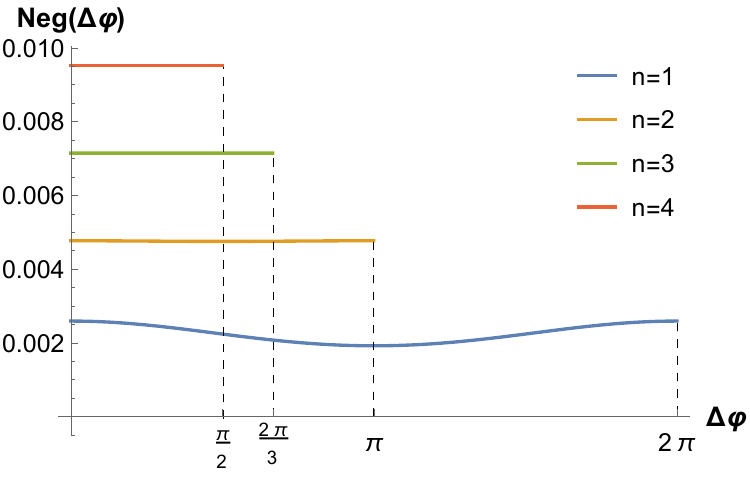}
          \caption{$r=1; j=0$}
\end{subfigure}
\begin{subfigure}{0.5\textwidth}
\centering
     \includegraphics[scale=0.58,frame]{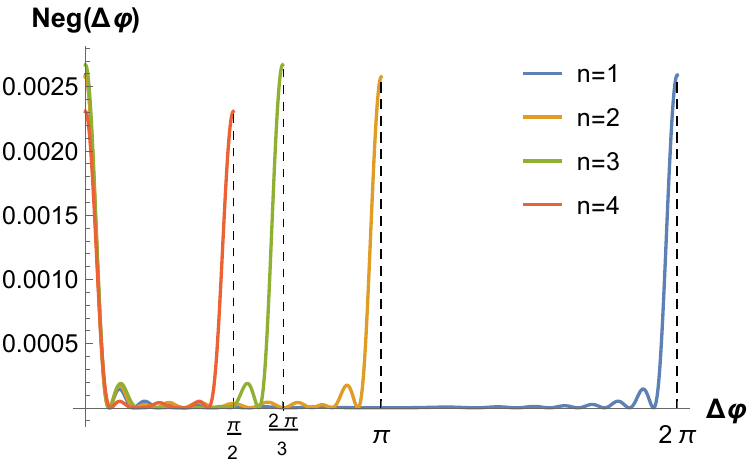}
          \caption{$r=25; j=0$}
\end{subfigure}
\begin{subfigure}{0.5\textwidth}
\centering
     \includegraphics[scale=0.58,frame]{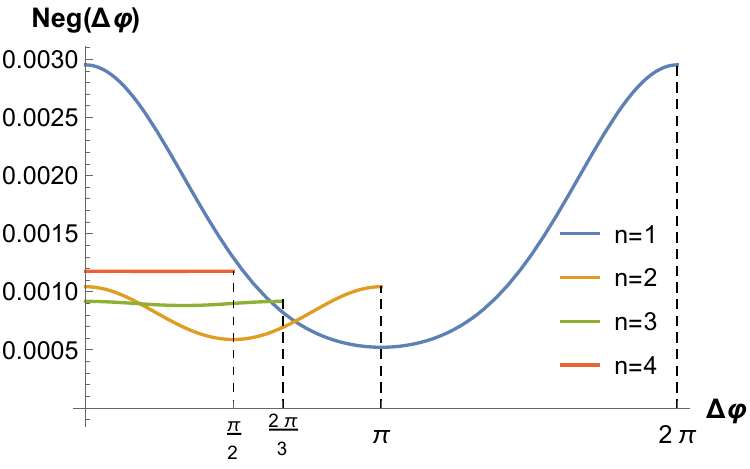}
          \caption{$r=1; j=1$}
\end{subfigure}
\begin{subfigure}{0.5\textwidth}
\centering
     \includegraphics[scale=0.58,frame]{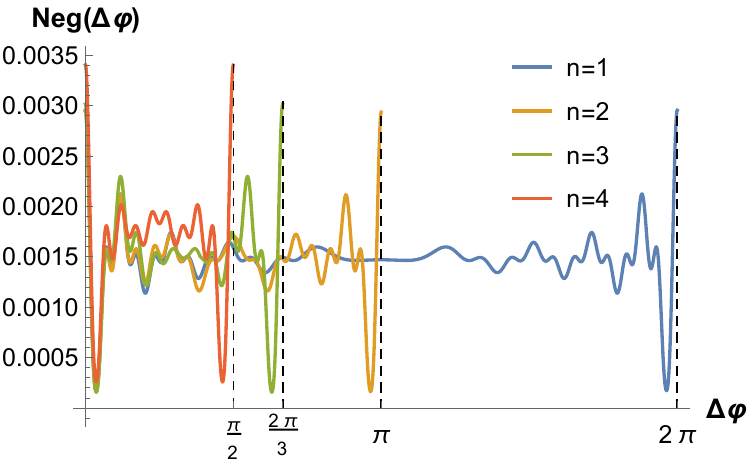}
          \caption{$r=25; j=1$}
\end{subfigure}
\caption{Negativity as a function of the angular difference between both qubits $\Delta\varphi$, evaluated with $\mathcal{T}=1000$ and $\omega=0.5$.}
    \label{T100_omega05_j_r_phi_n}
\end{figure}

\item 

\textit{\underline{Entanglement as function of angular separation}}\\
Here we analyze how the entanglement harvesting process depends on the angular distance between the qubits. In figure \ref{T100_omega05_j_r_phi_n}, one can see the behavior of the Negativity as a function of $\Delta\varphi$. There we obtain 
the same values for Negativity at angular difference $\Delta\varphi\in[0,\pi/n]$ and $\Delta\varphi\in[\pi/n,2\pi/n]$ because for both ranges of angular separation, equal qubit separations are obtained. It is worth recalling that the angular deficit of the conical spacetime induced by the cosmic string is $2\pi(1-b)=2\pi(n-1)/n$, then the range of values of $\Delta\varphi$ is smaller for large values of $n$.\\
From figures (\ref{T100_omega05_j_r_phi_n}a) and (\ref{T100_omega05_j_r_phi_n}b)  we observe the angular dependence for the ferromagnetic interaction $j<-\omega$.
There we note that for $r$ and $\Delta\varphi$ small, the position of both qubits coincides and the Negativity is approximately zero as described by the Property \ref{prop 7}. Indeed, since Negativity depends on functions of the type $\frac{\sin x}{x}$, with $x=r\sin(\pi k/n+\Delta\varphi/2)$, it is observed that when $j<-\omega$ and $r$ small, the Negativity goes from a value very close to zero, at $\Delta\varphi=0$, until reaching a maximum value at $\Delta\varphi=\pi/n$ as seen in figure (\ref{T100_omega05_j_r_phi_n}a). In figure (\ref{T100_omega05_j_r_phi_n}b) one obtains a different behavior from (\ref{T100_omega05_j_r_phi_n}a) this is because in this case the large value or $r$ one expects the oscillatory behavior of Negativity with the relative distance between the qubits as explain in the previous item and in the general Property \ref{prop 10}.
\\ On the other hand, for the case without a Heisenberg $XY$-interaction, $j=0$, we can see in figures (\ref{T100_omega05_j_r_phi_n}c) and (\ref{T100_omega05_j_r_phi_n}d) that although these graphs show us a very different behavior of Negativity, this result is still the same phenomena described in Property \ref{prop 10}, this means that, Negativity oscillates with the distance between the qubits (that depends on both $r$ and $\Delta\varphi$) and rapidly drops to zero with the distance between qubits, as shown in figure (\ref{T100_omega05_j_r_phi_n}d). In figure (\ref{T100_omega05_j_r_phi_n}c), the Negativity does not approach zero because the distance between qubits is still small and they are practically at the same position that coincides with the cosmic string location.\\
Finally, in figures (\ref{T100_omega05_j_r_phi_n}e) and (\ref{T100_omega05_j_r_phi_n}f) we see the angular dependence for the case of anti-ferromagnetic interaction $j>\omega$. As seen in figures (\ref{T100_omega1_j_r_phi_n}g), (\ref{T100_omega1_j_r_phi_n}h) and (\ref{T100_omega1_j_r_phi_n}j) the negativity has a maximum for the anti-ferromagnetic case when both qubits are at the same position. So the Negativity must be maximum, for small values of $r$, when the angular difference is $\Delta\varphi=0$ or $\Delta\varphi=2\pi/n$, as shown in (\ref{T100_omega05_j_r_phi_n}e) and (\ref{T100_omega05_j_r_phi_n}f). Also as seen from figures (\ref{T100_omega1_j_r_phi_n}g), (\ref{T100_omega1_j_r_phi_n}h) and (\ref{T100_omega1_j_r_phi_n}j) for large values of the distance between the qubits the Negativity tends asymptotically to a finite value. This value is obtained in the angular case for $\Delta\varphi=\pi/n$ where the distance between the qubits is the maximum possible.\\
\begin{figure}[b!]
\begin{subfigure}{0.3\textwidth}
\centering
     \includegraphics[scale=0.35]{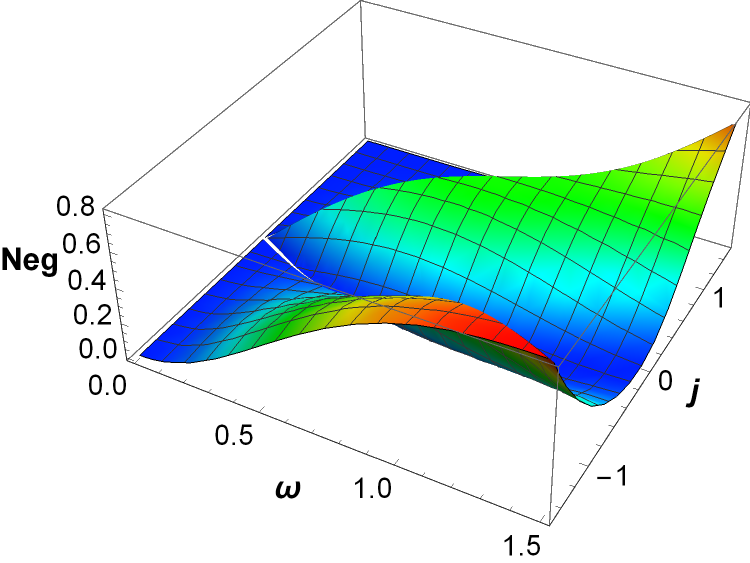}
          \caption{$n=1, r=1$}
\end{subfigure}
\begin{subfigure}{0.3\textwidth}
\centering
     \includegraphics[scale=0.35]{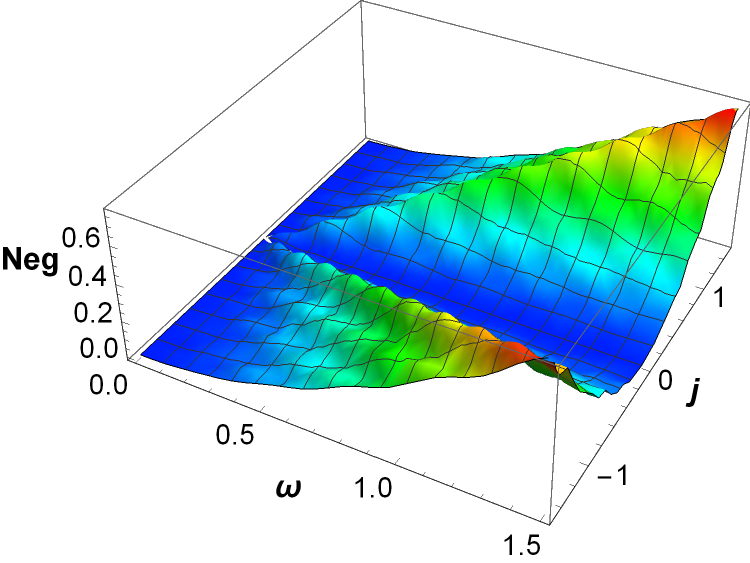}
        \caption{$n=1, r=10$}
\end{subfigure}
\begin{subfigure}{0.3\textwidth}
\centering
     \includegraphics[scale=0.35]{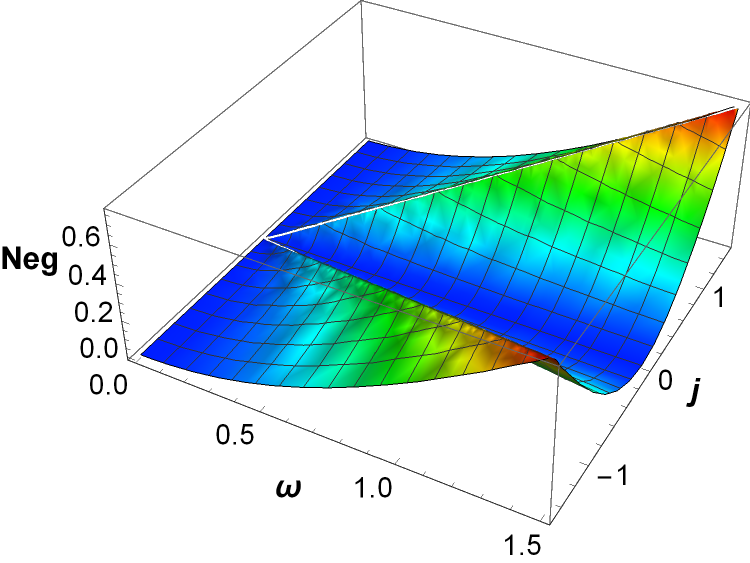}
          \caption{$n=1, r=100$}
\end{subfigure}

\begin{subfigure}{0.3\textwidth}
\centering
     \includegraphics[scale=0.35]{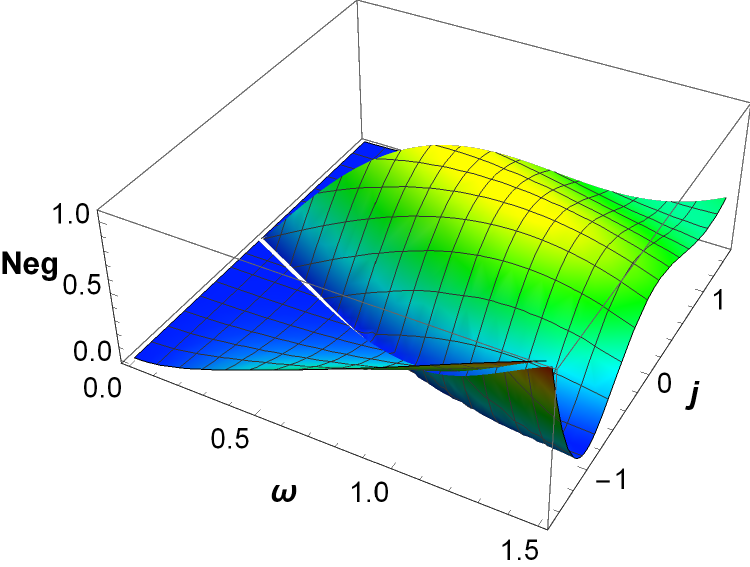}
        \caption{$n=2,r=1$}
\end{subfigure}
\begin{subfigure}{0.3\textwidth}
\centering
     \includegraphics[scale=0.35]{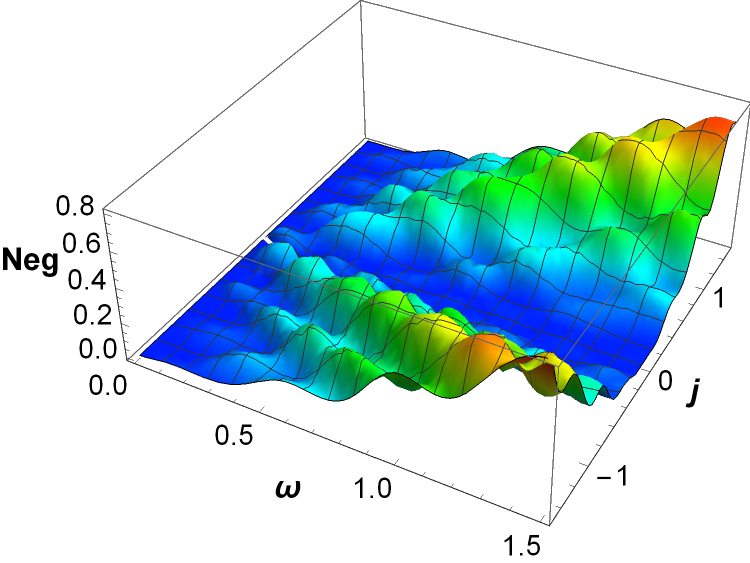}
        \caption{$n=2,r=10$}
\end{subfigure}
\begin{subfigure}{0.3\textwidth}
\centering
     \includegraphics[scale=0.35]{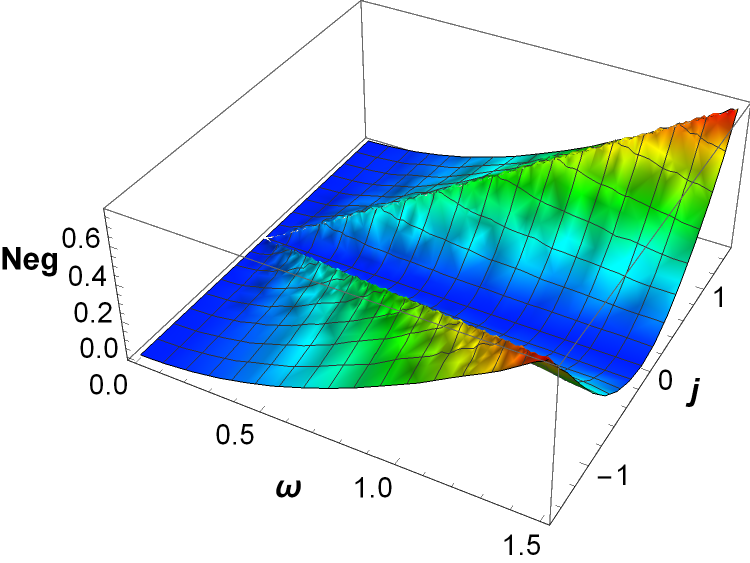}
        \caption{$n=2,r=100$}
\end{subfigure}

\begin{subfigure}{0.3\textwidth}
\centering
     \includegraphics[scale=0.35]{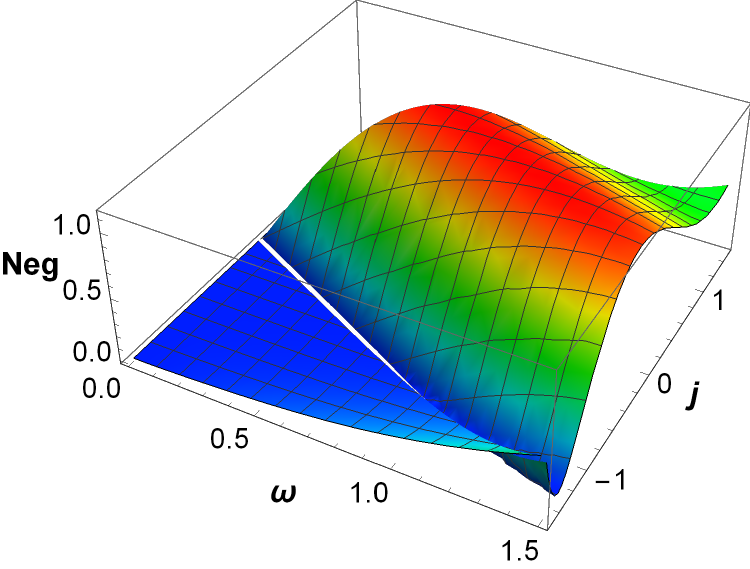}
        \caption{$n=3,r=1$}
\end{subfigure}
\begin{subfigure}{0.3\textwidth}
\centering
     \includegraphics[scale=0.35]{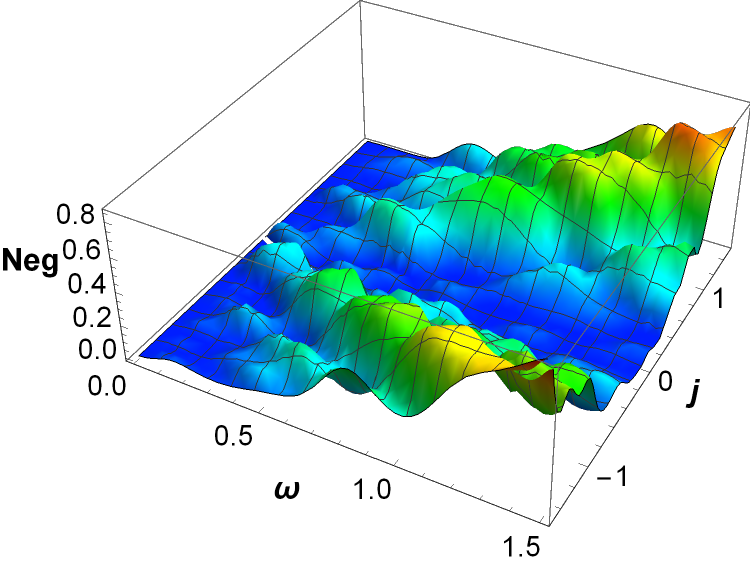}
        \caption{$n=3,r=10$}
\end{subfigure}
\begin{subfigure}{0.3\textwidth}
\centering
     \includegraphics[scale=0.35]{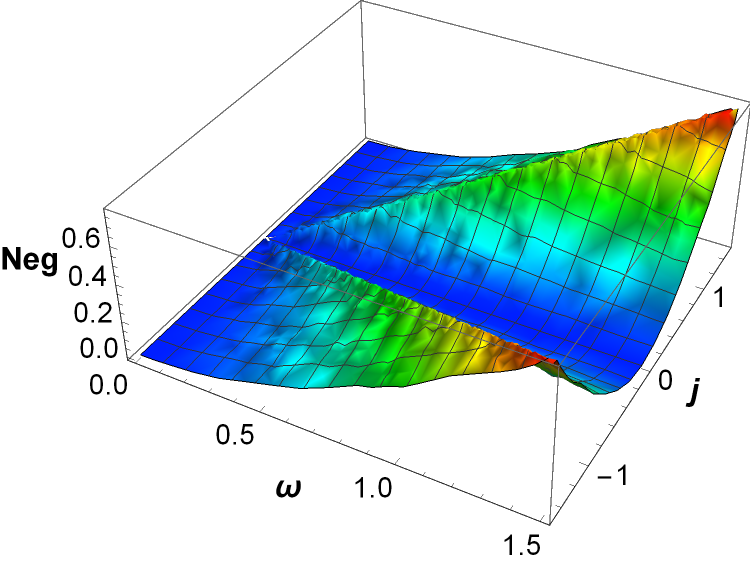}
        \caption{$n=3,r=100$}
\end{subfigure}

\begin{subfigure}{0.3\textwidth}
\centering
     \includegraphics[scale=0.35]{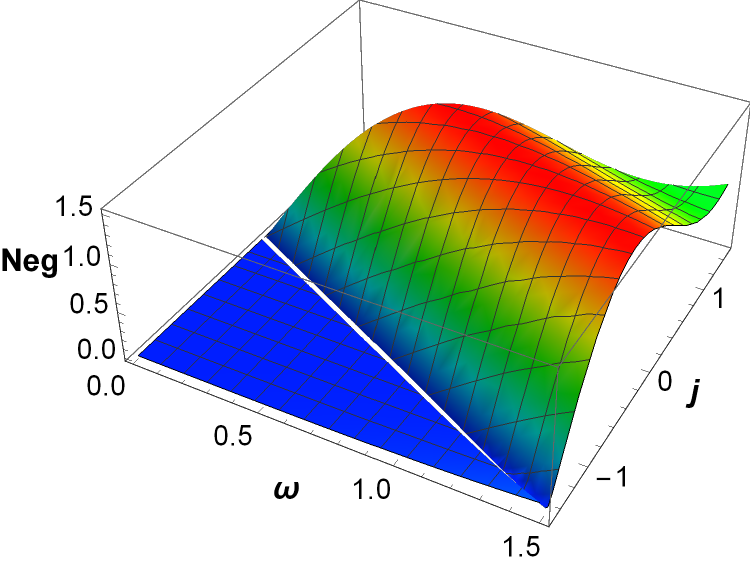}
        \caption{$n=4,r=1$}
\end{subfigure}
\begin{subfigure}{0.3\textwidth}
\centering
     \includegraphics[scale=0.35]{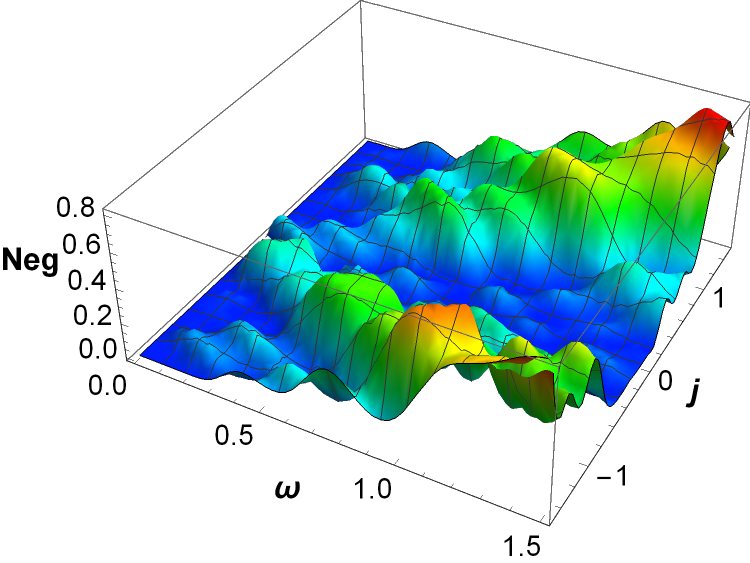}
        \caption{$n=4,r=10$}
\end{subfigure}
\begin{subfigure}{0.3\textwidth}
\centering
     \includegraphics[scale=0.35]{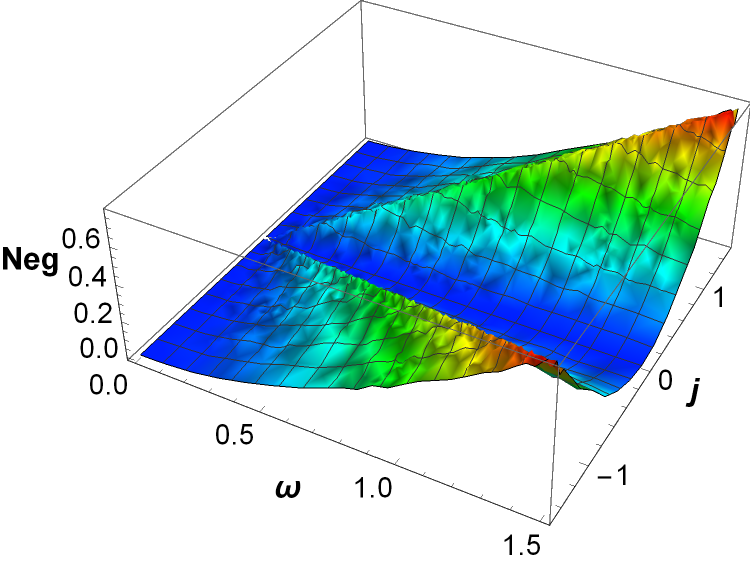}
        \caption{$n=4,r=100$}
\end{subfigure}

\caption{Tridimensional negativity graphs as a function of $\omega$ and $j$, evaluated with $\mathcal{T}=1000$, and $\Delta \varphi=\pi/n$. We can note that the scale of the Negativity in this graphics is of the order $10^{-2}$.}
\label{T100_omega_j_r_phipiovern_n_oscale}
\end{figure}

\item  \textit{\underline{Entanglement harvesting on phase space $(\omega, j)$}}\\
It is instructive to look at the dependence of Negativity on the phase space of energy parameters $(\omega,j)$, see figure \ref{T100_omega_j_r_phipiovern_n_oscale}. It shows that when $\omega<|j|$, the negativity increases, while when $\omega>|j|$, the negativity decreases, provided that the distance between the qubits is sufficiently large. Additionally, we also observe that when $j=0$ and $r$ is large enough, the negativity becomes zero.

\end{itemize} 

\subsection{THE AXIAL CASE}
\begin{figure}[h!]
\centering
     \begin{tikzpicture}[x=0.75pt,y=0.75pt,yscale=-1,xscale=1]

\draw    (303.55,8.99) -- (305.55,148.99) ;
\draw  [draw opacity=0][fill={rgb, 255:red, 74; green, 144; blue, 226 }  ,fill opacity=0.2 ] (267.46,73.99) -- (401.55,73.99) -- (344.08,126) -- (210,126) -- cycle ;
\draw  [draw opacity=0][fill={rgb, 255:red, 208; green, 2; blue, 27 }  ,fill opacity=0.2 ] (267.55,44.99) -- (400.55,44.99) -- (343.55,84.99) -- (210.55,84.99) -- cycle ;
\draw    (305.77,99.99) -- (359.55,100.95) ;
\draw [shift={(361.55,100.99)}, rotate = 181.02] [fill={rgb, 255:red, 0; green, 0; blue, 0 }  ][line width=0.08]  [draw opacity=0] (12,-3) -- (0,0) -- (12,3) -- cycle    ;
\draw    (305.55,64.99) -- (359.55,64.03) ;
\draw [shift={(361.55,63.99)}, rotate = 178.98] [fill={rgb, 255:red, 0; green, 0; blue, 0 }  ][line width=0.08]  [draw opacity=0] (12,-3) -- (0,0) -- (12,3) -- cycle    ;
\draw   (368.67,102) .. controls (373.34,101.95) and (375.65,99.6) .. (375.6,94.93) -- (375.58,93.24) .. controls (375.51,86.57) and (377.81,83.22) .. (382.48,83.17) .. controls (377.81,83.22) and (375.45,79.91) .. (375.38,73.24)(375.41,76.24) -- (375.36,71.55) .. controls (375.31,66.88) and (372.96,64.57) .. (368.29,64.62) ;

\draw (325.33,80.73) node [anchor=north west][inner sep=0.75pt]  [font=\Large]  {$r$};
\draw (324.67,40.07) node [anchor=north west][inner sep=0.75pt]  [font=\Large]  {$r$};
\draw (279.33,84.07) node [anchor=north west][inner sep=0.75pt]  [font=\Large]  {$z_{1}$};
\draw (279.33,47.73) node [anchor=north west][inner sep=0.75pt]  [font=\Large]  {$z_{2}$};
\draw (387.67,67.07) node [anchor=north west][inner sep=0.75pt]  [font=\Large]  {$\Delta z$};
\end{tikzpicture}
      \caption{Qubits spatial configuration for the axial case.}
      \label{axial case disposition}
\end{figure}
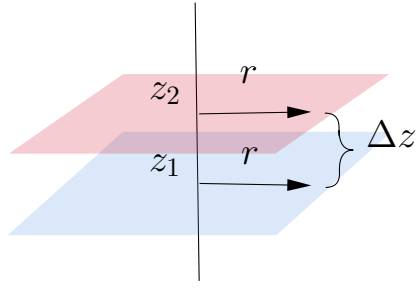

In this Section, we study the entanglement harvesting for an axial symmetric configuration of the qubits locations. We consider that the qubits' positions differ only by a distance $\Delta z$ along the cosmic string, see figure (\ref{axial case disposition}). Hence, the qubits' spacetime statics positions are given by
\begin{equation}
\begin{split}
\chi_1(\tau)&=(\tau,r,\varphi,z_1),\\
\chi_2(\tau)&=(\tau,r,\varphi,z_2).
\end{split}
\end{equation}

\noindent
Now, as indicated in previously, we denote the Wightman correlation function between the qubits positions at different proper times, as $G^+_{12}(\tau,\tau')$,  and in this specific case it is given by
\begin{equation}
G^+_{12}(\tau,\tau')=-\frac{1}{4\pi^2}\sum^{n-1}_{k=0}\frac{1}{(\Delta\tau-i\varepsilon)^2-d^2_{kn}}.
\end{equation} where, the distance $d_{kn}$ depends only on the distance from the qubits to the cosmic string, $r$, and on the relative distance between the qubits, $\Delta z$, and it is defined as
\begin{equation}
d^2_{kn}=4r^2\sin^2(\pi k/n)+\Delta z^2.
\end{equation}

\noindent
Now, we explore the entanglement harvesting dynamics in this scenario by graphically analyzing the Negativity with respect to all its variables. Ultimately, we will verify all the general properties established at the beginning of this Section for the specific case of axial symmetry. 

\begin{itemize}
\item 
\textit{\underline{Negativity dependence on the initial probability} }\\
The Negativity dependence on the initial probability $p$ that defined the initial density $X$-state of the qubits is given at 
figure \ref{T100_omega1_j05_z10_r10_vs_p}. There it is shown that the maximum value of Negativity occurs when $p=0$. We see that a sudden death of entanglement occurs after some maximum value of initial probability is exceeded. This is, in fact, a general property for entanglement harvesting in the cosmic string background with an initial density state of the qubits as given by eq. (\ref{initial density matrix}). This result for the axial case is in accordance with what is established by Property \ref{prop 2}.
    \begin{figure}[ht]
\centering
      \includegraphics[scale=0.58,frame]{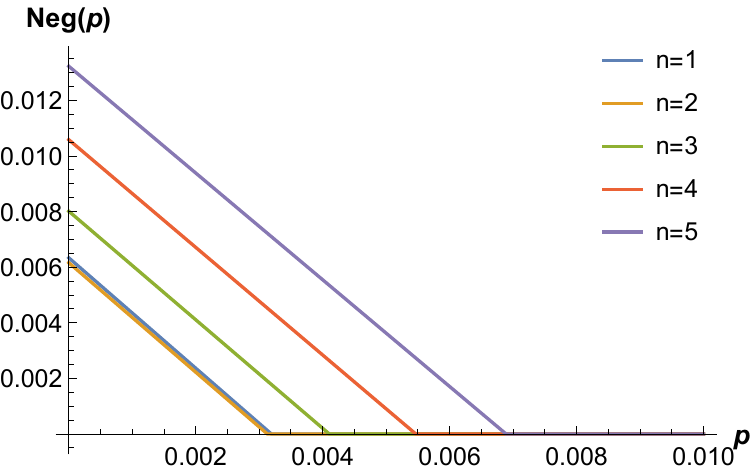}
\caption{Negativity as a function of initial probability. Here we use $\mathcal{T}=1000$, $\omega=1$, $j=0.5$, $\Delta z=1$, and $r=1$. 
}
\label{T100_omega1_j05_z10_r10_vs_p}
\end{figure}

\item 
  \textit{\underline{Time evolution of entanglement harvesting}}\\
  The time dependence of Negativity is presented in the figure
  \ref{T_omega1_jm05_r1_z10_}.
  One can see that the initial entanglement is zero for some finite amount of time until a sudden birth of entanglement occurs. From there on entanglement grows until a linear regime is reached. In order to avoid an exponential growth that invalidates the perturbation theory hypotheses, one must restrict our results to some finite maximum time.
  This behavior of negativity is in accordance with Property \ref{prop 5}, which describes the linear dependence with time of negativity. 

    \begin{figure}[b!]
\centering
     \includegraphics[scale=0.58,frame]{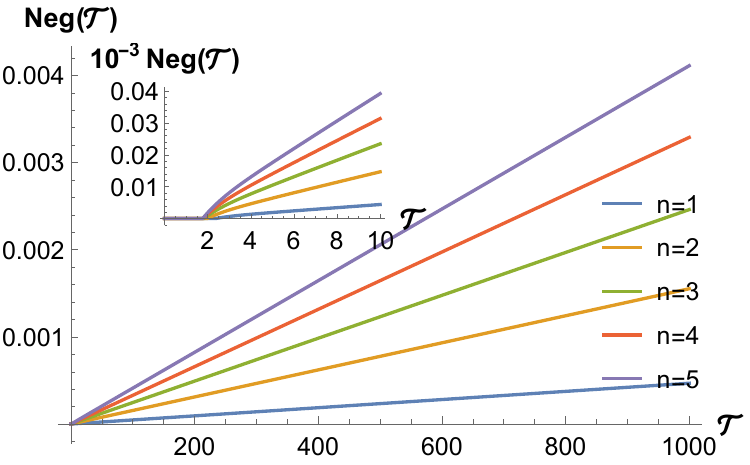}
\caption{Negativity as a function of interaction time $\mathcal{T}$, evaluated with $\omega=1$, $j=-0.5$, $\Delta z=1$, and $r=1$.}
    \label{T_omega1_jm05_r1_z10_}
\end{figure}

\item 
 \textit{\underline{Negativity as function of $XY$-Heinsenberg interaction coupling}}\\
 The effect of the Heisenberg $XY$-interaction between the qubits on the entanglement harvesting is shown in the figure
 \ref{T100_omega05_j_z100_r10_}. 
 There one can see that it is always possible to choose some finite value of $j$ (positive indicating an anti-ferromagnetic interaction or negative for a ferromagnetic term) that improves the entanglement harvesting when it is compared with the case without interaction, $j=0$. For small values of the distance between the qubits $\Delta z$, one sees that a maximum Negativity is reached at the resonance anti-ferromagnetic value $j=\omega$.
 As the distance between the qubits, $\Delta z$, increases there appears a symmetry between the ferromagnetic and antiferromagnetic interactions. For such large distances, the maximum Negativity is obtained for $j=\pm\omega$ while the minimum is for $j=0$.
 This is in accordance with the equivalence between ferromagnetic and antiferromagnetic $XY$-interaction, as stated in Property \ref{prop 6}.

\begin{figure}[t!]

\begin{subfigure}{0.325\textwidth}
\centering
     \includegraphics[scale=0.25,frame]{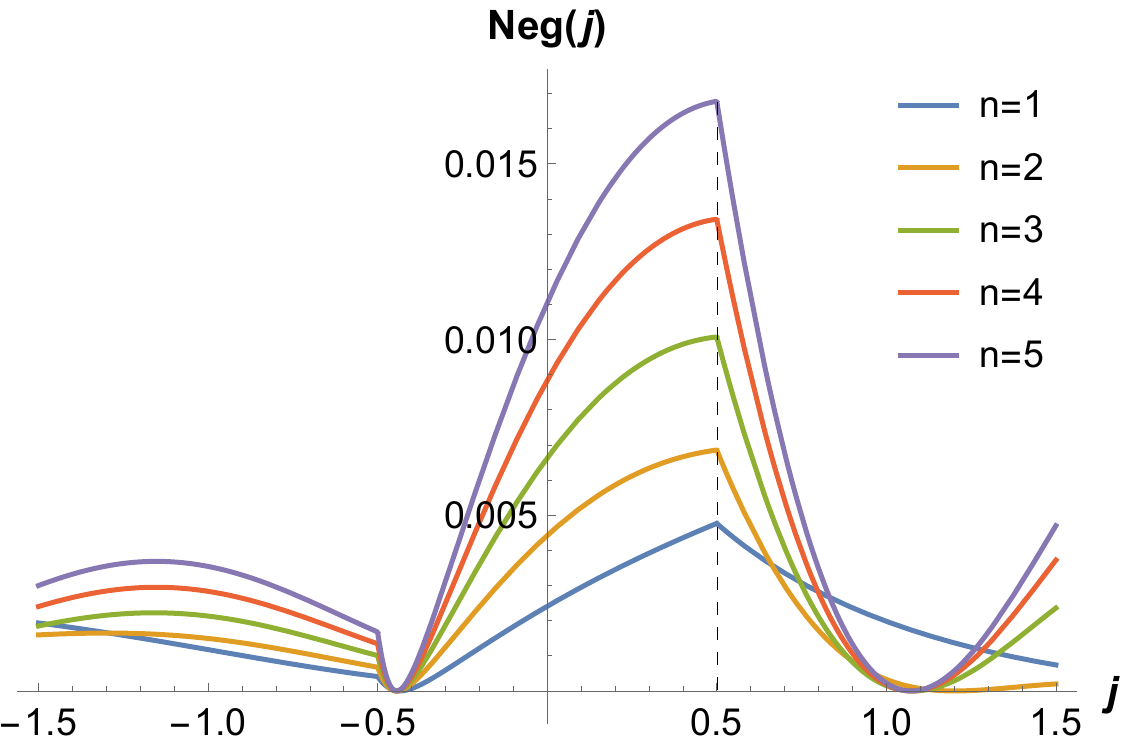}
          \caption{$r=1,\Delta z=1$}
\end{subfigure}
\begin{subfigure}{0.325\textwidth}
\centering
     \includegraphics[scale=0.375,frame]{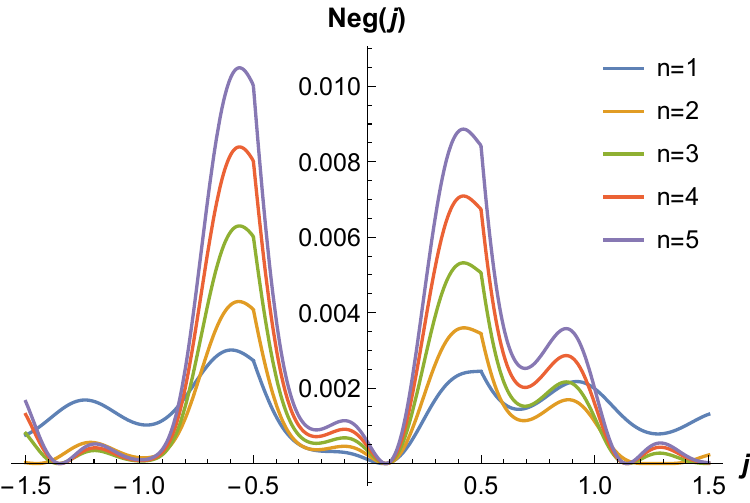}
        \caption{$r=1,\Delta z=10$}
\end{subfigure}
\begin{subfigure}{0.325\textwidth}
\centering
     \includegraphics[scale=0.25,frame]{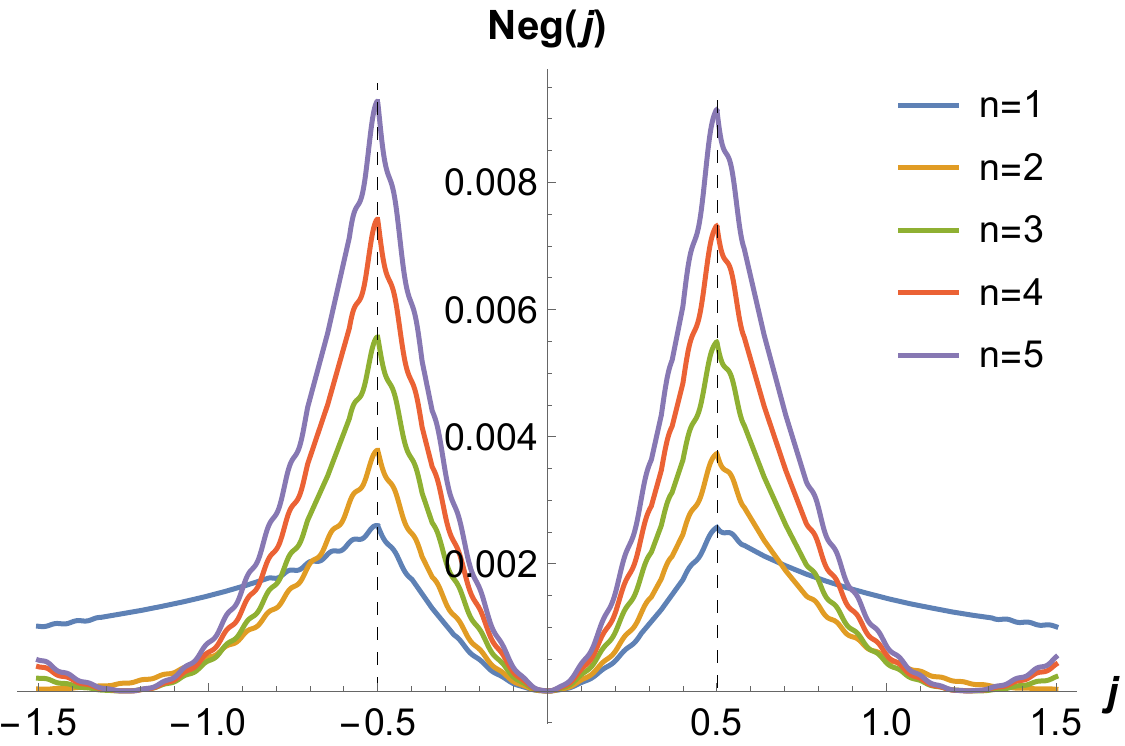}
        \caption{$r=1,\Delta z=100$}
\end{subfigure}

\begin{subfigure}{0.325\textwidth}
\centering
     \includegraphics[scale=0.38,frame]{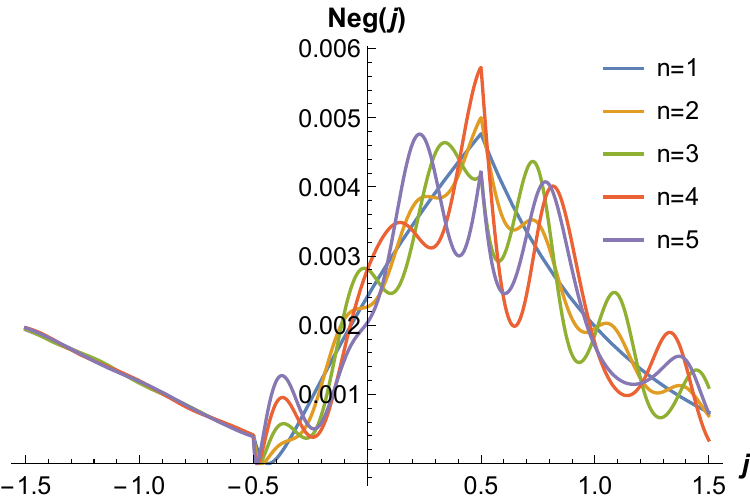}
        \caption{$r=10,\Delta z=1$}
\end{subfigure}
\begin{subfigure}{0.325\textwidth}
\centering
     \includegraphics[scale=0.38,frame]{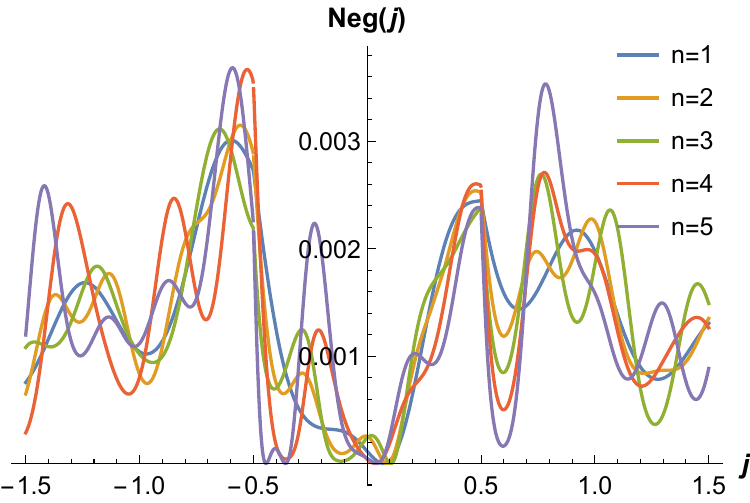}
        \caption{$r=10,\Delta z=10$}
\end{subfigure}
\begin{subfigure}{0.325\textwidth}
\centering
     \includegraphics[scale=0.38,frame]{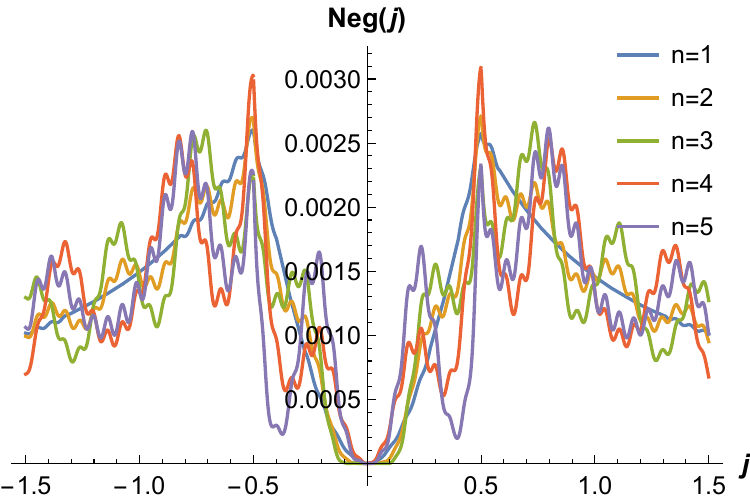}
        \caption{$r=10,\Delta z=100$}
\end{subfigure}

\begin{subfigure}{0.325\textwidth}
\centering
     \includegraphics[scale=0.25,frame]{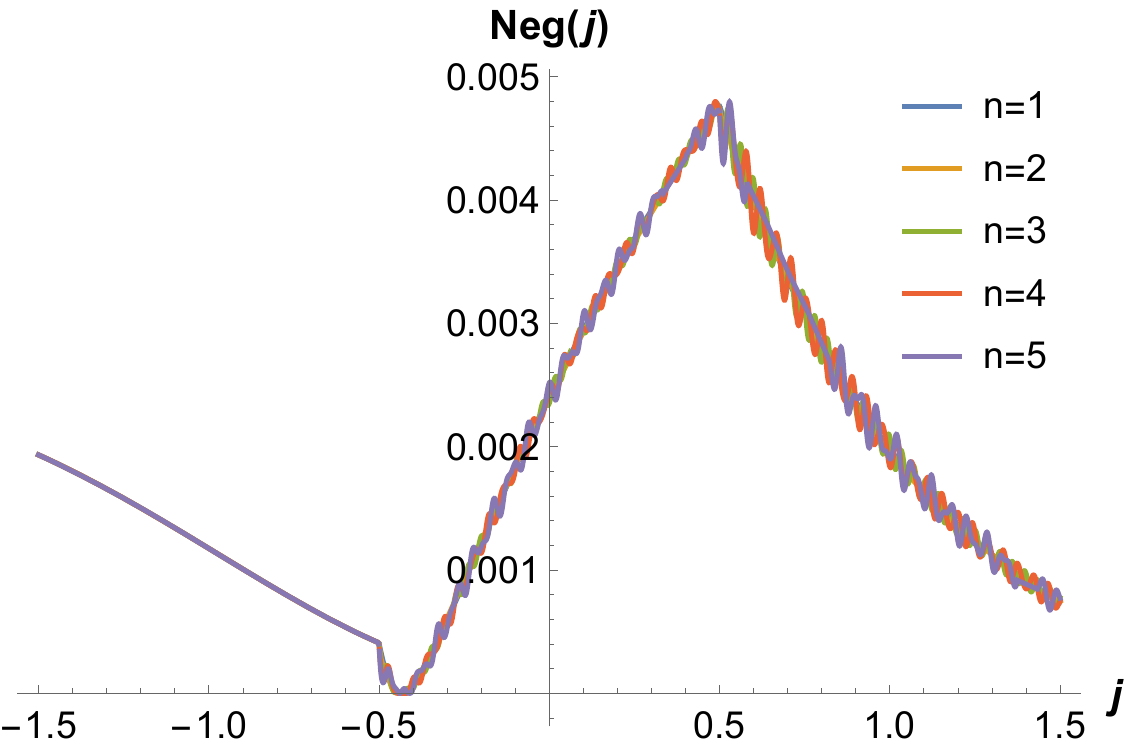}
        \caption{$r=100,\Delta z=1$}
\end{subfigure}
\begin{subfigure}{0.325\textwidth}
\centering
     \includegraphics[scale=0.25,frame]{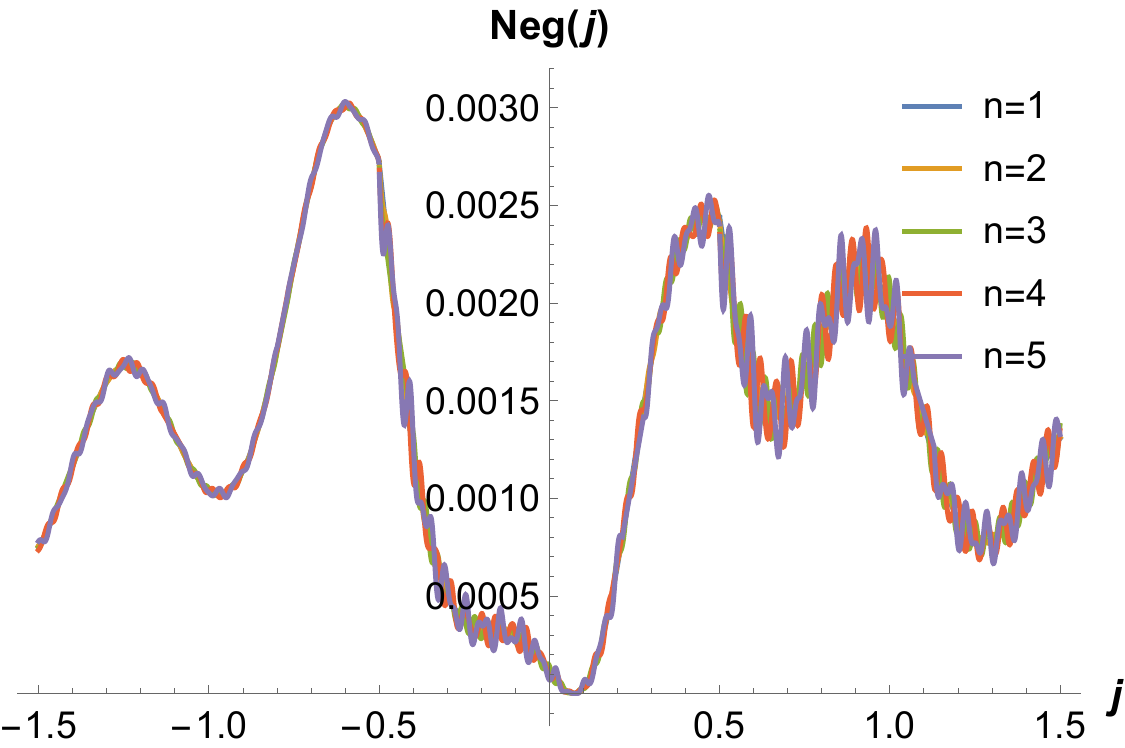}
        \caption{$r=100,\Delta z=10$}
\end{subfigure}
\begin{subfigure}{0.325\textwidth}
\centering
     \includegraphics[scale=0.25,frame]{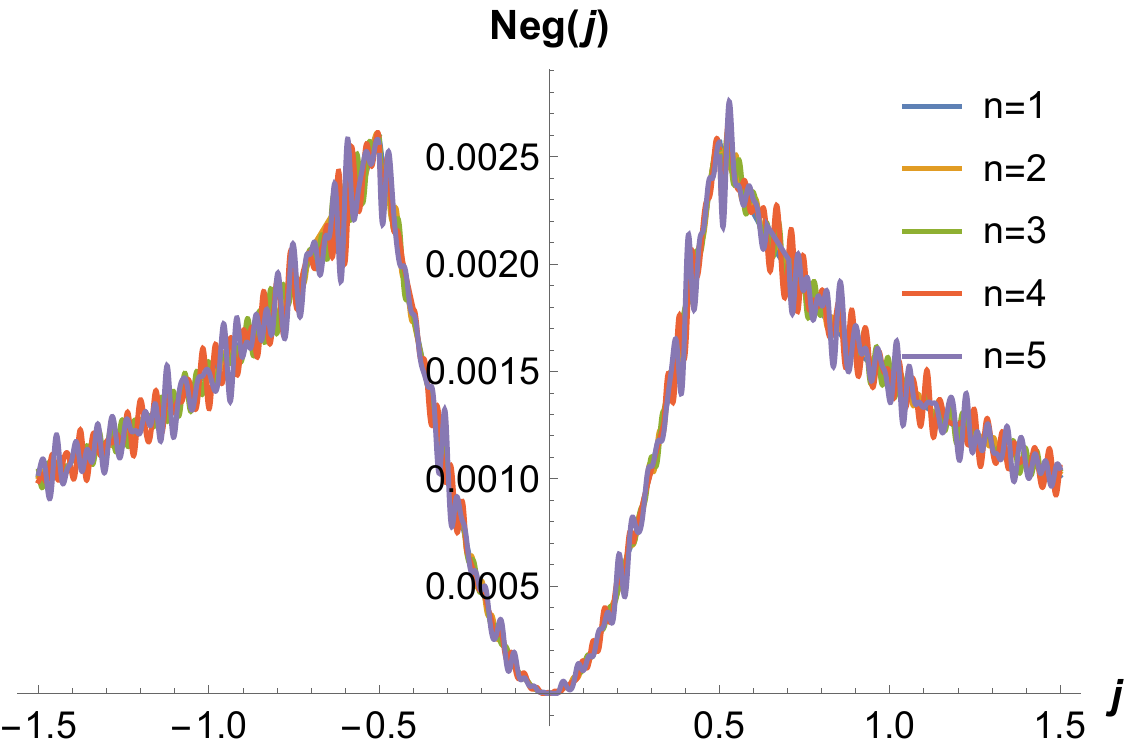}
        \caption{$r=100,\Delta z=100$}
\end{subfigure}
\caption{Negativity as a function of the interaction coupling constant $j$. These figures are given for $\mathcal{T}=1000$, $\omega=0.5$.}
\label{T100_omega05_j_z100_r10_}
\end{figure}

    \item 
    \textit{\underline{Entanglement harvesting as a function of the distance from cosmic string}}\\
    In the axial case, both qubits are separated from the cosmic string by the same radial distance $r$. On the other side, the qubits' relative distance $\Delta z$ is independent of $r$. Here we analyze how the distance to the cosmic string affects the entanglement generation. 
    In figure \ref{T100_omega15_j1_r_z_}, it can be noted that negativity is stronger and gets its maximum value when the qubits are located close to the cosmic string, $r\approx 0$.
    When the distance to the cosmic string increases Negativity decreases in an oscillatory way and tends asymptotically to the value expected for the case of Minkowski space without cosmic string, $n=1$.
    It is worth mentioning that this behavior is described in Property \ref{prop 3} corresponding to the Minkowski limit with finite distance between qubits.

    \begin{figure}[t!]

\begin{subfigure}{0.5\textwidth}
\centering
     \includegraphics[scale=0.36,frame]{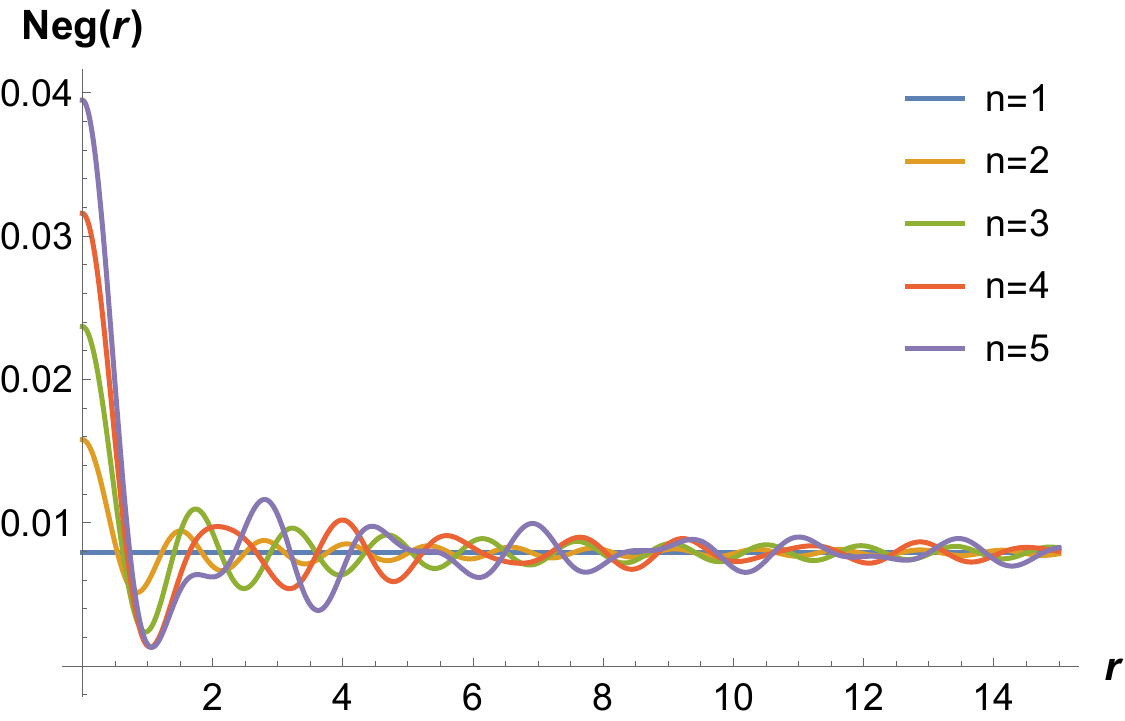}
          \caption{$\Delta z=1$}
\end{subfigure}
\begin{subfigure}{0.5\textwidth}
\centering
     \includegraphics[scale=0.37,frame]{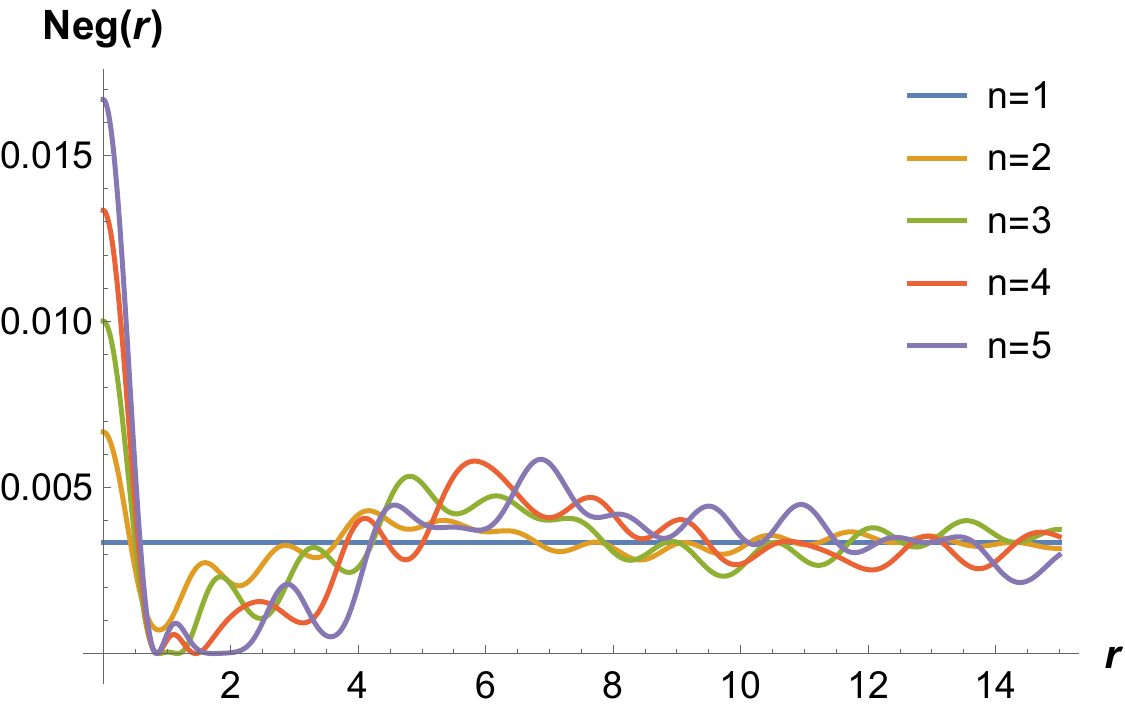}
        \caption{$\Delta z=10$}
\end{subfigure}

\centering

\begin{subfigure}{0.5\textwidth}
\centering
     \includegraphics[scale=0.37,frame]{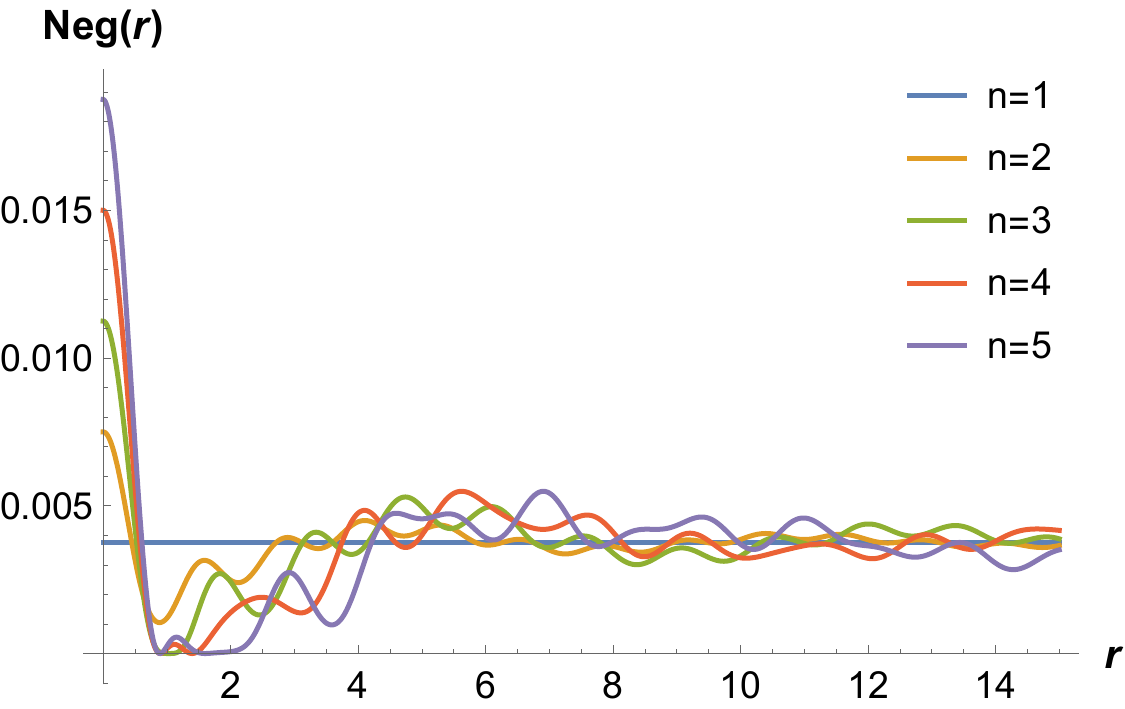}
          \caption{$\Delta z=100$}      
\end{subfigure}
\caption{Negativity as a function of the distance between the qubits to the cosmic string, $r$. Here it is use $\mathcal{T}=1000$, $\omega=1.5$, $j=1$.}
    \label{T100_omega15_j1_r_z_}
\end{figure}

\item 
\textit{\underline{Entanglement as function of qubits axial separation}}\\
Now we present the behavior of Negativity as the qubits' relative distance changes.
  Let us start with the case without Heisenberg $XY$-interaction, $j=0$, see figure \ref{T100_omega05_j0_z_r1_}. There we observe that when $j=0$, the negativity is maximum when both qubits are at the same position $\Delta z=0$ and tends to zero in an oscillatory manner as the qubits separation increases. This is in accordance with what was expressed in Property \ref{prop 10}.
\begin{figure}[hb!]
\centering
     \includegraphics[scale=0.58,frame]{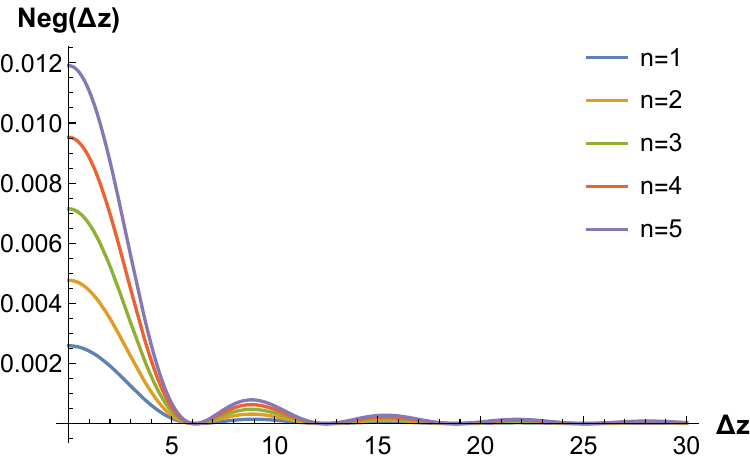}
\caption{Negativity as a function of the distance between the qubits $\Delta z$ for the case without $XY$-interaction. Here we evaluated with $\omega=0.5$, $j=0$, $\mathcal{T}=1000$, $r=1$.}
\label{T100_omega05_j0_z_r1_}
\end{figure}

\begin{figure}[ht!]
\begin{subfigure}{0.5\textwidth}
\centering
     \includegraphics[scale=0.58,frame]{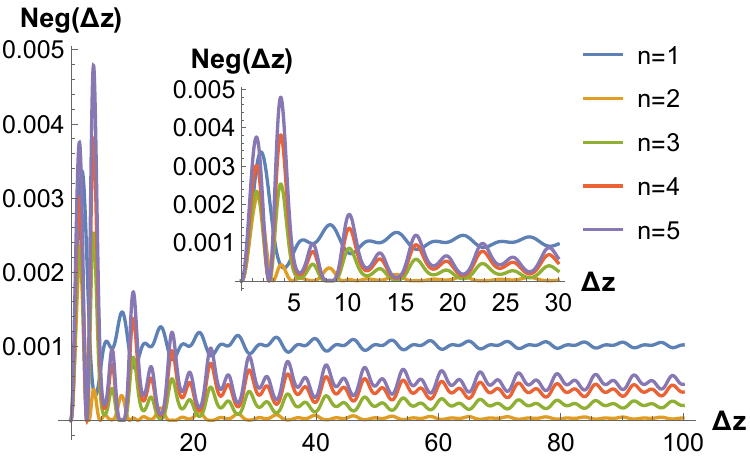}
          \caption{$r=1$}
\end{subfigure}
\begin{subfigure}{0.5\textwidth}
\centering
     \includegraphics[scale=0.58,frame]{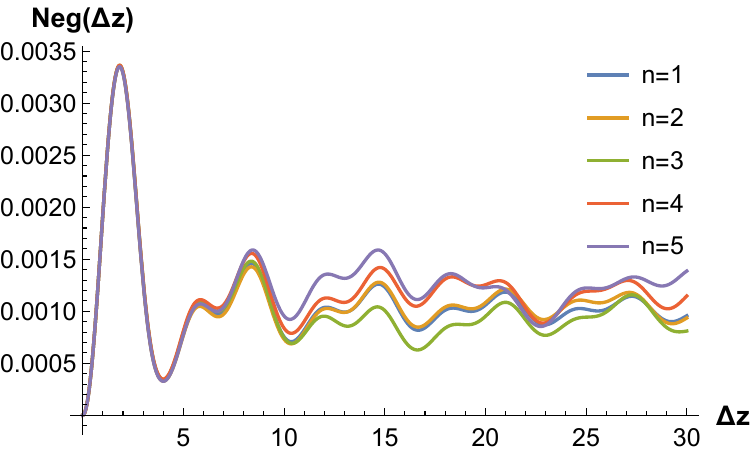}
        \caption{$r=30$}
\end{subfigure}
\caption{Negativity as a function of the distance between the qubits $\Delta z$, evaluated with $\omega=0.5$, $j=-1.5$, and $\mathcal{T}=1000$.}
\label{T100_omega05_jm15_z_r100_1}
\end{figure}
The case for a ferromagnetic interaction with $j<-\omega$ is shown in figure  \ref{T100_omega05_jm15_z_r100_1}. In contrast with the case without interaction, $j=0$, we see here that for a very small separation between the qubits, one gets that the Negativity is not maximum, but tends to zero when both qubits are located at the same position. When the spatial separation between the qubits increases the Negativity grows in an oscillatory manner and tends to a finite value when a large distance between the qubits is reached. This asymptotic value of Negativity depends on the distance between the cosmic string and the qubits, $r$, and when $r$ is sufficiently large, the asymptotic value of Negativity becomes unique and corresponds to the value it would take in Minkowski space. This agrees with what was stated in Property \ref{prop 7} and Property \ref{prop 3}, respectively. 

\item 
\textit{\underline{Entanglement harvesting on phase space $(\omega, j)$}}\\
It is useful to examine the Negativity as a function of the energetic parameters $\omega$ and $j$. In figure \ref{T100_omega_j_r1_z_n_} one can observe the behavior of the negativity with respect to these parameters in a global 3D graph. There one can see the symmetric behavior between ferromagnetic and antiferromagnetic interaction for the case of large separation between the qubits. This symmetry is highlighted by the maximum values of Negativity obtained at the resonance values $j=\pm\omega$.\\ 
We note also that for large separation between the qubits, the Negativity gets its minimum value at $j=0$ which indicates that a Heisenberg $XY$-interaction is necessary for obtaining effects of cosmic string entanglement generation. This result coincides with what was mentioned in Property \ref{prop 10}.
\\
These graphs show that for large values of the $XY$-interaction term,  $|j|>\omega$, the negativity increases as we expand the energy gap between the qubits $\omega$. For values of the interaction term inside the interval $|j|<\omega$, the negativity decreases as we augment the energy gap $\omega$ and as the distance between the qubits is sufficiently large. This aligns with what was mentioned in Property \ref{prop 9}. 
\begin{figure}[ht]
\begin{subfigure}{0.325\textwidth}
\centering
     \includegraphics[scale=0.35]{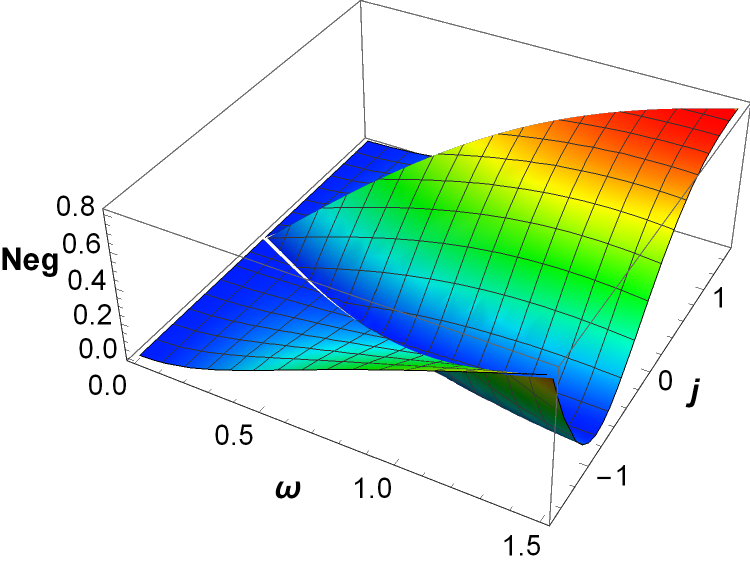}
          \caption{$n=1, \Delta z=1$}
\end{subfigure}
\begin{subfigure}{0.325\textwidth}
\centering
     \includegraphics[scale=0.35]{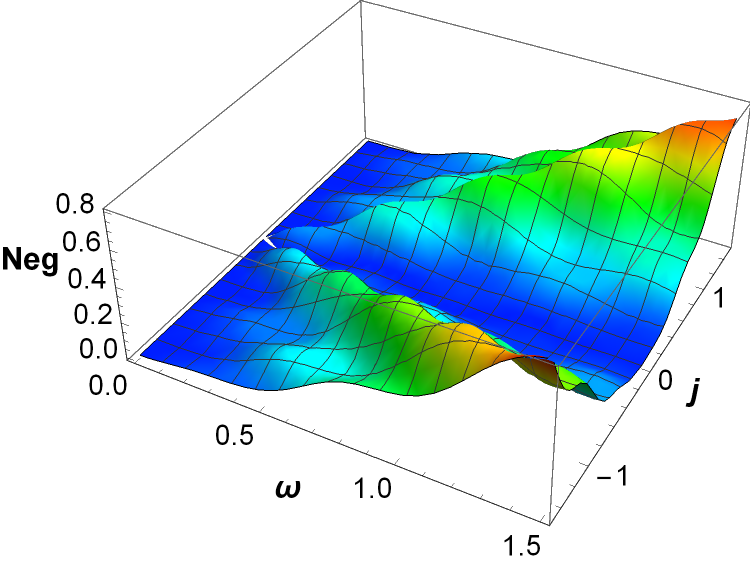}
        \caption{$n=1, \Delta z=10$}
\end{subfigure}
\begin{subfigure}{0.325\textwidth}
\centering
     \includegraphics[scale=0.35]{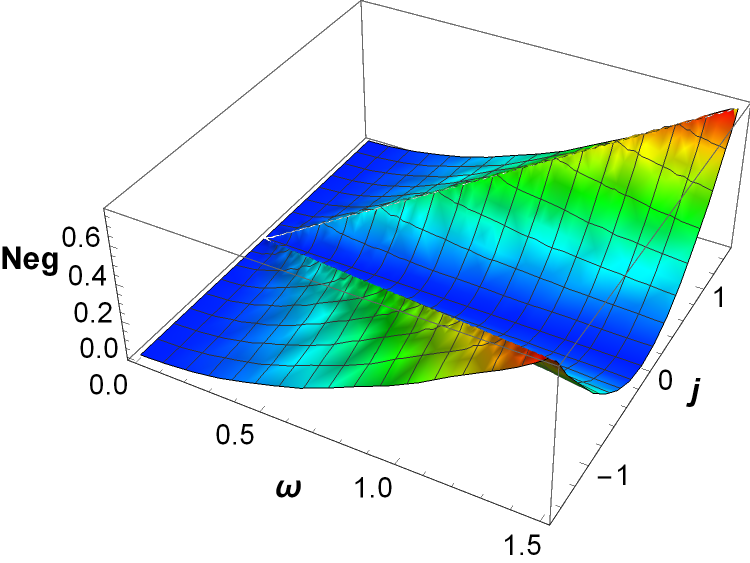}
          \caption{$n=1, \Delta z=100$}
\end{subfigure}
\begin{subfigure}{0.325\textwidth}
\centering
     \includegraphics[scale=0.35]{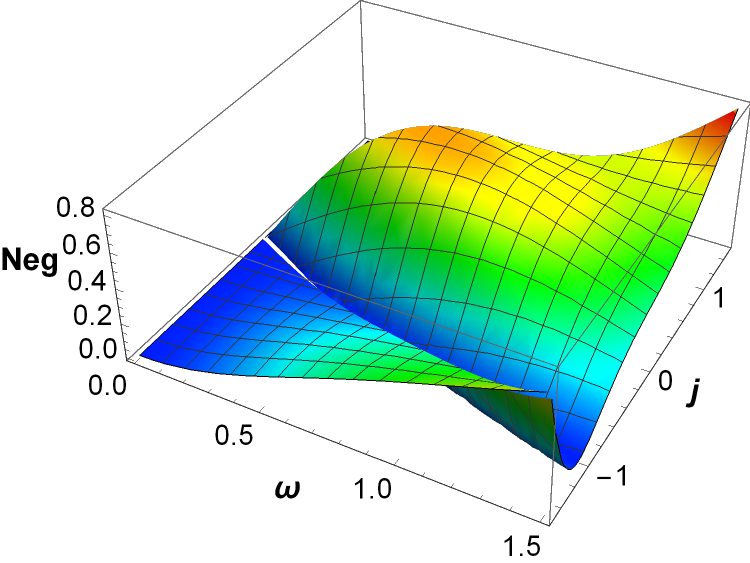}
          \caption{$n=2, \Delta z=1$}
\end{subfigure}
\begin{subfigure}{0.325\textwidth}
\centering
     \includegraphics[scale=0.35]{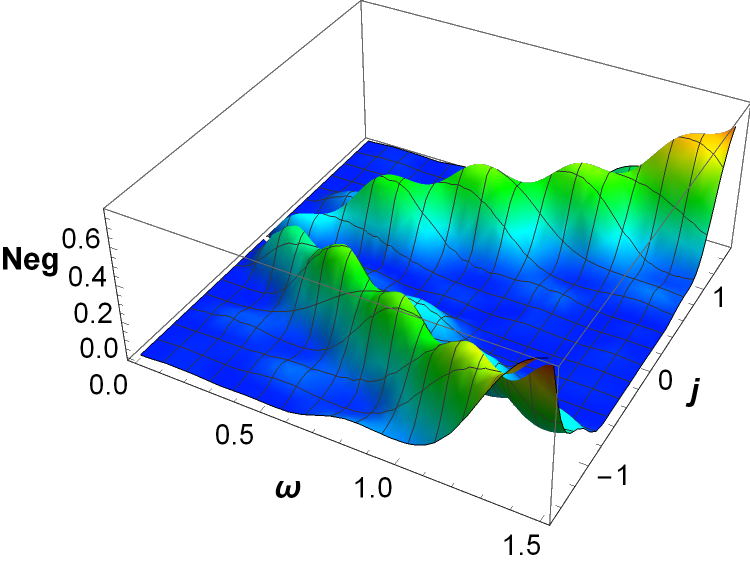}
        \caption{$n=2, \Delta z=10$}
\end{subfigure}
\begin{subfigure}{0.325\textwidth}
\centering
     \includegraphics[scale=0.35]{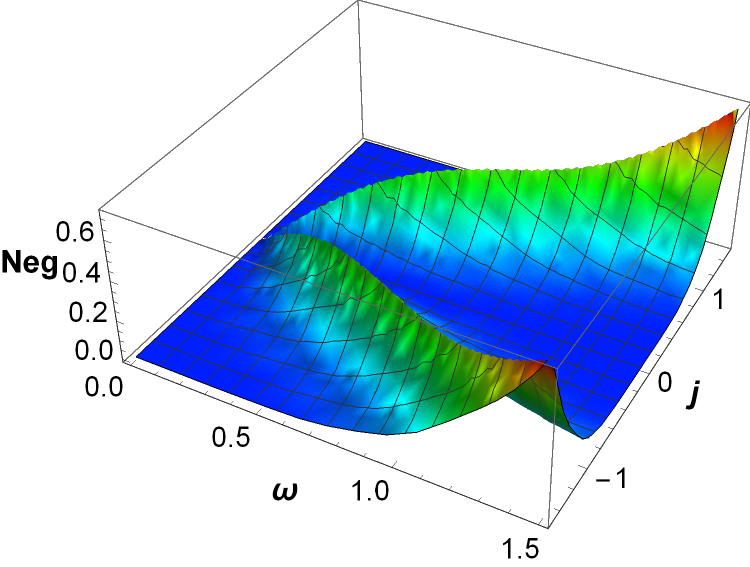}
          \caption{$n=2, \Delta z=100$}
\end{subfigure}
\begin{subfigure}{0.325\textwidth}
\centering
     \includegraphics[scale=0.35]{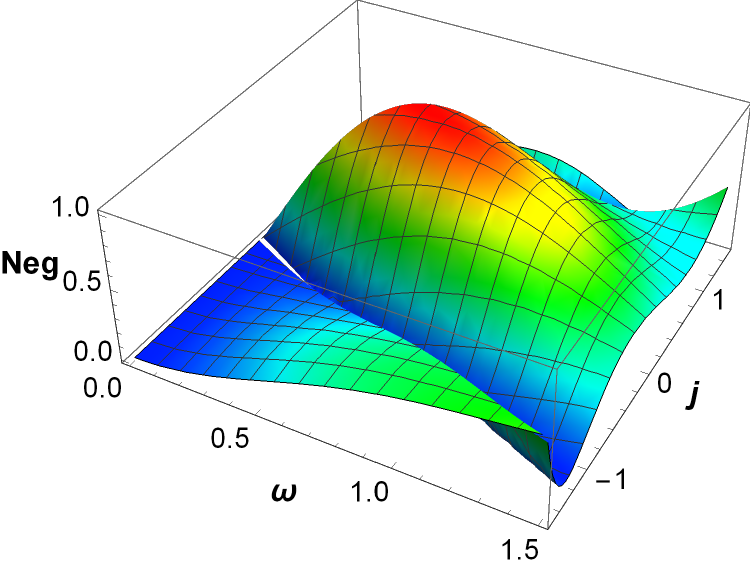}
          \caption{$n=3, \Delta z=1$}
\end{subfigure}
\begin{subfigure}{0.325\textwidth}
\centering
     \includegraphics[scale=0.35]{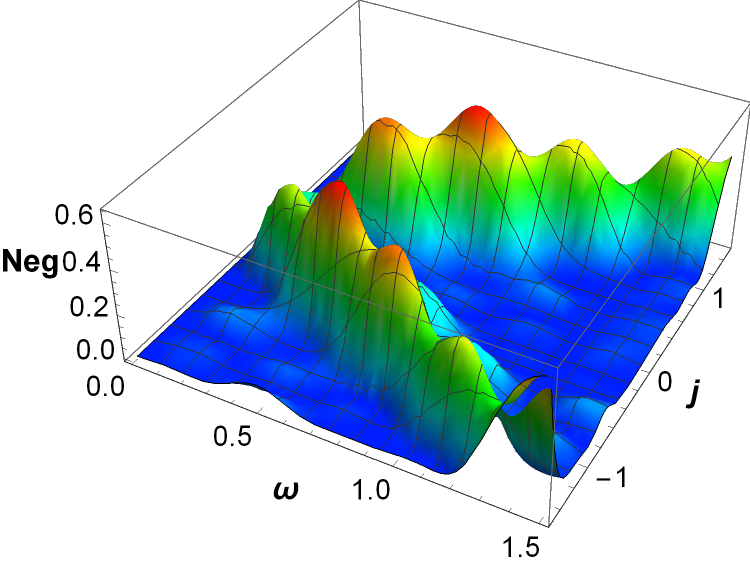}
        \caption{$n=3, \Delta z=10$}
\end{subfigure}
\begin{subfigure}{0.325\textwidth}
\centering
     \includegraphics[scale=0.35]{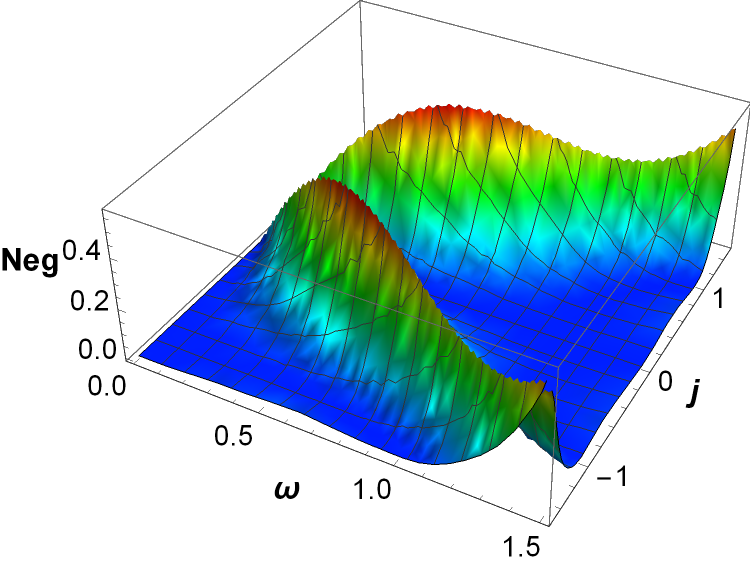}
          \caption{$n=3, \Delta z=100$}
\end{subfigure}
\begin{subfigure}{0.325\textwidth}
\centering
     \includegraphics[scale=0.35]{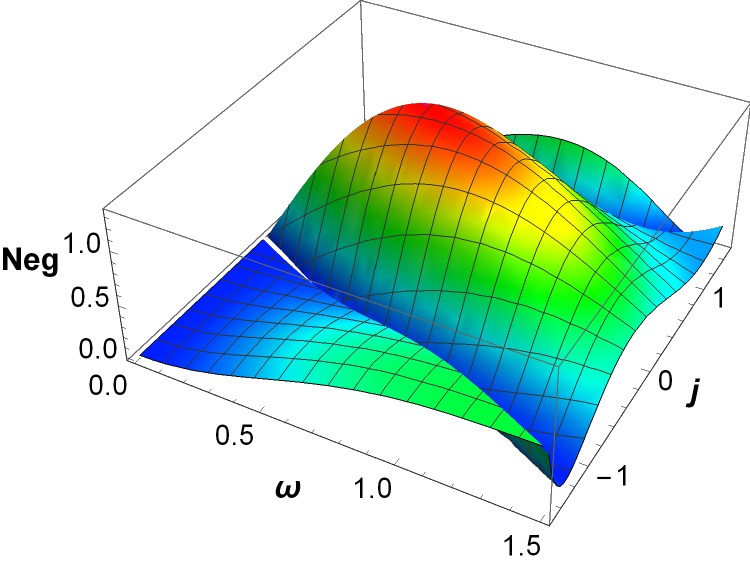}
          \caption{$n=4, \Delta z=1$}
\end{subfigure}
\begin{subfigure}{0.325\textwidth}
\centering
     \includegraphics[scale=0.35]{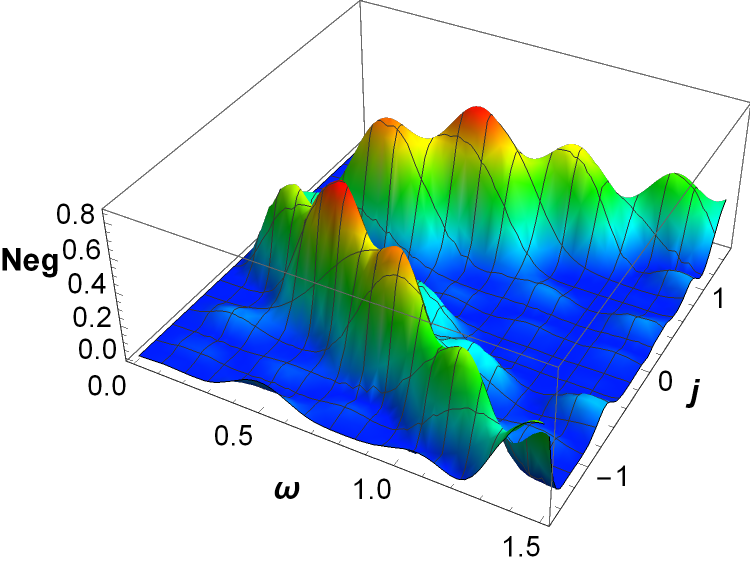}
        \caption{$n=4, \Delta z=10$}
\end{subfigure}
\begin{subfigure}{0.325\textwidth}
\centering
     \includegraphics[scale=0.35]{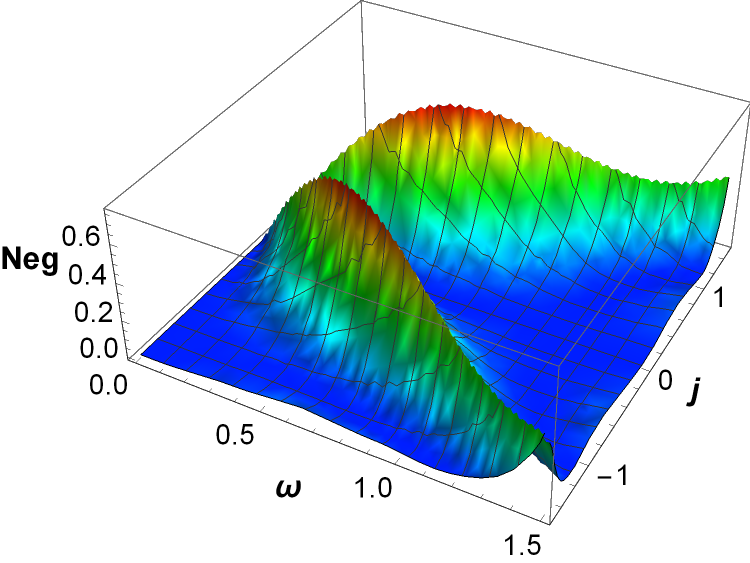}
          \caption{$n=4, \Delta z=100$}
\end{subfigure}

\caption{Negativity behavior on the phase space of the energy parameter $\omega$ and $j$. In all these graphs we use $\mathcal{T}=1000$, and $r=1$ and the scale of the Negativity is given in order of $10^{-2}$.}
\label{T100_omega_j_r1_z_n_}
\end{figure}

\newpage
\item
\underline{\textit{Non-local nature of entanglement harvesting}}\\
It is worthy to present, in a single graph, the dependence of Negativity on time and on the spatial separation between the qubits. This dependence of Negativity on spacetime coordinates $({\cal T}, \Delta z)$ is shown in figure \ref{T_omega05_j05_z_r1_}. There one observes that the negativity gets its maximum value for small distances, it shows an oscillatory behavior and tends to an asymptotic value for large distances, all in agreement with Property \ref{prop 10}. Furthermore, from figures (\ref{T_omega05_j05_z_r1_}b) and (\ref{T_omega05_j05_z_r1_}c), it is important to highlight the non-local nature of entanglement. We see that even when the qubits are causally disconnected and are separated by a distance much greater than the speed of light can travel in a given time, the existence of entanglement between these two qubits is evident. The entanglement generation can occur even when one qubit is outside the light cone of the other, this frontier is given by the white dashed line in figure (\ref{T_omega05_j05_z_r1_}c).
\begin{figure}[t!]
\begin{subfigure}{0.5\textwidth}
\centering
     \includegraphics[scale=0.6, frame]{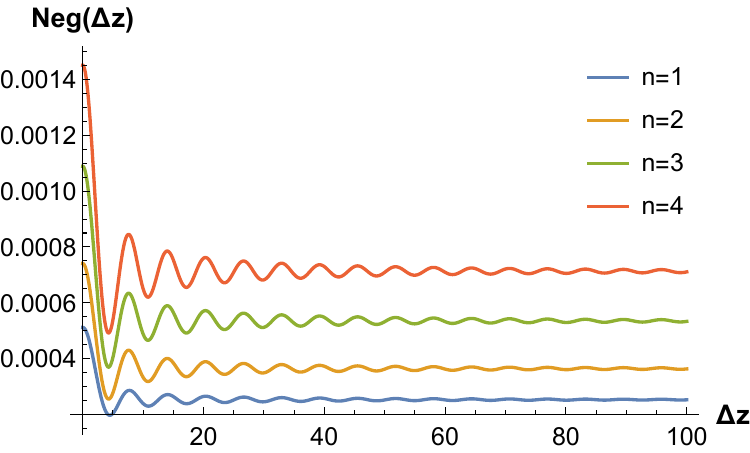}
          \caption{$\mathcal{T}=100$}
\end{subfigure}
\begin{subfigure}{0.5\textwidth}
\centering
     \includegraphics[scale=0.3]{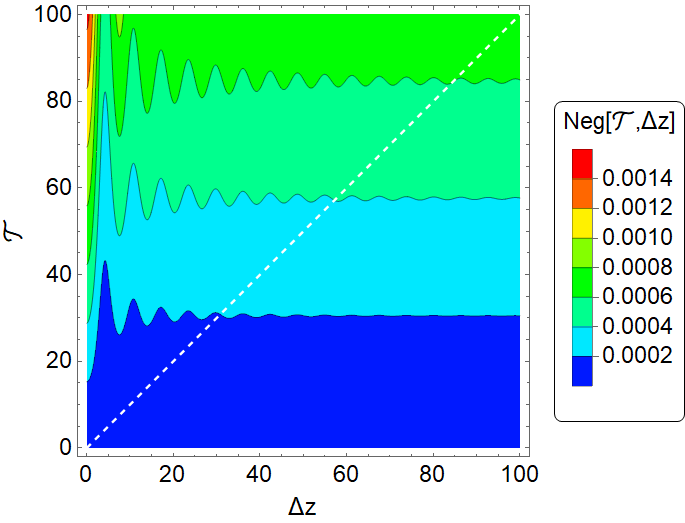}
          \caption{$n=4$}
\end{subfigure}

\caption{(b) and (c) are contour and three-dimensional plots of $T$ vs $\Delta z$ of the negativity with $n=4$, respectively, while (a) shows the behavior of the negativity at an interaction time of $\mathcal{T}=100$ with different values of $n$. All of these graphs are evaluated with $\omega=0.5$, $j=0.5$, $r=1$, $\Delta\varphi=\Delta r=p=0$, and $g=0.01$.}
\label{T_omega05_j05_z_r1_}
\end{figure}

\end{itemize}

\subsection{THE RADIAL CASE}
\begin{figure}[ht]
\centering
     \begin{tikzpicture}[x=0.75pt,y=0.75pt,yscale=-1,xscale=1]

\draw    (330.63,14.99) -- (330.08,155) ;
\draw  [draw opacity=0][fill={rgb, 255:red, 74; green, 144; blue, 226 }  ,fill opacity=0.2 ] (284.13,64.99) -- (447,64.99) -- (377.2,104.99) -- (214.33,104.99) -- cycle ;
\draw    (330.67,84.99) -- (363,85.59) ;
\draw [shift={(365,85.62)}, rotate = 181.05] [fill={rgb, 255:red, 0; green, 0; blue, 0 }  ][line width=0.08]  [draw opacity=0] (12,-3) -- (0,0) -- (12,3) -- cycle    ;
\draw    (365,85.62) -- (404.33,85.62) ;
\draw [shift={(406.33,85.62)}, rotate = 180] [fill={rgb, 255:red, 0; green, 0; blue, 0 }  ][line width=0.08]  [draw opacity=0] (12,-3) -- (0,0) -- (12,3) -- cycle    ;
\draw   (405,76.29) .. controls (404.93,71.62) and (402.56,69.33) .. (397.89,69.4) -- (393.56,69.47) .. controls (386.89,69.58) and (383.52,67.3) .. (383.45,62.63) .. controls (383.52,67.3) and (380.23,69.68) .. (373.56,69.78)(376.56,69.74) -- (369.22,69.85) .. controls (364.55,69.92) and (362.26,72.29) .. (362.33,76.96) ;

\draw (332.67,88.39) node [anchor=north west][inner sep=0.75pt]  [font=\Large]  {$r_{1}$};
\draw (378.34,87.4) node [anchor=north west][inner sep=0.75pt]  [font=\Large]  {$r_{2}$};
\draw (366.34,37.73) node [anchor=north west][inner sep=0.75pt]  [font=\Large]  {$\Delta r$};

\end{tikzpicture}

      \caption{Qubits locations for the radial case. }
      \label{radial case disposition}
\end{figure}
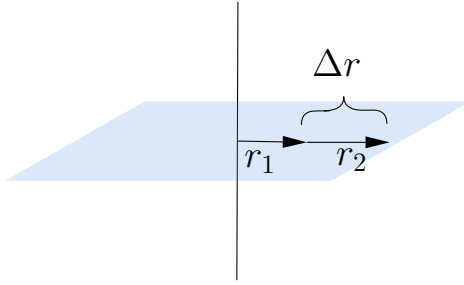

For completeness, we analyze here the spatial configuration of the statics qubits where both are located at a plane perpendicular to the cosmic string at the same angular position, but with different radial coordinates.
In this case, the detector positions differ only by some radial distance $\Delta r$, see figure \ref{radial case disposition}, and are given explicitly by
\begin{equation}
    \begin{split}
        \chi_1(\tau)&=(\tau,r_1,\varphi,z),\\
\chi_2(\tau)&=(\tau,r_2,\varphi,z).
    \end{split}
\end{equation}

For this case the Wightman function evaluated at the qubits location at different proper times, $G^+_{21}(\tau,\tau')$, is given by 
\begin{equation}
    G^+_{12}(\tau,\tau')=-\frac{1}{4\pi^2}\sum^{n-1}_{k=0}\frac{1}{(\Delta\tau-i\varepsilon)^2-d^2_{kn}},
\end{equation}
where the distance $d_{kn}$ in the cosmic string background is defined as \begin{equation}
    d^2_{kn}=\Delta r^{2}+4r(r+\Delta r)\sin^{2}(\pi k/n),
\end{equation}
and we have use $r_1=r$ and $r_2=r+\Delta r$ as the qubits radial coordinates.                
Now, we proceed to explore the harvesting entanglement properties in this special radial configuration.

\begin{itemize}
\item
    \textit{\underline{Negativity dependence on the initial probability} }\\
 As stated in Property \ref{prop 2}, one can verify that the Negativity is non-null only for values of initial probability very close to zero. For the radial configuration case this is confirmed at figure \ref{T100_omega1_j05_r10_deltar10_vs_p_1}. There one can see that for high values of initial probability, there is no entanglement generation, but for small values, a non-trivial entanglement is obtained. Furthermore, according to these results, the maximum degree of entanglement is generated for the case of initial probability $p=0$.
  Hence, in order to obtain maximum harvesting of entanglement, we use in all the following analyses $p=0$. 
    \begin{figure}[b!]
\centering
      \includegraphics[scale=0.58,frame]{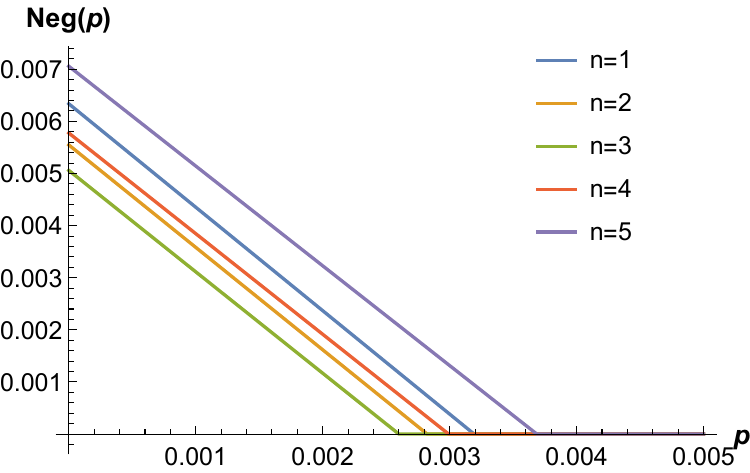}
\caption{Negativity as a function of initial probability for the radial case. We use $\mathcal{T}=1000$, $\omega=1$, $j=0.5$, $r=1$ and $\Delta r=1$.}
\label{T100_omega1_j05_r10_deltar10_vs_p_1}
\end{figure}
       \begin{figure}[ht!]
\centering
     \includegraphics[scale=0.58,frame]{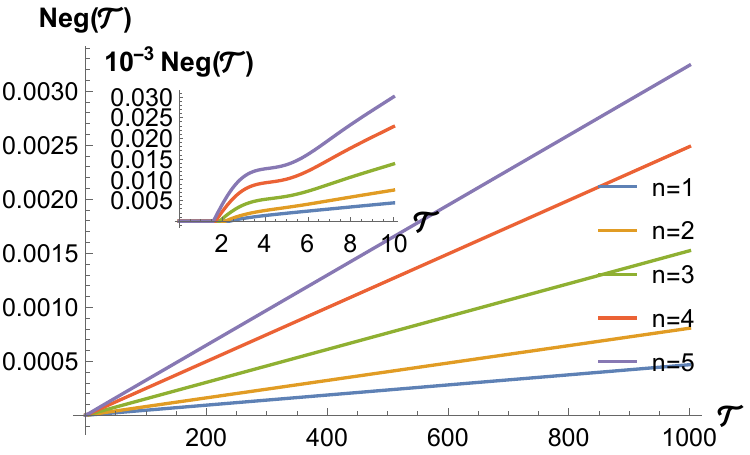}       
\caption{Negativity as a function of interaction time $\mathcal{T}$, evaluated with $\omega=1$, $j=-0.5$, $\Delta r=1$, and $r=1$.}
    \label{T_omega1_jm05_r1_deltar10_n}
\end{figure}
 \item 
 \textit{\underline{Time evolution of entanglement harvesting}}\\
 Now we inspect the entanglement generation over time in the qubits system obtained from the vacuum in the cosmic string background, see
 figure \ref{T_omega1_jm05_r1_deltar10_n}. There it is shown that for small times the Negativity remains zero until a sudden birth of entanglement occurs leading to a linear regime for large interaction times. This behavior is in accordance with what was stated in the Property \ref{prop 5}.
   \begin{figure}[hb!]

\begin{subfigure}{0.5\textwidth}
\centering
     \includegraphics[scale=0.58,frame]{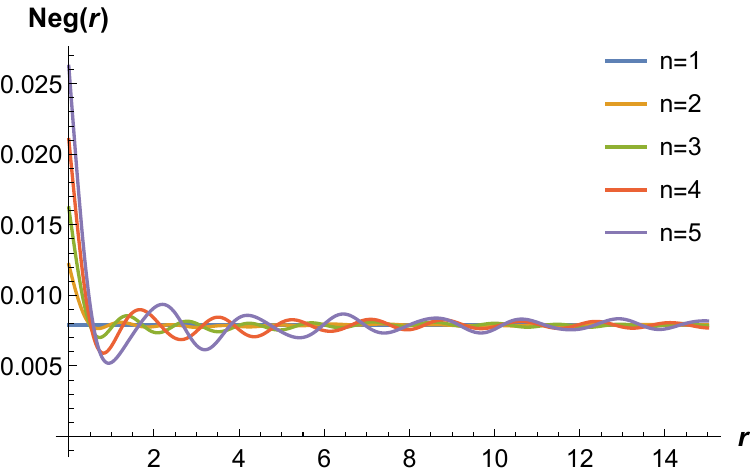}
          \caption{$\Delta r=1$}
\end{subfigure}
\begin{subfigure}{0.5\textwidth}
\centering
     \includegraphics[scale=0.585,frame]{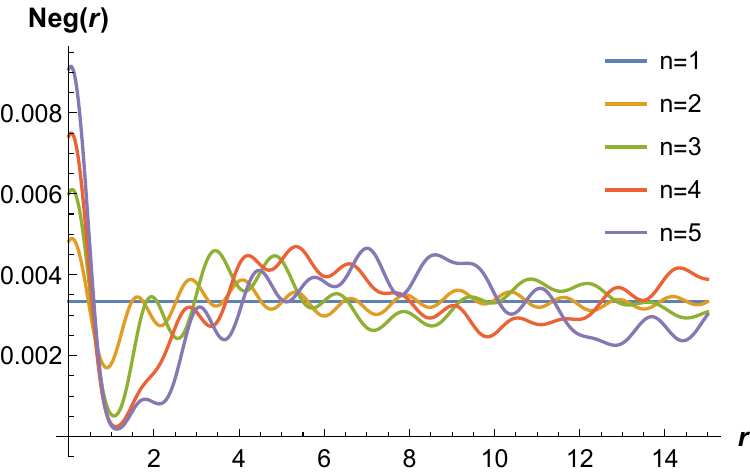}
        \caption{$\Delta r=10$}
\end{subfigure}

\centering

\begin{subfigure}{0.5\textwidth}
\centering
     \includegraphics[scale=0.58,frame]{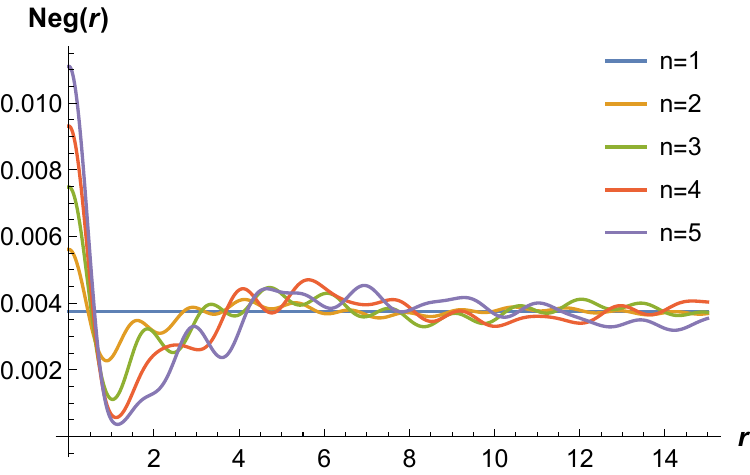}
          \caption{$\Delta r=100$}      
\end{subfigure}
\caption{Negativity as a function of the distance between the qubits and the cosmic string $r$, evaluated with $\mathcal{T}=1000$, $\omega=1.5$, $j=1$.}
    \label{T100_omega15_j1_r_deltar_n}
\end{figure}

    \item 
    \underline{\textit{Entanglement dependence on the distance to the cosmic string}}\\
    One can show that the relative distance to the cosmic string influences the entanglement harvesting. In figure \ref{T100_omega15_j1_r_deltar_n} we present the Negativity as a function of the distance $r$ to the cosmic string. One can see that when both qubits are located near the cosmic string one gets a maximum value of Negativity, as established in Property \ref{prop 3}. The same behavior is observed for both ferromagnetic and antiferromagnetic cases.\\
    As the system of qubits moves away from the cosmic string one obtains that the Negativity departs from its maximum values and decreases in an oscillatory manner until an asymptotic finite value is obtained in the limit of infinite distance from the cosmic string. This asymptotic value coincides with what is expected to be the entanglement negativity between the qubits generated by vacuum fluctuations in the absence of cosmic string $n=1$ and single  Minkowski spacetime. 
It is good to remember that even though the distance to the cosmic string is infinite, we are always considering a finite distance between qubits.   

\begin{figure}[ht!]
\begin{subfigure}{0.325\textwidth}
\centering
     \includegraphics[scale=0.38,frame]{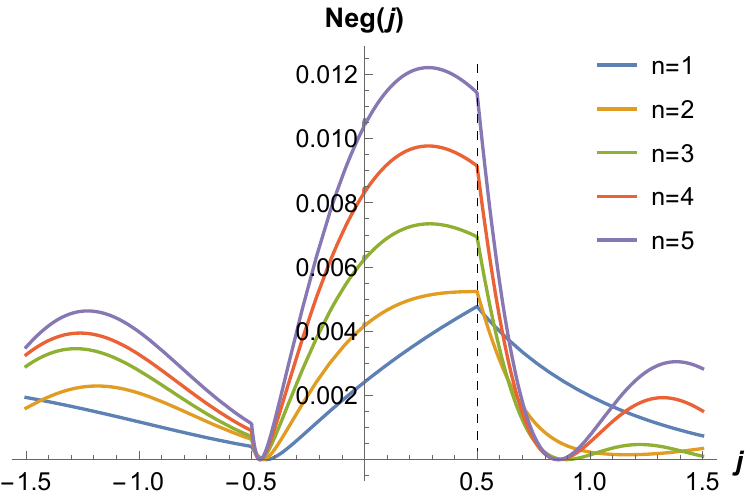}
          \caption{$r=1,\Delta r=1$}
\end{subfigure}
\begin{subfigure}{0.325\textwidth}
\centering
     \includegraphics[scale=0.38,frame]{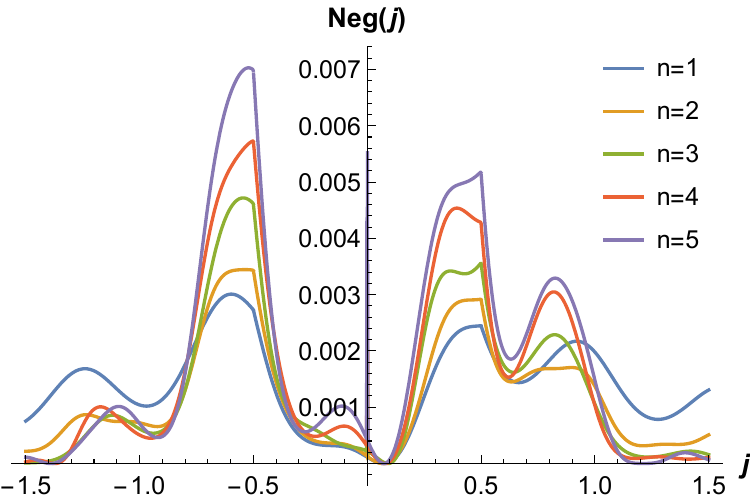}
        \caption{$r=1,\Delta r=10$}
\end{subfigure}
\begin{subfigure}{0.325\textwidth}
\centering
     \includegraphics[scale=0.38,frame]{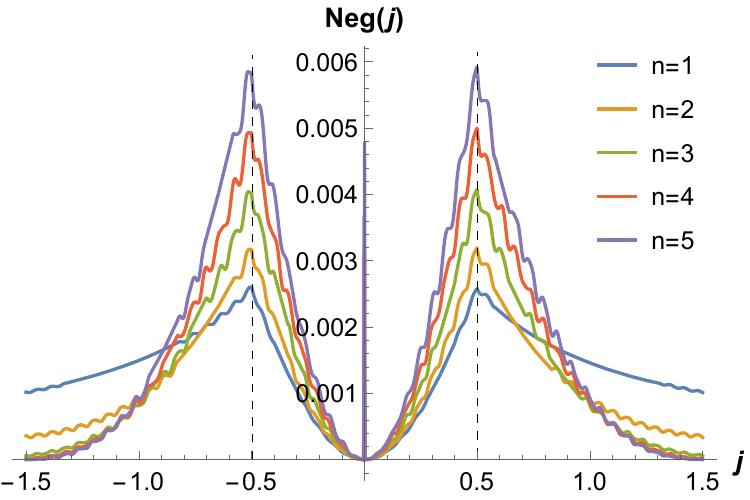}
        \caption{$r=1,\Delta r=100$}
\end{subfigure}
\begin{subfigure}{0.325\textwidth}
\centering
     \includegraphics[scale=0.38,frame]{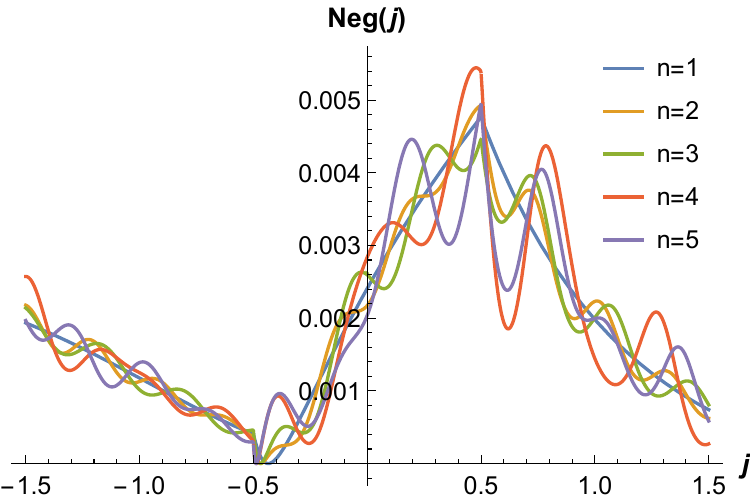}
        \caption{$r=10,\Delta r=1$}
\end{subfigure}
\begin{subfigure}{0.325\textwidth}
\centering
     \includegraphics[scale=0.38,frame]{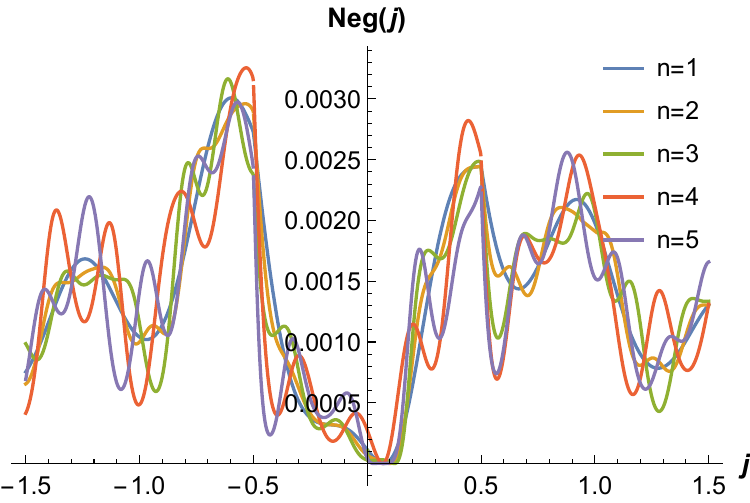}
        \caption{$r=10,\Delta r=10$}
\end{subfigure}
\begin{subfigure}{0.325\textwidth}
\centering
     \includegraphics[scale=0.38,frame]{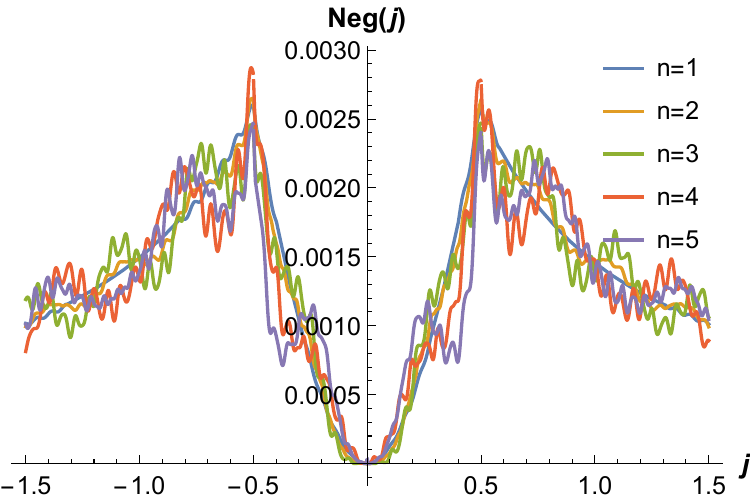}
        \caption{$r=10,\Delta r=100$}
\end{subfigure}
\begin{subfigure}{0.325\textwidth}
\centering
     \includegraphics[scale=0.38,frame]{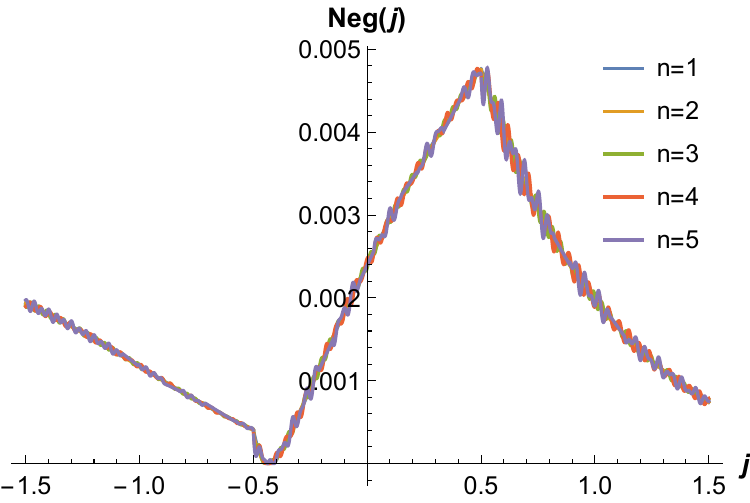}
        \caption{$r=100,\Delta r=1$}
\end{subfigure}
\begin{subfigure}{0.325\textwidth}
\centering
     \includegraphics[scale=0.38,frame]{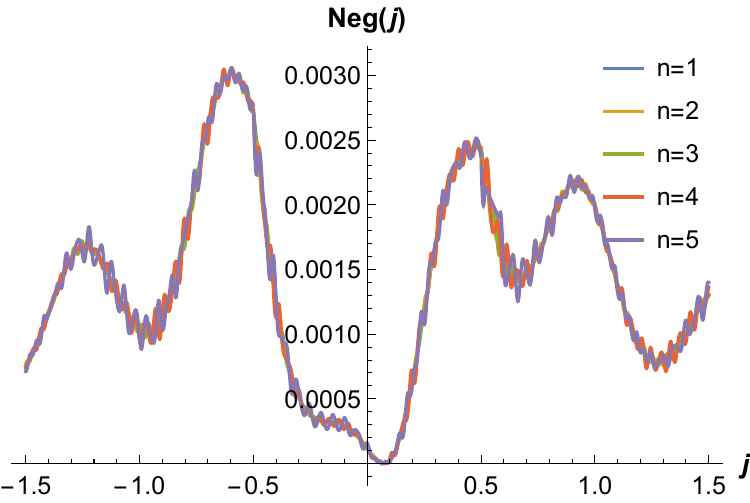}
        \caption{$r=100,\Delta r=10$}
\end{subfigure}
\begin{subfigure}{0.325\textwidth}
\centering
     \includegraphics[scale=0.38,frame]{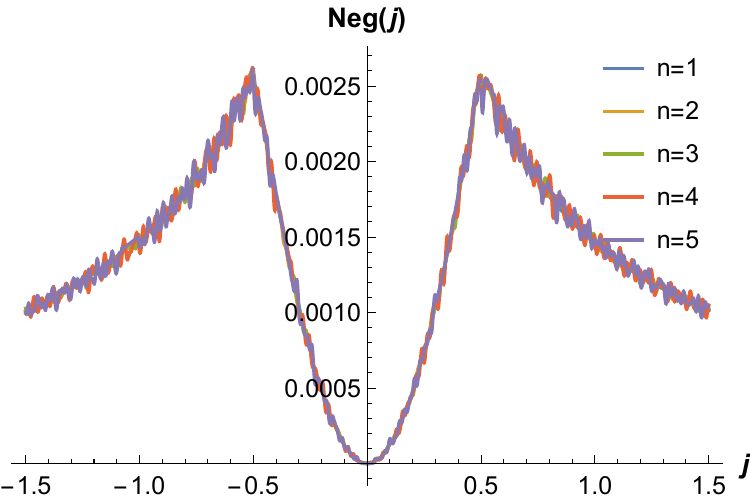}
        \caption{$r=100,\Delta r=100$}
\end{subfigure}
\caption{Negativity as a function of the interaction coupling constant $j$, evaluated with $\mathcal{T}=1000$, $\omega=0.5$.}
\label{T100_omega05_j_r_deltar_n}
\end{figure}

\item 
\underline{\textit{Negativity as function of $XY$-Heisenberg interaction coupling}}\\
The effect of the direct $XY$-Heisenberg interaction between the qubits is analyzed in the figure
\ref{T100_omega05_j_r_deltar_n}. There it is shown that it is always possible to find some finite value of the interaction constant, $j\neq 0$, that overcomes and enhances the entanglement harvesting with respect to the case without interaction $j=0$. Furthermore, the Negativity exhibits a parity symmetry with respect to $j$ when the distance between the qubits $\Delta r$ is sufficiently large, regardless of their position relative to the cosmic string, $r$. This symmetry is marked by the maximum degree of entanglement obtained at the resonance point $j=\pm\omega$. This is in accordance with what was established in Property \ref{prop 6}.

\begin{figure}[t!]
\centering
     \includegraphics[scale=0.58,frame]{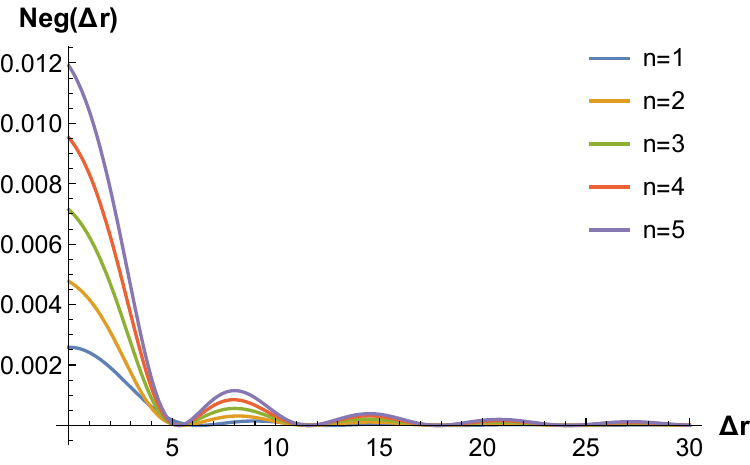}

\caption{Negativity as a function of the distance between the qubits $\Delta z$, evaluated with $\omega=0.5$, $j=0$, $\mathcal{T}=1000$, $r=1$.}
\label{T100_omega05_j0_r1_deltar_n}
\end{figure}

\item 
\textit{\underline{Entanglement as a function of qubits radial separation}}\\
 In the case without direct Heisenberg $XY$-interaction, $j=0$, we have that Negativity is maximum when both qubits are located at the same position and it decreases and oscillates as the radial relative distance between the qubits increase, see figure \ref{T100_omega05_j0_r1_deltar_n}. All these general behaviors are in agreement with Property \ref{prop 10}. It can be shown that the antiferromagnetic case, $j>0$, presents a similar behavior where entanglement departs from a maximum value at $\Delta r=0$, and shows a decreased oscillatory behavior towards a finite non-null value of entanglement. In this respect, the presence of a direct Heisenberg interaction becomes necessary in order to detect a cosmic string through the use of entanglement harvesting.\\
 Figure \ref{T100_omega05_j_r_deltar_n2} shows the dependence of Negativity with the radial distance between the qubits for the ferromagnetic case $j<0$.
In the ferromagnetic case, one sees a different behavior when both qubits are located at the same position. In this case when $\Delta r=0$ one finds a zero Negativity, which is in contrast with the maximum degree of entanglement found for the antiferromagnetic case.\\
 One can see in figure \ref{T100_omega05_j_r_deltar_n2}, that when $\Delta r$ increases, the negativity departs from zero, oscillates, and tends to a certain finite value that depends on the minimum distance of the qubits to the cosmic string $r$. In this way when $r$ is sufficiently large, the asymptotic value of Negativity becomes unique and corresponds to the value it would take in Minkowski space, i.e. as if there were no cosmic string at all. These results agree and confirm what was stated in Property \ref{prop 7} and Property \ref{prop 3}. 

\begin{figure}[h!]

\begin{subfigure}{0.5\textwidth}
\centering
     \includegraphics[scale=0.58,frame]{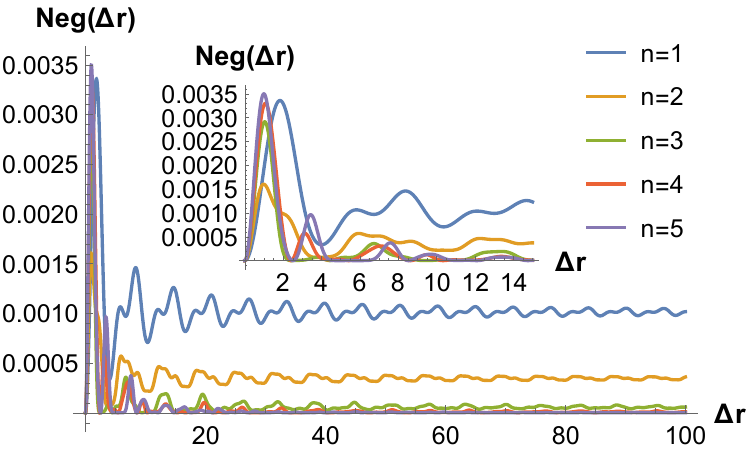}
          \caption{$r=1$}
\end{subfigure}
\begin{subfigure}{0.5\textwidth}
\centering
     \includegraphics[scale=0.58,frame]{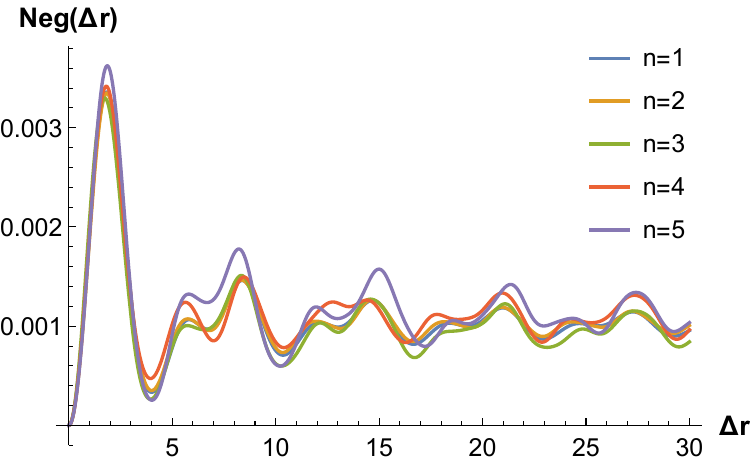}
        \caption{$r=30$}
\end{subfigure}

\caption{Negativity as a function of the distance between the qubits $\Delta r$, evaluated with $\omega=0.5$, $j=-1.5$, and $\mathcal{T}=1000$.}
\label{T100_omega05_j_r_deltar_n2}
\end{figure}

\item 
\underline{\textit{Non-local nature of entanglement harvesting}}\\
By examining the Negativity as a function of the relative distance between the qubits $\Delta r$ and the interaction time $\cal T$ one can manifest the non-local nature of entanglement harvesting in this system, see figure \ref{T_omega05_j05_r1_deltar_n}.
There we can also observe the oscillatory behavior of the negativity with respect to the relative distance between the qubits $\Delta z$. This is a general behavior of entanglement harvesting in the cosmic string background and is established in this work as Property \ref{prop 10}.\\
Furthermore, from figures (\ref{T_omega05_j05_r1_deltar_n}b) and (\ref{T_omega05_j05_r1_deltar_n}c), it is evident the non-local nature of the entanglement harvesting process.
In (\ref{T_omega05_j05_r1_deltar_n}c) one has the dashed white line that represents the boundary of the light cone of one qubit relative to the other. Then, one can see that even when the qubits are causally disconnected (outside the light cone), there is some amount of entanglement harvesting due to the vacuum fluctuations. 

\begin{figure}[ht]
\begin{subfigure}{0.5\textwidth}
\centering
     \includegraphics[scale=0.6,frame]{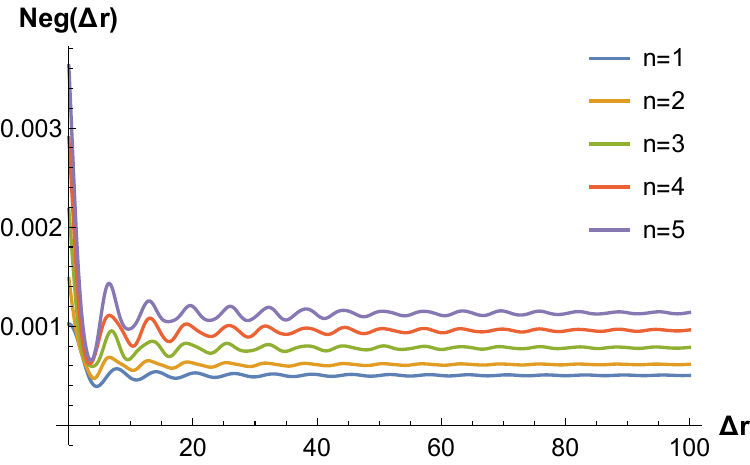}
          \caption{$\mathcal{T}=100$}
\end{subfigure}
\begin{subfigure}{0.5\textwidth}
\centering
     \includegraphics[scale=0.45]{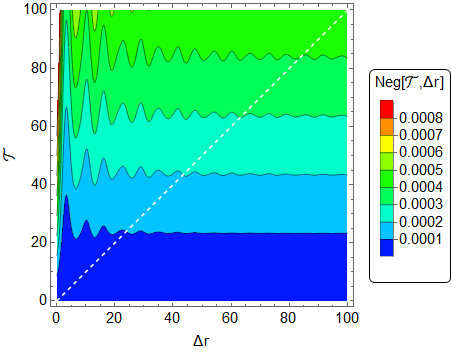}
          \caption{$n=4$}
\end{subfigure}

\caption{(b) and (c) are contour and three-dimensional plots of $T$ vs $\Delta r$ of the negativity with $n=4$, respectively, while (a) shows the behavior of the negativity at an interaction time of $\mathcal{T}=100$ with different values of $n$. All these graphs are evaluated with $\omega=0.5$, $j=0.5$, $r=1$, $\Delta\varphi=\Delta z=p=0$, and $g=0.01$.}
\label{T_omega05_j05_r1_deltar_n}
\end{figure}

\end{itemize} 

\newpage

\section{Conclusions}

In this work, the generation of entanglement between two qubits experiencing vacuum fluctuations of a massless scalar field in the cosmic string spacetime was investigated. The two qubits were modeled as Unruh-De Witt detectors coupled to the massless scalar field. In this study, we introduce a Heisenberg $XY$-interaction as a direct coupling between the qubits that can enhance entanglement harvesting. The dynamics of the qubits were determined perturbatively, assuming that the coupling constant $g$ with the scalar field is very small and that the initial density state of the system has null entanglement. General analyses of the entanglement between the qubits, measured by the Negativity function, were conducted. It was found that under our given initial conditions, entanglement can only be generated when the system of two qubits starts in a mixed state of states $|E_2\rangle$ and $|E_3\rangle$, with equal probabilities, see eq. (\ref{qubits energy states}).
Additionally, when both qubits are far from the cosmic string, the generation of entanglement is the same as that found in Minkowski space without the cosmic string. However, when both qubits are very close to the cosmic string, the generation of entanglement is maximum, being explicitly $n$ times the value found in Minkowski space, where $n$ is the cosmic string topological charge that depends on the energy density of the string. It was also found that the Negativity increases linearly with the interaction time with the scalar field until reaching an appropriate time that still satisfies the validity conditions of perturbation theory.

Regarding the description of the Negativity with respect to the coupling constant $j$, of the XY-Heisenberg interaction, it was observed that when both qubits are far apart, the ferromagnetic interaction ($j<0$) is equivalent to the antiferromagnetic interaction ($j>0$), reaching a maximum entanglement at the resonance points $j=\pm\omega$. Now, if both qubits are very close to each other and $j<-\omega$, the entanglement is approximately zero, being maximum only at $j=\omega$. The dependence of negativity on the distance between qubits will reach a maximum at a certain point with a sufficiently small distance between qubits and then decay, oscillating towards a negativity value that depends on a fixed $n$ and $r$, where $r$ is the distance between the cosmic string and the qubit closest to it. Finally, we found that when the separation between qubits is sufficiently large, the Negativity monotonically increases with $j$, as long as $j$ belongs to the intervals $[-\infty, -\omega]$ and $[0,\omega]$, while if $j$ belongs to the intervals $[-\omega,0]$ and $[\omega,\infty]$, then the Negativity monotonically decreases with $j$.

In particular, entanglement was analyzed for three specific cases of spatial symmetry: the axial case, the radial case, and the angular case, where each of the general properties obtained could be verified. Finally, it is worth mentioning that the results obtained in this work complement those obtained by \cite{he_yu_hu_2020, huang_he_2020}, where the entanglement behavior of two qubits without interaction between them was analyzed using the Kossakowski-Lindblad master equation regarding the study of open systems interacting with an external field. Additionally, given the advantages of simplicity, flexibility, and physical intuition offered by the perturbation theory compared to the Kossakowski-Lindblad equation, we were able to obtain general results on the behavior of entanglement generation, something that is very complicated to accomplish with the open quantum system approach.

\appendix

\section{Calculation of probability response functions $H$ and $F$}

Given the importance of $H$ and $F$ to obtain the density matrix corrections, let us calculate $H$ from \eqref{H function} with the Wightman function given by \eqref{whithmann in cosmic string}. Thus \begin{equation}\label{compute of J0}
    H({\cal T}, \Omega, \Delta r,\Delta\varphi,\Delta z)=\frac{-1}{4\pi^2}\int_{-{\cal T}/2}^{{\cal T}/2}d\uptau\int_{-{\cal T}/2}^{{\cal T}/2}d\uptau'\, e^{i\Omega\Delta\uptau} \sum^{n-1}_{k=0}\frac{1}{(\Delta \uptau-i\epsilon)^2-g^2_{kn}(r,\varphi,z)},
\end{equation} where \begin{equation}\label{gkn definition}
    g^2_{kn}(r,\varphi,z)=\Delta r^2+\Delta z^2+4rr'\sin^2(\pi k/n+\Delta\varphi/2).
\end{equation}

Now, to integrate this, we can use the method used by \cite{Svaiter_1994}, however here we will use the regularization method used by \cite{Enrike}. For this we introduce the following function of regularization \begin{equation}\label{regularization function}
    \xi_{\mathcal{T}}(\uptau)=\frac{(\mathcal{T}/2)^2}{\uptau^2+(\mathcal{T}/2)^2},
\end{equation} then, since $\xi_{\mathcal{T}}$ falls rapidly when $\uptau> \mathcal{T}$. we can introduce these functions in the finite integrals \eqref{compute of J0} in order to make it an infinite integral and in this way use the residue method to make the integration of said integral. In this way \begin{equation}
    H({\cal T}, \Omega, \Delta r,\Delta\varphi,\Delta z)=\frac{-1}{4\pi^2}\sum^{n-1}_{k=0}\int_{-\infty}^{\infty}d\uptau\int_{-\infty}^{\infty}d\uptau'\, \xi_{\mathcal{T}}(\uptau)\xi_{\mathcal{T}}(\uptau') \frac{e^{i\Omega\Delta\uptau}}{(\Delta \uptau-i\epsilon)^2-g^2_{kn}(r,\varphi,z)}.
\end{equation}

The product yield \begin{equation}
    \xi_{\mathcal{T}}(\uptau)\xi_{\mathcal{T}}(\uptau')=\frac{\mathcal{T}^4}{\left[(\uptau+\uptau')^2-(\Delta\tau+i\mathcal{T})^2\right]\left[(\uptau+\uptau')^2-(\Delta\uptau-i\mathcal{T})^2\right]}.
\end{equation}

Also, making the change of variables in the integral from $\uptau,\uptau'$ to $T=\uptau+\uptau'$ and $\Delta\uptau=\uptau-\uptau'$, we have \begin{align}\label{compute of J1}
    H({\cal T}, \Omega, \Delta r,\Delta\varphi,\Delta z)=\frac{-1}{8\pi^2}\sum^{n-1}_{k=0}&\int_{-\infty}^{\infty}d(\Delta\uptau)\, \frac{e^{i\Omega\Delta\uptau}}{(\Delta \uptau-i\epsilon)^2-g^2_{kn}(r,\varphi,z)}\nn\\
    &\times\int_{-\infty}^{\infty}dT\frac{\mathcal{T}^4}{\left[T^2-(\Delta\uptau+i\mathcal{T})^2\right]\left[T^2-(\Delta\uptau-i\mathcal{T})^2\right]}.
\end{align}

Integrating by residue method in the integral in $T$, we have\footnote{See, eq.(B.8) in \cite{Enrike}.} \begin{equation}\label{compute of J2}
    H({\cal T}, \Omega, \Delta r,\Delta\varphi,\Delta z)=\frac{-\mathcal{T}^3}{16\pi}\sum^{n-1}_{k=0}I_{kn}(\mathcal{T},\Omega),
\end{equation} where, $I_{kn}(\mathcal{T},\Omega)$ is the integral \begin{equation}\label{I function definition}
    I_{kn}(\mathcal{T},\Omega)=\int^\infty_{-\infty}d(\Delta \uptau)\frac{e^{i\Omega\Delta\uptau}}{[\Delta\uptau^2+\mathcal{T}^2][(\Delta\uptau-i\varepsilon)^2-g^2_{kn}(r,\varphi,z)]}.
\end{equation}

Using the remainder method, we replace $\Delta\uptau$ with a complex, so we see that the choice to integrate in the upper or lower complex plane depends on the sign of $\Omega$. Therefore, we divide this integral according to the sign of $\Omega$ with the help of the Heaviside function \begin{align*}
    I_{kn}(\mathcal{T},\Omega)=&\int_{-\infty}^{\infty}d(\Delta\uptau)\, \frac{e^{i\Omega\Delta\uptau}}{\left(\Delta\uptau+i\mathcal{T}\right)(\Delta\uptau-i\mathcal{T})\left(\Delta \uptau+g_{kn}(r,\varphi,z)-i\varepsilon\right)(\Delta\uptau-g_{kn}(r,\varphi,z)-i\varepsilon)}\Theta(-\Omega)\\
    &+\int_{-\infty}^{\infty}d(\Delta\uptau)\, \frac{e^{i\Omega\Delta\uptau}}{\left(\Delta\uptau+i\mathcal{T}\right)(\Delta\uptau-i\mathcal{T})\left(\Delta \uptau+g_{kn}(r,\varphi,z)-i\varepsilon\right)(\Delta\uptau-g_{kn}(r,\varphi,z)-i\varepsilon)}\Theta(\Omega).
\end{align*}  

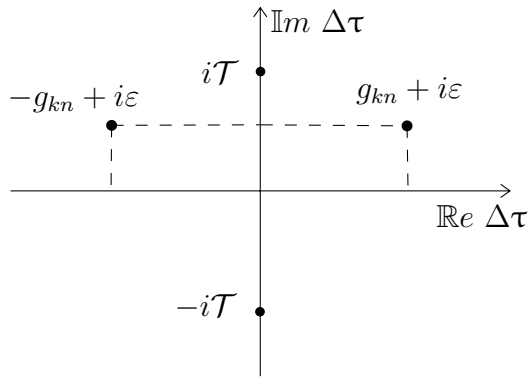
\begin{figure}[!ht]
        \begin{center}
          \begin{tikzpicture}[x=0.5pt,y=0.5pt,yscale=-1,xscale=1]

\draw  (144,149.91) -- (518,149.91)(331.01,11.71) -- (331.01,289.71) (511,144.91) -- (518,149.91) -- (511,154.91) (326.01,18.71) -- (331.01,11.71) -- (336.01,18.71)  ;
\draw  [fill={rgb, 255:red, 0; green, 0; blue, 0 }  ,fill opacity=1 ] (327.6,241.09) .. controls (327.6,239.33) and (329.03,237.9) .. (330.8,237.9) .. controls (332.56,237.9) and (333.99,239.33) .. (333.99,241.09) .. controls (333.99,242.85) and (332.56,244.28) .. (330.8,244.28) .. controls (329.03,244.28) and (327.6,242.85) .. (327.6,241.09) -- cycle ;
\draw  [dash pattern={on 4.5pt off 4.5pt}]  (219.72,100.45) -- (219.27,150.62) ;
\draw  [dash pattern={on 4.5pt off 4.5pt}]  (219.72,100.45) -- (331.37,100.06) ;
\draw [shift={(219.72,100.45)}, rotate = 359.8] [color={rgb, 255:red, 0; green, 0; blue, 0 }  ][fill={rgb, 255:red, 0; green, 0; blue, 0 }  ][line width=0.75]      (0, 0) circle [x radius= 3.35, y radius= 3.35]   ;
\draw  [dash pattern={on 4.5pt off 4.5pt}]  (440.92,100.85) -- (441.68,151.42) ;
\draw [shift={(440.92,100.85)}, rotate = 89.15] [color={rgb, 255:red, 0; green, 0; blue, 0 }  ][fill={rgb, 255:red, 0; green, 0; blue, 0 }  ][line width=0.75]      (0, 0) circle [x radius= 3.35, y radius= 3.35]   ;
\draw  [fill={rgb, 255:red, 0; green, 0; blue, 0 }  ,fill opacity=1 ] (328,59.89) .. controls (328,58.13) and (329.43,56.7) .. (331.2,56.7) .. controls (332.96,56.7) and (334.39,58.13) .. (334.39,59.89) .. controls (334.39,61.65) and (332.96,63.08) .. (331.2,63.08) .. controls (329.43,63.08) and (328,61.65) .. (328,59.89) -- cycle ;
\draw  [dash pattern={on 4.5pt off 4.5pt}]  (331.37,100.06) -- (440.92,100.85) ;

\draw (140.13,68.04) node [anchor=north west][inner sep=0.75pt]  [font=\large]  {$-g_{kn} +i\varepsilon $};
\draw (400.93,62.04) node [anchor=north west][inner sep=0.75pt]  [font=\large]  {$g_{kn} +i\varepsilon $};
\draw (285.22,50.11) node [anchor=north west][inner sep=0.75pt]  [font=\large]  {$i\mathcal{T}$};
\draw (265.22,226.04) node [anchor=north west][inner sep=0.75pt]  [font=\large]  {$-i\mathcal{T}$};
\draw (461,158.4) node [anchor=north west][inner sep=0.75pt]  [font=\large]  {$\mathbb{R} e\ \Delta \uptau $};
\draw (337,11.4) node [anchor=north west][inner sep=0.75pt]  [font=\large]  {$\mathbb{I} m\ \Delta \uptau $};

\end{tikzpicture}

          \caption{Four poles of integration.}
          \label{poles of integration}
         \end{center}
   \end{figure}
   
By the residue theorem, as long as $g^2_{kn}\neq 0$, we have \begin{align*}
    I_{kn}(\mathcal{T},\Omega)=&-2\pi i\left[\frac{e^{\Omega\mathcal{T}}}{(-2i\mathcal{T})(g_{kn}(r,\varphi,z)-i(\mathcal{T}+\varepsilon))(-g_{kn}(r,\varphi,z)-i(\mathcal{T}+\varepsilon))}\right]\Theta(-\Omega)\\
&+2\pi i\Bigg[\frac{e^{-\Omega\mathcal{T}}}{(2i\mathcal{T})(g_{kn}(r,\varphi,z)+i(\mathcal{T}-\varepsilon))(-g_{kn}(r,\varphi,z)+i(\mathcal{T}-\varepsilon))}\\
&\qquad\; + \frac{e^{-i\Omega g_{kn}(r,\varphi,z)}e^{-\Omega\varepsilon}}{(-2g_{kn}(r.\varphi.z))(-g_{kn}(r,\varphi,z)+i(\mathcal{T}+\varepsilon))(-g_{kn}(r,\varphi,z)-i(\mathcal{T}-\varepsilon))}\\
&\qquad\; + \frac{e^{i\Omega g_{kn}(r,\varphi,z)}e^{-\Omega\varepsilon}}{(2g_{kn}(r.\varphi.z))(g_{kn}(r,\varphi,z)+i(\mathcal{T}+\varepsilon))(g_{kn}(r,\varphi,z)-i(\mathcal{T}-\varepsilon))}\Bigg]\Theta(\Omega).
\end{align*}

Simplifying and making $\varepsilon\rightarrow 0$, yield \begin{align}
    \nn I_{kn}(\mathcal{T},\Omega)=-\frac{\pi}{\mathcal{T}}\frac{e^{\Omega\mathcal{T}}}{g^2_{kn}(r,\varphi,z)+\mathcal{T}^2}\Theta(-\Omega)+\Bigg[&-\frac{\pi}{\mathcal{T}}\frac{e^{-\Omega\mathcal{T}}}{g^2_{kn}(r,\varphi,z)+\mathcal{T}^2}+\frac{-\pi i}{g_{kn}(r,\varphi,z)} \frac{e^{-i\Omega g_{kn}(r,\varphi,z)}}{g^2_{kn}(r,\varphi,z)+\mathcal{T}^2}\\
&+\frac{\pi i}{g_{kn}(r,\varphi,z)} \frac{e^{i\Omega g_{kn}(r,\varphi,z)}}{g^2_{kn}(r,\varphi,z)+\mathcal{T}^2}\Bigg]\Theta(\Omega).
\end{align}

Rearranging and grouping \begin{align}
     \nn I_{kn}(\mathcal{T},\Omega)=&-\frac{\pi}{\mathcal{T}}\frac{1}{g^2_{kn}(r,\varphi,z)+\mathcal{T}^2}\left[e^{\Omega\mathcal{T}}\Theta(-\Omega)+e^{-\Omega\mathcal{T}}\Theta(\Omega)\right]-\frac{2\pi\sin[\Omega g_{kn}(r,\varphi,z)]}{g_{kn}(r,\varphi,z)\left[g^2_{kn}(r,\varphi,z)+\mathcal{T}^2\right]}\Theta(\Omega)\\\label{compute of J3}
     =&-\frac{\pi}{\mathcal{T}}\frac{e^{-|\Omega|\mathcal{T}}}{g^2_{kn}(r,\varphi,z)+\mathcal{T}^2}-\frac{2\pi\sin[\Omega g_{kn}(r,\varphi,z)]}{g_{kn}(r,\varphi,z)\left[g^2_{kn}(r,\varphi,z)+\mathcal{T}^2\right]}\Theta(\Omega).
\end{align}

When the condition of $g^2_{kn}(r,\varphi,z)= 0$ is fulfilled, then \eqref{I function definition} will be \begin{align}\label{I function definition2}
\nn I_{kn}(\mathcal{T},\Omega)=&\int^\infty_{-\infty}d(\Delta \uptau)\frac{e^{i\Omega\Delta\uptau}}{(\Delta\uptau+i\mathcal{T})(\Delta \uptau-i\mathcal{T})(\Delta\uptau-i\varepsilon)^2}\Theta(-\Omega)\\
&+\int^\infty_{-\infty}d(\Delta \uptau)\frac{e^{i\Omega\Delta\uptau}}{(\Delta\uptau+i\mathcal{T})(\Delta \uptau-i\mathcal{T})(\Delta\uptau-i\varepsilon)^2}\Theta(\Omega).
\end{align}

\begin{figure}[!ht]
        \begin{center}
          \begin{tikzpicture}[x=0.5pt,y=0.5pt,yscale=-1,xscale=1]

\draw  (200,150.2) -- (502,150.2)(351.01,12) -- (351.01,290) (495,145.2) -- (502,150.2) -- (495,155.2) (346.01,19) -- (351.01,12) -- (356.01,19)  ;
\draw  [fill={rgb, 255:red, 0; green, 0; blue, 0 }  ,fill opacity=1 ] (347.6,241.38) .. controls (347.6,239.61) and (349.03,238.18) .. (350.8,238.18) .. controls (352.56,238.18) and (353.99,239.61) .. (353.99,241.38) .. controls (353.99,243.14) and (352.56,244.57) .. (350.8,244.57) .. controls (349.03,244.57) and (347.6,243.14) .. (347.6,241.38) -- cycle ;
\draw  [fill={rgb, 255:red, 0; green, 0; blue, 0 }  ,fill opacity=1 ] (348,60.18) .. controls (348,58.41) and (349.43,56.98) .. (351.2,56.98) .. controls (352.96,56.98) and (354.39,58.41) .. (354.39,60.18) .. controls (354.39,61.94) and (352.96,63.37) .. (351.2,63.37) .. controls (349.43,63.37) and (348,61.94) .. (348,60.18) -- cycle ;
\draw  [fill={rgb, 255:red, 0; green, 0; blue, 0 }  ,fill opacity=1 ] (348.18,121.68) .. controls (348.18,119.92) and (349.61,118.49) .. (351.37,118.49) .. controls (353.14,118.49) and (354.57,119.92) .. (354.57,121.68) .. controls (354.57,123.45) and (353.14,124.88) .. (351.37,124.88) .. controls (349.61,124.88) and (348.18,123.45) .. (348.18,121.68) -- cycle ;

\draw (315.93,106.33) node [anchor=north west][inner sep=0.75pt]  [font=\large]  {$i\varepsilon $};
\draw (305.22,50.4) node [anchor=north west][inner sep=0.75pt]  [font=\large]  {$i\mathcal{T}$};
\draw (285.22,226.33) node [anchor=north west][inner sep=0.75pt]  [font=\large]  {$-i\mathcal{T}$};
\draw (433,160.69) node [anchor=north west][inner sep=0.75pt]  [font=\large]  {$\mathbb{R} e\ \Delta \uptau $};
\draw (357,11.69) node [anchor=north west][inner sep=0.75pt]  [font=\large]  {$\mathbb{I} m\ \Delta \uptau $};

\end{tikzpicture}

          \caption{three poles of integration.}
          \label{three poles of integration}
         \end{center}
   \end{figure}
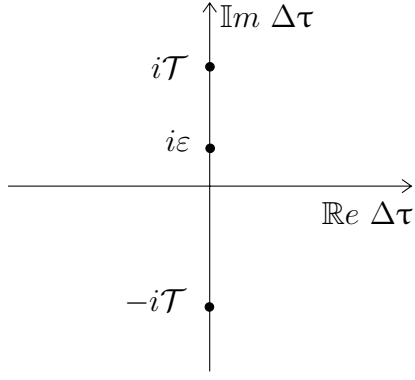

Therefore we have three poles, two simple poles and one pole of second order, such as\\ Figure \ref{three poles of integration} show us.

Thus, using the residue method, we have \begin{align*}
    I_{kn}(\mathcal{T},\Omega)=&-2\pi i\left[\frac{e^{\Omega\mathcal{T}}}{(-2i\mathcal{T})(-i(\mathcal{T}+\varepsilon))^2}\right]\Theta(-\Omega)\\
    &+2\pi i\Bigg[\frac{e^{-\Omega\mathcal{T}}}{(2i\mathcal{T})(i(\mathcal{T}-\varepsilon))^2}+i\Omega\frac{e^{-\Omega\varepsilon}}{i(\mathcal{T}+\varepsilon)(-i(\mathcal{T}-\varepsilon))}\\
    &\qquad\quad -\frac{e^{-\Omega\varepsilon}}{(i(\mathcal{T}+\varepsilon))^2(-i(\mathcal{T}-\varepsilon))}-\frac{e^{-\Omega\varepsilon}}{(i(\mathcal{T}+\varepsilon))(-i(\mathcal{T}-\varepsilon))^2}\Bigg]\Theta(\Omega)
\end{align*}

Simplifying and making $\varepsilon\rightarrow 0$, yield \begin{align}
    \nn I_{kn}(\mathcal{T},\Omega)=&-\frac{\pi}{\mathcal{T}^3}e^{\Omega\mathcal{T}}\Theta(-\Omega)-\left[\frac{\pi}{\mathcal{T}^3}e^{-\Omega\mathcal{T}}+\frac{2\pi\Omega}{\mathcal{T}^2}\right]\Theta(\Omega)\\
   \nn =&-\frac{\pi}{\mathcal{T}^3}\left[e^{\Omega\mathcal{T}}\Theta(-\Omega)+e^{-\Omega\mathcal{T}}\Theta(\Omega)\right]-\frac{2\pi\Omega}{\mathcal{T}^2}\Theta(\Omega)\\\label{calculation of I2}
   =&-\frac{\pi}{\mathcal{T}^3}e^{-|\Omega|\mathcal{T}}-\frac{2\pi\Omega}{\mathcal{T}^2}\Theta(\Omega).
\end{align}

Here it is worth noting that if we calculate the limit $g_{kn}\rightarrow 0$ at \eqref{compute of J3}, we note that the result is \eqref{calculation of I2}, therefore the most general result for $I_{kn}$ is \eqref{compute of J3} since we can include the condition $g_{kn}=0$ simply making the limit $g_{kn}\rightarrow 0$.

Now, substituting $I_{kn}$ \eqref{compute of J3} into $H$ \eqref{compute of J2} we have \begin{align}\label{function H}
     H(\mathcal{T},\Omega,\Delta r,\Delta\varphi, \Delta z)=&\frac{\mathcal{T}^2}{16}e^{-|\Omega|\mathcal{T}}\sum^{n-1}_{k=0}\frac{1}{g^2_{kn}(r,\varphi,z)+\mathcal{T}^2}+\Theta(\Omega)\frac{\mathcal{T}^3}{8}\sum^{n-1}_{k=0}\frac{\sin[\Omega g_{kn}(r,\varphi,z)]}{g_{kn}(r,\varphi,z)\left[g^2_{kn}(r,\varphi,z)+\mathcal{T}^2\right]}.
\end{align} 

The function $F$ is the particular case when $g_{kn}=2r\sin(\pi k/n)$ in $H$ so the function $F$ will be \begin{equation}\label{function F}
    F(\mathcal{T},\Omega,r)=\frac{\mathcal{T}^2}{64r^2}e^{-|\Omega|\mathcal{T}}\sum^{n-1}_{k=0}\frac{1}{\sin^2(\pi k/n)+\mathcal{T}^2/4r^2}+\Theta(\Omega)\frac{\mathcal{T}^3}{64r^3}\sum^{n-1}_{k=0}\frac{\sin[2\Omega r \sin(\pi k/n)]}{\sin(\pi k/n)\left[\sin^2(\pi k/n)+\mathcal{T}^2/4r^2\right]}
\end{equation}

\section{Calculation of coherence response functions $\mathcal{Y}_0$ and $\mathcal{Y}_I$}

Now, secondly let us calculate $\mathcal{Y}_0$ from \eqref{Y0 function} with the Wightman function given by \eqref{whithmann in cosmic string}, when $g_{kn}=2r\sin(\pi k/n)$. Thus \begin{equation}
    \mathcal{Y}_0(\omega, j, {\cal T}, r)=\frac{-1}{4\pi^2}\int_{-{\cal T}/2}^{{\cal T}/2}d\uptau\int_{-{\cal T}/2}^{{\cal T}/2}d\uptau'\, e^{i\omega T}\left(e^{ij\Delta\uptau}-e^{-ij\Delta\uptau}\right) \sum^{n-1}_{k=0}\frac{1}{(\Delta \uptau-i\epsilon)^2-g^2_{kn}}.
\end{equation}

Again, we use the function of regularization \eqref{regularization function}. Therefore, we have to calculate \begin{align}\label{compute of Y1}
    \mathcal{Y}_0(\omega, j, {\cal T}, r)=\frac{-\mathcal{T}^4}{8\pi^2}\sum^{n-1}_{k=0}&\int_{-\infty}^{\infty}d(\Delta\uptau)\, \frac{\left(e^{ij\Delta\uptau}-e^{-ij\Delta\uptau}\right)}{(\Delta \uptau-i\epsilon)^2-g^2_{kn}}\nn\\
    &\times\int_{-\infty}^{\infty}dT\frac{e^{i\omega T}}{\left[T^2-(\Delta\uptau+i\mathcal{T})^2\right]\left[T^2-(\Delta\uptau-i\mathcal{T})^2\right]}.
\end{align}

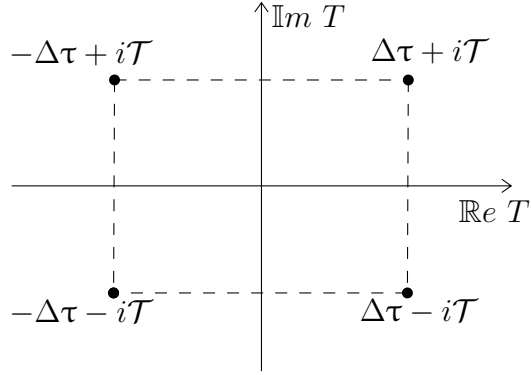
\begin{figure}[!ht]
        \begin{center}
          \begin{tikzpicture}[x=0.5pt,y=0.5pt,yscale=-1,xscale=1]

\draw  (164,150.2) -- (538,150.2)(351.01,12) -- (351.01,290) (531,145.2) -- (538,150.2) -- (531,155.2) (346.01,19) -- (351.01,12) -- (356.01,19)  ;
\draw  [dash pattern={on 4.5pt off 4.5pt}]  (241,70.3) -- (240.29,230.99) ;
\draw [shift={(240.29,230.99)}, rotate = 90.25] [color={rgb, 255:red, 0; green, 0; blue, 0 }  ][fill={rgb, 255:red, 0; green, 0; blue, 0 }  ][line width=0.75]      (0, 0) circle [x radius= 3.35, y radius= 3.35]   ;
\draw [shift={(241,70.3)}, rotate = 90.25] [color={rgb, 255:red, 0; green, 0; blue, 0 }  ][fill={rgb, 255:red, 0; green, 0; blue, 0 }  ][line width=0.75]      (0, 0) circle [x radius= 3.35, y radius= 3.35]   ;
\draw  [dash pattern={on 4.5pt off 4.5pt}]  (241,70.3) -- (460.96,70.33) ;
\draw [shift={(460.96,70.33)}, rotate = 0.01] [color={rgb, 255:red, 0; green, 0; blue, 0 }  ][fill={rgb, 255:red, 0; green, 0; blue, 0 }  ][line width=0.75]      (0, 0) circle [x radius= 3.35, y radius= 3.35]   ;
\draw  [dash pattern={on 4.5pt off 4.5pt}]  (460.29,230.99) -- (240.29,230.99) ;
\draw [shift={(240.29,230.99)}, rotate = 180] [color={rgb, 255:red, 0; green, 0; blue, 0 }  ][fill={rgb, 255:red, 0; green, 0; blue, 0 }  ][line width=0.75]      (0, 0) circle [x radius= 3.35, y radius= 3.35]   ;
\draw [shift={(460.29,230.99)}, rotate = 180] [color={rgb, 255:red, 0; green, 0; blue, 0 }  ][fill={rgb, 255:red, 0; green, 0; blue, 0 }  ][line width=0.75]      (0, 0) circle [x radius= 3.35, y radius= 3.35]   ;
\draw  [dash pattern={on 4.5pt off 4.5pt}]  (460.96,70.33) -- (460.29,230.99) ;

\draw (430.93,37.67) node [anchor=north west][inner sep=0.75pt]  [font=\large]  {$\Delta \uptau +i\mathcal{T}$};
\draw (160.22,39.07) node [anchor=north west][inner sep=0.75pt]  [font=\large]  {$-\Delta \uptau +i\mathcal{T}$};
\draw (160.56,234.33) node [anchor=north west][inner sep=0.75pt]  [font=\large]  {$-\Delta \uptau -i\mathcal{T}$};
\draw (497,158.69) node [anchor=north west][inner sep=0.75pt]  [font=\large]  {$\mathbb{R} e\ T$};
\draw (357,11.69) node [anchor=north west][inner sep=0.75pt]  [font=\large]  {$\mathbb{I} m\ T$};
\draw (422.67,234.21) node [anchor=north west][inner sep=0.75pt]  [font=\large]  {$\Delta \uptau -i\mathcal{T}$};

\end{tikzpicture}

          \caption{Four poles of integration.}
          \label{Four poles of integration Y}
         \end{center}
   \end{figure}

First, let us to calculate the integral on $T$, where we observe that there are four poles, as shown in Figure \ref{Four poles of integration Y}

In this way, using the residue method, this integral becomes \begin{align}
   \nn \int_{-\infty}^{\infty}dT\frac{e^{i\omega T}}{\left[T^2-(\Delta\uptau+i\mathcal{T})^2\right]\left[T^2-(\Delta\uptau-i\mathcal{T})^2\right]}=&\frac{\pi e^{\omega\mathcal{T}}}{4\mathcal{T}}\left[\frac{e^{-i\omega\Delta\uptau}}{\Delta\uptau(\Delta\uptau+i\mathcal{T})}+\frac{e^{i\omega\Delta\uptau}}{\Delta\uptau(\Delta\uptau-i\mathcal{T})}\right]\Theta(-\omega)\\
   \nn &+\frac{\pi e^{-\omega\mathcal{T}}}{4\mathcal{T}}\left[\frac{e^{-i\omega\Delta\uptau}}{\Delta\uptau(\Delta\uptau-i\mathcal{T})}+\frac{e^{i\omega\Delta\uptau}}{\Delta\uptau(\Delta\uptau+i\mathcal{T})}\right]\Theta(\omega)\\\label{result of T calculo}
   =&\frac{\pi e^{-|\omega|\mathcal{T}}}{4\mathcal{T}}\left[\frac{e^{-i|\omega|\Delta\uptau}}{\Delta\uptau(\Delta\uptau-i\mathcal{T})}+\frac{e^{i|\omega|\Delta\uptau}}{\Delta\uptau(\Delta\uptau+i\mathcal{T})}\right]
\end{align}

Therefore, we will have \begin{align}\label{Y0 sum}
     \mathcal{Y}_0(\omega, j, \mathcal{T}, r)=\frac{-\mathcal{T}^3e^{-|\omega|\mathcal{T}}}{32\pi}\sum^{n-1}_{k=0}\left[A_k(j+|\omega|)-B_k(-j-|\omega|)-\left(A_k(-j+|\omega|)-B_k(j-|\omega|)\right)\right]
\end{align} where \begin{align}
    A_k(\mathcal{T},\Omega)=\int^\infty_{-\infty}d(\Delta\uptau)\frac{e^{i\Omega\Delta\uptau}}{\Delta\uptau(\Delta\uptau+i\mathcal{T})(\Delta\uptau-g_{kn}-i\varepsilon)(\Delta\uptau+g_{kn}-i\varepsilon)}\\
    B_k(\mathcal{T},\Omega)=\int^\infty_{-\infty}d(\Delta\uptau)\frac{e^{i\Omega\Delta\uptau}}{\Delta\uptau(\Delta\uptau-i\mathcal{T})(\Delta\uptau-g_{kn}-i\varepsilon)(\Delta\uptau+g_{kn}-i\varepsilon)}
\end{align}

We must observe that the integrals $A_k$ and $B_k$ are divergent in $\Delta \uptau$, so to compute these integrals we use the Sokhotski-Plemelj formula \begin{equation}
    \lim_{\delta\rightarrow 0}\int^\infty_{-\infty} \frac{f(x)}{x\pm i\delta} dx=P\int^\infty_{-\infty}\frac{f(x)}{x}dx\mp i\pi f(0),
\end{equation} where $f(x)$ is any function that is smooth and non-singular in a neighborhood of the origin and said function fall to zero when $x\rightarrow\pm\infty$. Also, the principal value is defined by \begin{equation}
    P\int^\infty_{-\infty}\frac{f(x)}{x}dx=\lim_{\delta\rightarrow 0}\left(\int^{-\delta}_{-\infty}\frac{f(x)}{x}dx+\int^\infty_\delta\frac{f(x)}{x}dx\right).
\end{equation}

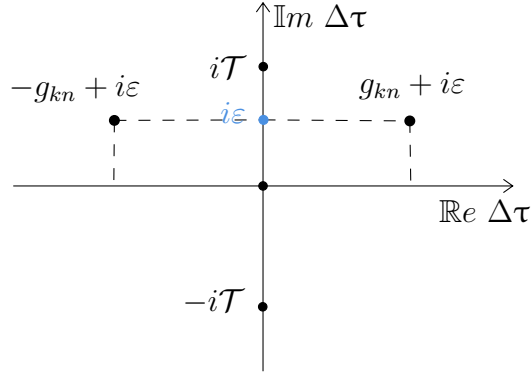
\begin{figure}[!ht]
        \begin{center}
          \begin{tikzpicture}[x=0.5pt,y=0.5pt,yscale=-1,xscale=1]

\draw  (145,150.2) -- (519,150.2)(332.01,12) -- (332.01,290) (512,145.2) -- (519,150.2) -- (512,155.2) (327.01,19) -- (332.01,12) -- (337.01,19)  ;
\draw  [fill={rgb, 255:red, 0; green, 0; blue, 0 }  ,fill opacity=1 ] (328.6,241.38) .. controls (328.6,239.61) and (330.03,238.18) .. (331.8,238.18) .. controls (333.56,238.18) and (334.99,239.61) .. (334.99,241.38) .. controls (334.99,243.14) and (333.56,244.57) .. (331.8,244.57) .. controls (330.03,244.57) and (328.6,243.14) .. (328.6,241.38) -- cycle ;
\draw  [dash pattern={on 4.5pt off 4.5pt}]  (220.72,100.74) -- (220.27,150.91) ;
\draw  [dash pattern={on 4.5pt off 4.5pt}]  (220.72,100.74) -- (332.37,100.35) ;
\draw [shift={(220.72,100.74)}, rotate = 359.8] [color={rgb, 255:red, 0; green, 0; blue, 0 }  ][fill={rgb, 255:red, 0; green, 0; blue, 0 }  ][line width=0.75]      (0, 0) circle [x radius= 3.35, y radius= 3.35]   ;
\draw  [dash pattern={on 4.5pt off 4.5pt}]  (441.92,101.14) -- (442.68,151.71) ;
\draw [shift={(441.92,101.14)}, rotate = 89.15] [color={rgb, 255:red, 0; green, 0; blue, 0 }  ][fill={rgb, 255:red, 0; green, 0; blue, 0 }  ][line width=0.75]      (0, 0) circle [x radius= 3.35, y radius= 3.35]   ;
\draw  [fill={rgb, 255:red, 0; green, 0; blue, 0 }  ,fill opacity=1 ] (329,60.18) .. controls (329,58.41) and (330.43,56.98) .. (332.2,56.98) .. controls (333.96,56.98) and (335.39,58.41) .. (335.39,60.18) .. controls (335.39,61.94) and (333.96,63.37) .. (332.2,63.37) .. controls (330.43,63.37) and (329,61.94) .. (329,60.18) -- cycle ;
\draw  [dash pattern={on 4.5pt off 4.5pt}]  (332.37,100.35) -- (441.92,101.14) ;
\draw  [color={rgb, 255:red, 74; green, 144; blue, 226 }  ,draw opacity=1 ][fill={rgb, 255:red, 74; green, 144; blue, 226 }  ,fill opacity=1 ] (328.98,100.35) .. controls (328.98,98.48) and (330.5,96.96) .. (332.37,96.96) .. controls (334.25,96.96) and (335.77,98.48) .. (335.77,100.35) .. controls (335.77,102.22) and (334.25,103.74) .. (332.37,103.74) .. controls (330.5,103.74) and (328.98,102.22) .. (328.98,100.35) -- cycle ;
\draw  [fill={rgb, 255:red, 0; green, 0; blue, 0 }  ,fill opacity=1 ] (328.82,150.2) .. controls (328.82,148.43) and (330.25,147) .. (332.01,147) .. controls (333.78,147) and (335.21,148.43) .. (335.21,150.2) .. controls (335.21,151.96) and (333.78,153.39) .. (332.01,153.39) .. controls (330.25,153.39) and (328.82,151.96) .. (328.82,150.2) -- cycle ;

\draw (140,65) node [anchor=north west][inner sep=0.75pt]  [font=\large]  {$-g_{kn} +i\varepsilon $};
\draw (401.93,62.33) node [anchor=north west][inner sep=0.75pt]  [font=\large]  {$g_{kn} +i\varepsilon $};
\draw (290,50.4) node [anchor=north west][inner sep=0.75pt]  [font=\large]  {$i\mathcal{T}$};
\draw (270,226.33) node [anchor=north west][inner sep=0.75pt]  [font=\large]  {$-i\mathcal{T}$};
\draw (462,158.69) node [anchor=north west][inner sep=0.75pt]  [font=\large]  {$\mathbb{R} e\ \Delta \uptau $};
\draw (338,11.69) node [anchor=north west][inner sep=0.75pt]  [font=\large]  {$\mathbb{I} m\ \Delta \uptau $};
\draw (298.33,84.73) node [anchor=north west][inner sep=0.75pt]  [font=\large,color={rgb, 255:red, 74; green, 144; blue, 226 }  ,opacity=1 ]  {$i\varepsilon $};

\end{tikzpicture}

          \caption{Integration poles of integrals $A_k$ and $B_k$, together with the delocalization of polo at the origin of the complex plane.}
          \label{Four poles of integration A and B}
         \end{center}
   \end{figure}

Thus, \begin{align}
   \nn A_k(\mathcal{T},\Omega)=&\lim_{\varepsilon\rightarrow 0}\int^\infty_{-\infty}d(\Delta\uptau)\frac{e^{i\Omega\Delta\uptau}}{(\Delta\uptau- i\varepsilon)(\Delta\uptau+i\mathcal{T})(\Delta\uptau-g_{kn}-i\varepsilon)(\Delta\uptau+g_{kn}-i\varepsilon)}\\\label{Sokhotski-Plemej for A}
     &- \frac{-\pi}{\mathcal{T}(g_{kn}+i\varepsilon)(g_{kn}-i\varepsilon)}\\
   \nn B_k(\mathcal{T},\Omega)=&\lim_{\varepsilon\rightarrow 0}\int^\infty_{-\infty}d(\Delta\uptau)\frac{e^{i\Omega\Delta\uptau}}{(\Delta\uptau- i\varepsilon)(\Delta\uptau-i\mathcal{T})(\Delta\uptau-g_{kn}-i\varepsilon)(\Delta\uptau+g_{kn}-i\varepsilon)}\\\label{Sokhotski-Plemej for B}
     &- \frac{\pi}{\mathcal{T}(g_{kn}+i\varepsilon)(g_{kn}-i\varepsilon)},
\end{align} here we have delocalized the pole at zero to $i\varepsilon$, see Figure \ref{Four poles of integration A and B}. 

Using the residue method, first we integrate $A_k$, considering the cases when $g_{kn}\neq 0$ \begin{align}
   \nn A_k(\mathcal{T},\Omega)=&\lim_{\varepsilon\rightarrow 0}\Bigg[2\pi i\left(\frac{e^{i\Omega g_{kn}}e^{-\Omega\varepsilon}}{2g^2_{kn}(g_{kn}+i(\mathcal{T}+\varepsilon))}-\frac{e^{-i\Omega g_{kn}}e^{-\Omega\varepsilon}}{2g^2_{kn}(g_{kn}-i(\mathcal{T}+\varepsilon))}-\frac{ e^{-\Omega\varepsilon}}{i(\mathcal{T}+\varepsilon)g^2_{kn}}\right)\Theta(\Omega)\\
   \nn &+\left(\frac{-2\pi e^{\Omega\mathcal{T}}}{(\mathcal{T}+\varepsilon)(g^2_{kn}+(\mathcal{T}+\varepsilon)^2)}\right)\Theta(-\Omega)\Bigg]+\frac{\pi}{\mathcal{T}(g_{kn}+i\varepsilon)(g_{kn}-i\varepsilon)},
\end{align} where if we apply the limit $\varepsilon\rightarrow 0$, we have \begin{align}
 \nn A_k(\mathcal{T},\Omega)=&\frac{-2\pi}{\mathcal{T}g^2_{kn}(g^2_{kn}+\mathcal{T}^2)}\Big[\mathcal{T}g_{kn}\sin(\Omega g_{kn})-\mathcal{T}^2\cos(\Omega g_{kn})+g^2_{kn}+\mathcal{T}^2\Big]\Theta(\Omega)\\\label{funtion of integral A}
  &+\Bigg[\frac{-2\pi e^{\Omega\mathcal{T}}}{\mathcal{T}(g^2_{kn}+\mathcal{T}^2)}\Bigg]\Theta(-\Omega)+\frac{\pi}{\mathcal{T}(g_{kn}+i\varepsilon)(g_{kn}-i\varepsilon)}.
\end{align}

Now, we use again the residue method to calculate $B_k$. Thus when $g_{kn}\neq 0$, we have \begin{align}
   \nn B_k(\mathcal{T},\Omega)=&\lim_{\varepsilon\rightarrow 0}\Bigg[2\pi i\left(\frac{e^{i\Omega g_{kn}}e^{-\Omega\varepsilon}}{2g^2_{kn}(g_{kn}-i(\mathcal{T}-\varepsilon))}-\frac{e^{-i\Omega g_{kn}}e^{-\Omega\varepsilon}}{2g^2_{kn}(g_{kn}+i(\mathcal{T}-\varepsilon))}+\frac{ e^{-\Omega\varepsilon}}{i(\mathcal{T}-\varepsilon)g^2_{kn}}\right)\Theta(\Omega)\\
   \nn &+\left(\frac{-2\pi e^{-\Omega\mathcal{T}}}{(\mathcal{T}-\varepsilon)(g^2_{kn}+(\mathcal{T}-\varepsilon)^2)}\right)\Theta(\Omega)\Bigg]-\frac{\pi}{\mathcal{T}(g_{kn}+i\varepsilon)(g_{kn}-i\varepsilon)},
\end{align} and making the limit $\varepsilon\rightarrow 0$, yield \begin{align}
   \nn B_k(\mathcal{T},\Omega)=&\frac{-2\pi}{\mathcal{T}g^2_{kn}(g^2_{kn}+\mathcal{T}^2)}\Big[\mathcal{T}g_{kn}\sin(\Omega g_{kn})+\mathcal{T}^2\cos(\Omega g_{kn})-(g^2_{kn}+\mathcal{T}^2)\Big]\Theta(\Omega)\\\label{funtion of integral B}
  &+\Bigg[\frac{-2\pi e^{-\Omega\mathcal{T}}}{\mathcal{T}(g^2_{kn}+\mathcal{T}^2)}\Bigg]\Theta(\Omega)-\frac{\pi}{\mathcal{T}(g_{kn}+i\varepsilon)(g_{kn}-i\varepsilon)}.
\end{align}

 Now we calculate both $A_k$ and $B_k$ for the case $g_{kn}=0$, that is, from \eqref{Sokhotski-Plemej for A} and \eqref{Sokhotski-Plemej for B} we must calculate \begin{align}
    A_k(\mathcal{T},\Omega)=&\lim_{\varepsilon\rightarrow 0}\int^\infty_{-\infty}d(\Delta\uptau)\frac{e^{i\Omega\Delta\uptau}}{(\Delta\uptau- i\varepsilon)^3(\Delta\uptau+i\mathcal{T})}- \frac{-\pi}{\mathcal{T}\varepsilon^2}\\
    B_k(\mathcal{T},\Omega)=&\lim_{\varepsilon\rightarrow 0}\int^\infty_{-\infty}d(\Delta\uptau)\frac{e^{i\Omega\Delta\uptau}}{(\Delta\uptau- i\varepsilon)^3(\Delta\uptau-i\mathcal{T})}- \frac{\pi}{\mathcal{T}\varepsilon^2},
\end{align} where we observe that the $i\varepsilon$ pole is a triple pole.

Using the residue method, we calculate $A_k$ when $g_{kn}=0$, so that \begin{align*}
A_k(\mathcal{T},\Omega)=&\lim_{\varepsilon \rightarrow 0}\Bigg[2\pi i\bigg(\frac{1}{2!}\lim_{\Delta\uptau\rightarrow i\varepsilon}\frac{d^2}{d(\Delta\uptau)^2}\left(\frac{e^{i\Omega\Delta\uptau}}{\Delta\uptau+i\mathcal{T}}\right)\bigg)\Theta(\Omega)-2\pi i\frac{e^{\Omega \mathcal{T}}}{i(\mathcal{T}+\varepsilon)^3}\Theta(-\Omega)\Bigg]+\frac{\pi}{\mathcal{T}\varepsilon^2}\\
=&\lim_{\varepsilon \rightarrow 0}\Bigg[2\pi i\bigg(\frac{1}{2}\lim_{\Delta\uptau\rightarrow i\varepsilon}\left(\frac{-\Omega^2e^{i\Omega\Delta\uptau}}{\Delta\uptau+i\mathcal{T}}-\frac{2i\Omega e^{i\Omega\Delta\uptau}}{(\Delta\uptau+i\mathcal{T})^2}+\frac{2e^{i\Omega\Delta\uptau}}{(\Delta\uptau+i\mathcal{T})^3}\right)\bigg)\Theta(\Omega)\\
&-\frac{2\pi e^{\Omega \mathcal{T}}}{(\mathcal{T}+\varepsilon)^3}\Theta(-\Omega)\Bigg]+\frac{\pi}{\mathcal{T}\varepsilon^2}\\
=&\lim_{\varepsilon \rightarrow 0}\Bigg[\pi i\bigg(\frac{-\Omega^2e^{-\Omega\varepsilon}}{i(\mathcal{T}+\varepsilon)}+\frac{2i\Omega e^{-\Omega\varepsilon}}{(\mathcal{T}+\varepsilon)^2}-\frac{2e^{-\Omega\varepsilon}}{i(\mathcal{T}+\varepsilon)^3}\bigg)\Theta(\Omega)-\frac{2\pi e^{\Omega \mathcal{T}}}{(\mathcal{T}+\varepsilon)^3}\Theta(-\Omega)\Bigg]+\frac{\pi}{\mathcal{T}\varepsilon^2},
\end{align*} and applying the limit $\varepsilon\rightarrow 0$, we have \begin{align}\label{funtion of integral A when g equal 0}
   A_k(\mathcal{T},\Omega)= \frac{-2\pi}{\mathcal{T}^3}\left[\frac{(\Omega\mathcal{T}+1)^2+1}{2}\Theta(\Omega)+e^{\Omega\mathcal{T}}\Theta(-\Omega)\right]+\frac{\pi}{\mathcal{T}\varepsilon^2}.
\end{align}

Again using the remainder method we calculate $B_k$, when $g_{kn}=0$ \begin{align*}
B_k(\mathcal{T},\Omega)=&2\pi i\lim_{\varepsilon \rightarrow 0}\Bigg[\frac{1}{2!}\lim_{\Delta\uptau\rightarrow i\varepsilon}\frac{d^2}{d(\Delta\uptau)^2}\left(\frac{e^{i\Omega\Delta\uptau}}{\Delta\uptau-i\mathcal{T}}\right)-\frac{e^{-\Omega \mathcal{T}}}{i(\mathcal{T}-\varepsilon)^3}\Bigg]\Theta(\Omega)-\frac{\pi}{\mathcal{T}\varepsilon^2}\\
=&2\pi i\lim_{\varepsilon \rightarrow 0}\Bigg[\frac{1}{2}\lim_{\Delta\uptau\rightarrow i\varepsilon}\left(\frac{-\Omega^2e^{i\Omega\Delta\uptau}}{\Delta\uptau-i\mathcal{T}}-\frac{2i\Omega e^{i\Omega\Delta\uptau}}{(\Delta\uptau-i\mathcal{T})^2}+\frac{2e^{i\Omega\Delta\uptau}}{(\Delta\uptau-i\mathcal{T})^3}\right)-\frac{e^{\Omega \mathcal{T}}}{i(\mathcal{T}-\varepsilon)^3}\Bigg]\Theta(\Omega)-\frac{\pi}{\mathcal{T}\varepsilon^2}\\
=&\lim_{\varepsilon \rightarrow 0}\Bigg[\pi i\bigg(\frac{\Omega^2e^{-\Omega\varepsilon}}{i(\mathcal{T}-\varepsilon)}+\frac{2i\Omega e^{-\Omega\varepsilon}}{(\mathcal{T}-\varepsilon)^2}+\frac{2e^{-\Omega\varepsilon}}{i(\mathcal{T}-\varepsilon)^3}\bigg)-\frac{2\pi e^{\Omega \mathcal{T}}}{(\mathcal{T}-\varepsilon)^3})\Bigg]\Theta(\Omega)-\frac{\pi}{\mathcal{T}\varepsilon^2},
\end{align*} and applying the limit $\varepsilon\rightarrow 0$, we have \begin{align}\label{funtion of integral B when g equal 0}
    B_k(\mathcal{T},\Omega)=\frac{2\pi}{\mathcal{T}^3}\left[\frac{(\Omega\mathcal{T}-1)^2+1}{2}-e^{-\Omega\mathcal{T}}\right]\Theta(\Omega)-\frac{\pi}{\mathcal{T}\varepsilon^2}.
\end{align}

It is worth mentioning that when we calculate the limit $g_{kn}\rightarrow 0$ in both \eqref{funtion of integral A} and \eqref{funtion of integral B}, we obtain \eqref{funtion of integral A when g equal 0} and \eqref{funtion of integral B when g equal 0} respectively. So for this reason we generally consider \eqref{funtion of integral A} and \eqref{funtion of integral B} as results of $A_k$ and $B_k$ respectively for any value of $g_{kn}$.

On the other hand, we calculate \begin{align}
   \nn A_k(\mathcal{T},\Omega)-B_k(\mathcal{T},-\Omega)=&\frac{-2\pi }{\mathcal{T}g^2_{kn}(g^2_{kn}+\mathcal{T}^2)}\left[\mathcal{T}g_{kn}\sin(\Omega g_{kn})-\mathcal{T}^2\cos(\Omega g_{kn})+g^2_{kn}+\mathcal{T}^2\right]\\\label{A menos B}
    &+\frac{2\pi}{\mathcal{T}(g_{kn}+i\varepsilon)(g_{kn}-i\varepsilon)}
\end{align}

Furthermore, we also calculate \begin{align}
    \nn A_k(\mathcal{T},\Omega)+B_k(\mathcal{T},-\Omega)=&\frac{-2\pi }{\mathcal{T}g^2_{kn}(g^2_{kn}+\mathcal{T}^2)}\left[\mathcal{T}g_{kn}\sin(\Omega g_{kn})-\mathcal{T}^2\cos(\Omega g_{kn})+g^2_{kn}+\mathcal{T}^2\right]\sign(\Omega)\\\label{A plus B}
     &-\Bigg[\frac{4\pi e^{\Omega\mathcal{T}}}{\mathcal{T}(g^2_{kn}+\mathcal{T}^2)}\Bigg]\Theta(-\Omega),
\end{align} where we have introduced the sign function, $\sign(\Omega)=\Theta(\Omega)-\Theta(-\Omega)$.

Using \eqref{A menos B} in \eqref{Y0 sum} we can calculate $\mathcal{Y}_0$. In this way \begin{align}\label{Y0 function result1}
    \mathcal{Y}_0(\omega, j,\mathcal{T}, r)=&\frac{\mathcal{T}^3e^{-|\omega|\mathcal{T}}}{8}\sum^{n-1}_{k=0}\frac{\sin(j g_{kn})}{g^2_{kn}(g^2_{kn}+\mathcal{T}^2)}\Big[g_{kn}\cos(|\omega|g_{kn})+\mathcal{T}\sin(|\omega|g_{kn})\Big]
\end{align} and if we analyze the expression for $g_{kn}=2r\sin(\pi k/n)$, we see that it vanishes only when $k=0$, which corresponds to the first term of the sum. So for this first term we can apply the limit $g_{kn}\rightarrow 0$, therefore \begin{equation}\label{Y0 function result2}
    \mathcal{Y}_0=\frac{1}{8}\mathcal{T}j(|\omega|\mathcal{T}+1)e^{-|\omega|\mathcal{T}}+\frac{\mathcal{T}^3e^{-|\omega|\mathcal{T}}}{8}\sum^{n-1}_{k=1}\frac{\sin(j g_{kn})}{g^2_{kn}(g^2_{kn}+\mathcal{T}^2)}\Big[g_{kn}\cos(|\omega|g_{kn})+\mathcal{T}\sin(|\omega|g_{kn})\Big].
\end{equation}

Lastly, we are going to calculate the function $\mathcal{Y}_I$ from \eqref{Yi integral definition}. In this way  \begin{equation}
    \mathcal{Y}_I(\omega, j, {\cal T},\Delta r,\Delta z, \Delta\varphi)=\frac{-1}{4\pi^2}\int_{-{\cal T}/2}^{{\cal T}/2}d\uptau\int_{-{\cal T}/2}^{{\cal T}/2}d\uptau'\, e^{i\omega T}\left(e^{ij\Delta\uptau}+e^{-ij\Delta\uptau}\right) \sum^{n-1}_{k=0}\frac{1}{(\Delta \uptau-i\epsilon)^2-g^2_{kn}},
\end{equation} where $g_{kn}$ again becomes \eqref{gkn definition}.

From the function of regularization \eqref{regularization function}. Therefore, we have to calculate \begin{align}\label{compute of Yi}
    \mathcal{Y}_I(\omega, j, {\cal T},\Delta r,\Delta z, \Delta\varphi)=\frac{-\mathcal{T}^4}{8\pi^2}\sum^{n-1}_{k=0}&\int_{-\infty}^{\infty}d(\Delta\uptau)\, \frac{\left(e^{ij\Delta\uptau}+e^{-ij\Delta\uptau}\right)}{(\Delta \uptau-i\epsilon)^2-g^2_{kn}}\nn\\
    &\times\int_{-\infty}^{\infty}dT\frac{e^{i\omega T}}{\left[T^2-(\Delta\uptau+i\mathcal{T})^2\right]\left[T^2-(\Delta\uptau-i\mathcal{T})^2\right]}.
\end{align}

Thus, from the result \eqref{result of T calculo}, we have \begin{align}\label{Yi serie}
     \mathcal{Y}_I(\omega, j, \mathcal{T},\Delta r,\Delta z,\Delta \varphi)=\frac{-\mathcal{T}^3e^{-|\omega|\mathcal{T}}}{32\pi}\sum^{n-1}_{k=0}\left[A_k(j+|\omega|)+B_k(-j-|\omega|)+A_k(-j+|\omega|)+B_k(j-|\omega|)\right].
\end{align}

Using \eqref{A plus B} we can calculate \eqref{Yi serie}. In this way \begin{align}
   \nn \mathcal{Y}_I(\omega,j, \mathcal{T},\Delta r, \Delta z, \Delta \varphi)=&\frac{\mathcal{T}^2e^{-|\omega|\mathcal{T}}}{16}\sum^{n-1}_{k=0}\frac{1}{g^2_{kn}(g^2_{kn}+\mathcal{T}^2)}\Bigg\{\bigg[\mathcal{T}\cos(j g_{kn})\Big(g_{kn}\sin(|\omega|g_{kn})-\mathcal{T}\cos(|\omega|g_{kn})\Big)\\
   \nn &+g^2_{kn}+\mathcal{T}^2\bigg]\gamma_++\mathcal{T}\sin(j g_{kn})\Big[g_{kn}\cos(|\omega|g_{kn})+\mathcal{T}\sin(|\omega|g_{kn})\Big]\gamma_-\Bigg\}\\\label{YI function result1}
    &+\frac{\mathcal{T}^2}{8}\left(e^{j\mathcal{T}}\Theta(-j-|\omega|)+e^{-j\mathcal{T}}\Theta(j-|\omega|)\right)\sum^{n-1}_{k=0}\frac{1}{g^2_{kn}+\mathcal{T}^2},
\end{align} where we have define \begin{align}
    \gamma_+=\sign(j+|\omega|)+\sign(-j+|\omega|)\\
    \gamma_-=\sign(j+|\omega|)-\sign(-j+|\omega|)
\end{align} and analyzing the expression for $g_{kn}$ in \eqref{gkn definition}, we see that it vanishes when $\Delta r^2=\Delta z^2=0$ and $\Delta\varphi/2+\pi k/n$ is multiple of $\pi$ and if we take into account that $\varphi\in [0,2\pi/n]$ then $\pi k/n+\Delta\varphi/2\in [\pi(k-1)/n,\pi(k+1)/n]$, so that three possible conditions are obtained in which $g_{kn}$ vanishes. These are when $k=0$ then $\Delta \varphi$ has to be $0$ or when $k=1$ then $\Delta \phi= -2\pi/n$ or also when $k=n-1$ then $\Delta\varphi=2\pi/n$. Thus, making the limit $g_{kn}\rightarrow 0$ for any of the terms in the sum that satisfies these 3 conditions, we have  \begin{align}\label{YI function result2}
   \frac{e^{-|\omega|\mathcal{T}}}{32}\bigg[\Big(\mathcal{T}^2j^2+(|\omega|\mathcal{T}+1)^2+1\Big)\gamma_+ +2j\mathcal{T}(|\omega|\mathcal{T}+1)\gamma_-\bigg]+\frac{1}{8}\Big(e^{j\mathcal{T}}\Theta(-j-|\omega|)+e^{-j\mathcal{T}}\Theta(j-|\omega|)\Big)
\end{align}

\acknowledgments
W. I. would like to thank professor G. Krein for useful discussions and conversations. The authors would like to thank Coordena\c{c}\~ao de Aperfei\c{c}oamento de Pessoal de N\'ivel Superior CAPES (Brazilian agency) for financial support. 
J. B. would like to thank Vicerrectorado de Investigación Universidad Nacional de Ingeniería (VRI-UNI) for financial support Project number FC-PF-34-2021.



 \bibliographystyle{JHEP}
 \bibliography{biblio.bib}







\end{document}